\documentclass[9pt]{elife} 

\usepackage{lipsum} 
\usepackage[version=4]{mhchem}
\usepackage{siunitx}
\usepackage{color,soul} 
\usepackage{todonotes}
\usepackage{multirow}
\usepackage{pdflscape}

\title{Temporal processing and context dependency in  \textit{C. elegans} mechanosensation}

\author[1]{Mochi Liu}
\author[2]{Anuj K Sharma}
\author[1,2]{Joshua W Shaevitz}
\author[2,3*]{Andrew M Leifer}

\affil[1]{Lewis-Sigler Institute for Integrative Genomics, Princeton University, USA}
\affil[2]{Department of Physics, Princeton University, Princeton University, USA}
\affil[3]{Princeton Neuroscience Institute, Princeton University, USA}

\corr{leifer@princeton.edu}{AML}

\begin{document}
\maketitle

\begin{abstract}
A quantitative understanding of how  sensory signals are transformed into motor outputs places useful constraints on brain function and helps reveal the brain's underlying computations. We investigate how the nematode \textit{C. elegans} responds to time-varying mechanosensory signals using a high-throughput optogenetic assay and automated behavior quantification. In the prevailing picture of the touch circuit, the animal's  behavior is  determined  by which neurons are stimulated and by the stimulus amplitude. 
In contrast, we find that  the behavioral response is tuned to temporal properties of mechanosensory signals, like its integral and derivative, that extend over many seconds. Mechanosensory signals, even in the same neurons, can be tailored to elicit different behavioral responses. Moreover, we find that the animal's response also depends on its behavioral context. Most dramatically, the animal ignores all tested mechanosensory stimuli during turns. Finally, we present a linear-nonlinear model that predicts the animal's behavioral response to stimulus.
\end{abstract}

\section{Introduction}
An animal's nervous system interprets sensory signals to guide behavior, including to evade predation.  Temporal properties of a stimulus can be important for determining an animal's behavioral response. For example, mice exhibit defensive behaviors in response to looming visual stimuli that increasein size, but not to similar stimuli that decrease in size \citep{yilmaz_rapid_2013}. Investigating how the nervous system processes these  signals is a critical  step towards understanding neural function. 

Mechanosensation in the  nematode \textit{Caenorhabditis elegans} is an attractive platform for investigating  sensorimotor processing. Six soft-touch mechanosensory neurons arranged throughout the body detect mechanical stimuli delivered by a small probe in what is called a touch, or by striking the petri dish containing the animal  in what is called a tap \citep{chalfie_developmental_1981}.
In the prevailing picture of the touch circuit, the animal's  behavior is  determined entirely  by which neurons are stimulated and by the stimulus amplitude. 
 Despite decades of  investigation, however, the behavioral response to dynamic time-varying mechanosensory signals  has not been fully explored. 
 
 Here we revise the prevailing picture of the mechanosensory response system by quantitatively exploring the animal's detailed behavioral response to rich, dynamically varying signals. 
 In contrast to the prevailing picture, we find that the animal responds to temporal features of signals in its mechanosensory neurons, like its time-derivative, that extend over many seconds.   Moreover, we find evidence that the animal's sensorimotor response  depends on the  animal's current behavior state. 
 That we find temporal processing and context dependency even in the nematode's relatively simple touch circuit, raises the possibility that these features could be ubiquitous across sensory systems.  
  Finally, we present a simple quantitative model that predicts the animal's response to novel mechanosensory signals.

Mechanosensation is important for \textit{C. elegans} survival. \textit{C. elegans} are preyed upon by nematophagous fungi,  and touch-defective animals fail to detect and escape from the fungus \citep{maguire_c._2011}. Much is already known about this critical  circuit. The six soft-touch mechanosensory neurons detect both spatially localized and non-localized stimuli. Anterior touches are detected by anterior neurons  ALML, ALMR and AVM and evoke reversal behaviors while posterior touches are detected by posterior neurons PLML and PLMR and evoke forward sprints \citep{chalfie_developmental_1981,chalfie_neural_1985, mazzochette_tactile_2018}.
 Non-spatially localized plate taps are detected by  both anterior and posterior soft-touch neurons and evoke reversals  \citep{chalfie_developmental_1981, rankin_caenorhabditis_1990} and, on rare occasions, forward acceleration  \citep{wicks_integration_1995, chiba_developmental_1990}.  Due in part to its ease of delivery, and its inherent compatibility with high-throughput methods \citep{swierczek_high-throughput_2011}, plate tap has emerged as an assay for studying sensitization and habituation \citep{rankin_caenorhabditis_1990} and the development, circuitry \citep{chalfie_developmental_1981}, genes, molecules and receptors \citep{sanyal_dopamine_2004, kindt_dopamine_2007} of the mechanosensory system. 

When the animal interacts with its environment or brushes up against a nematophagous fungi's constricting ring, it inherently receives time-varying stimuli.   
An individual touch receptor neuron's response to force is  well characterized \citep{ohagan_mec-4_2005}, including to time varying stimuli \citep{eastwood_tissue_2015}. The onset and offset of an applied force evokes strong excitatory currents that adapt on a few tens of milliseconds timescale and have a frequency response  thought to peak in the 100 Hz regime \citep{eastwood_tissue_2015}. In contrast to the detailed  understanding at the single neuron level,  the animal's downstream response to rich temporally varying mechanosensory signals has not been explored. 
 
The animal's  behavior response to mechanosensory stimuli has primarily been studied in response to impulse stimuli. These are brief applications of touch, tap or optogenetic stimulation whose most salient feature is the stimulus amplitude, not its temporal profile.  In the classical touch assay, for example, a saturating force is applied  lasting a few tenths of a second  \citep{nekimken_forces_2017}. Tap stimuli are even shorter in duration. In impulse-like experiments, it is the stimulus amplitude (force, indentation or optogenetic intensity) that determines  behavioral response \citep{petzold_mems-based_2013, stirman_real-time_2011, mazzochette_tactile_2018}. 
To our knowledge,  the only work investigating behavioral responses to temporally varying stimuli  involves  trains of  taps or touches \citep{chiba_developmental_1990,kitamura_contribution_2001}, trains of optogenetic pulses \citep{porto_reverse-correlation_2017, leifer_optogenetic_2011} or the delivery of a 1 kHz  acoustic vibration \citep{nagy_measurements_2014, nagy_homeostasis_2014}.

  
Similarly, the quantification of behavior responses to these stimuli have focused on a few behaviors that were chosen \textit{a priori}, and usually related to reversals.  Early work  scored the animal's reversals \citep{chiba_developmental_1990} and accelerations \citep{wicks_integration_1995}  and more recent work includes reversal distance \citep{kitamura_contribution_2001},  rate of reversals \citep{swierczek_high-throughput_2011} or pauses, reversal duration and reversal latency \citep{ardiel_habituation_2017}.  Yet the animal's repertoire of behavior is known to be larger \citep{stephens_dimensionality_2008, brown_dictionary_2013}.

The  picture that emerges from these studies is one where  behavior depends solely on the set of neurons stimulated and the stimulus strength. The  location of an applied force determines which  touch receptor neurons  are activated and thus whether the animal accelerates or reverses, while the amplitude of the applied stimulus determines the probability that the animal responds at all \citep{driscoll_mechanotransduction_1997, petzold_mems-based_2013, mazzochette_tactile_2018}. 
Now, however, this picture is coming under greater scrutiny. 
Recently, Porto et al.~report the use of reverse correlation and a binary optogenetic stimulus to present evidence that temporal processing is  important for the animal's behavioral response  \citep{porto_reverse-correlation_2017}.
In our work here, we show that the nervous system processes  signals from the mechanosensory neurons as a timeseries over many seconds. We find that the animal's  behavior response depends on higher order temporal features like the derivative of those mechanosensory signal, and also depends on the animal's own behavioral context.

Here we revisit the animal's behavioral response to mechanosensory stimulation armed with high-throughput optogenetic methods for delivering time varying stimuli and  improved techniques for measuring  animal posture  \citep{stephens_dimensionality_2008} and behavior  \citep{berman_mapping_2014}. Using reverse correlation \citep{ringach_reverse_2004, schwartz_spike-triggered_2006, gepner_computations_2015} we analyze over 8,000 animal-hours of  recordings (three orders of magnitude greater  than previous investigations) and find 
new insights into the interplay between sensory processing and behavior.


\section{Results}

\begin{figure}

\includegraphics[width=.8\textwidth]{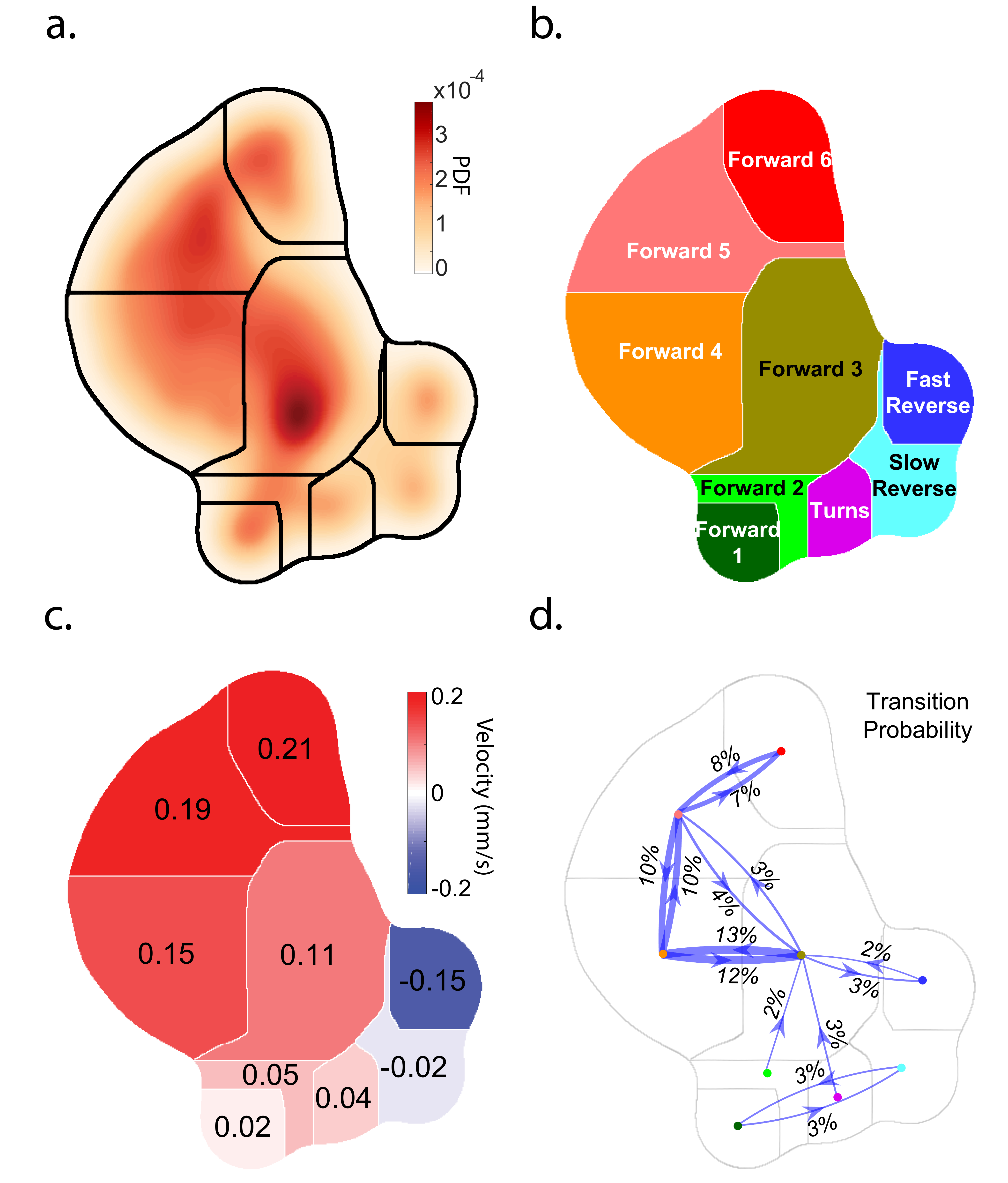}
\caption{\textit{C. elegans} behavior quantification. \textbf{a.)} Behavior map shows the probability density of  posture dynamics observed during 2,284 animal-hours of behavior (``Random Noise'' row in \autoref{tab:experimentsTable}). Posture dynamics are high dimensional but are projected down into a low-dimensional space using the t-SNE method as in \citep{berman_mapping_2014}. Peaks indicate stereotyped postures.   Discrete behavior states are defined by dividing the posture map into nine regions via  a watershedding algorithm. \textbf{b.)} Human-readable behavior names are provided by the experimenters. \textbf{c.)} Mean center of mass velocity for animals in each region is shown. Positive velocity is in the direction of the animal's head. 
\textbf{d.)} Probability of transitioning between behaviors is shown. Thickness of lines scales with probability. Transition probabilities < 2\% were omitted.}
\label{fig:introBehavior}
\figsupp{Analysis pipeline for classifying behavior.}{\includegraphics[width=\textwidth]{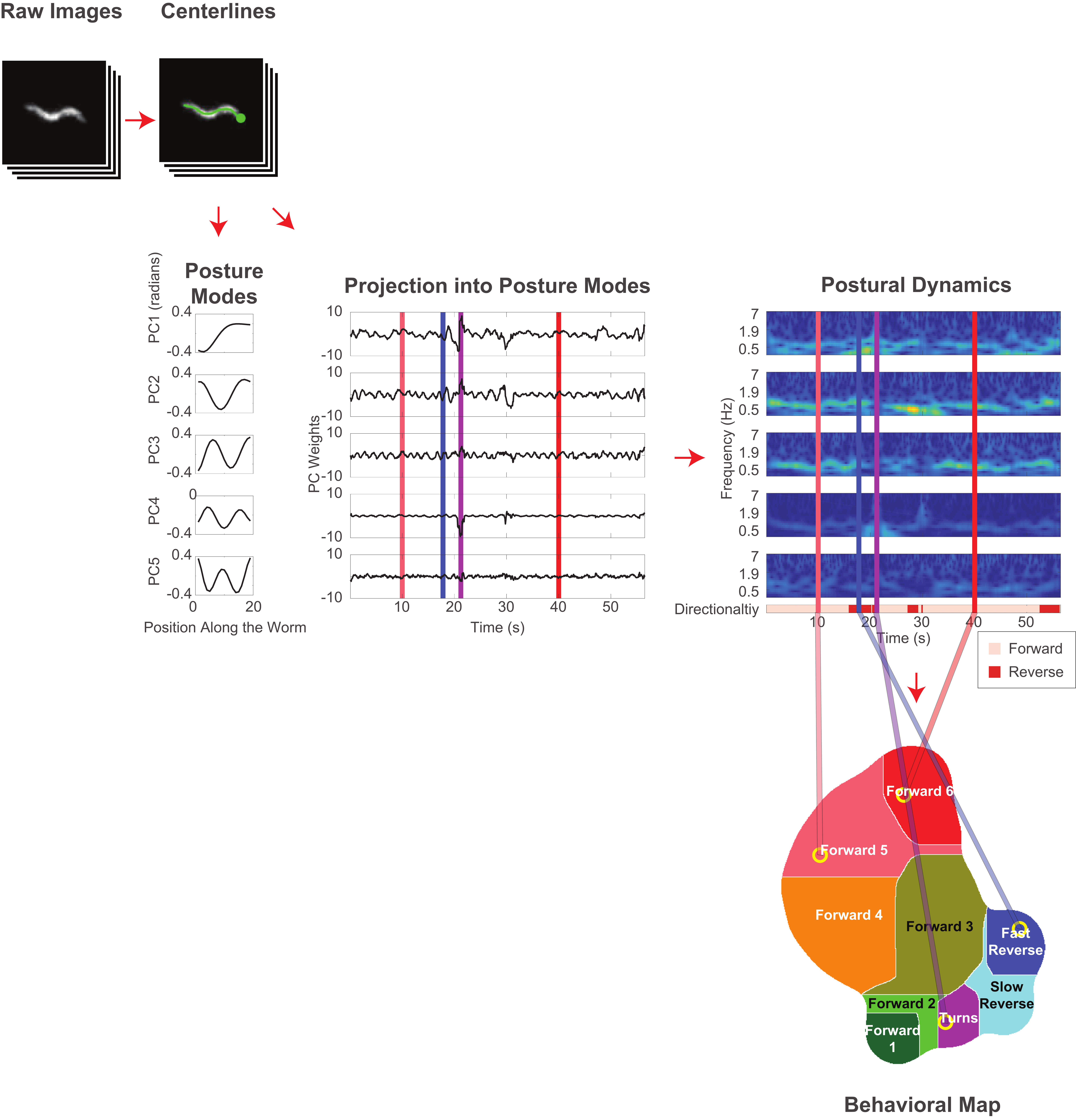}}{Behavior is mapped and classified according to the animal's posture dynamics, similar to in  \citep{berman_mapping_2014}. Images of \textit{C. elegans} are segmented to extract the animal's centerline.  Each centerline is  projected into a linear combination of  posture modes. The animal's time-varying posture  is represented as a time-series of corresponding weights. Spectrograms of these time-series  describe the animal's postural dynamics at each point in time.  Posture dynamics are mapped into a two-dimensional plane using t-SNE. The animal occupies a different point on the behavior map depending on its postural dynamics, and its placement in this map determines the behavioral classification.}  \label{figs:pipeline}
\figsupp{Behavior map details.}{\includegraphics[width=\textwidth]{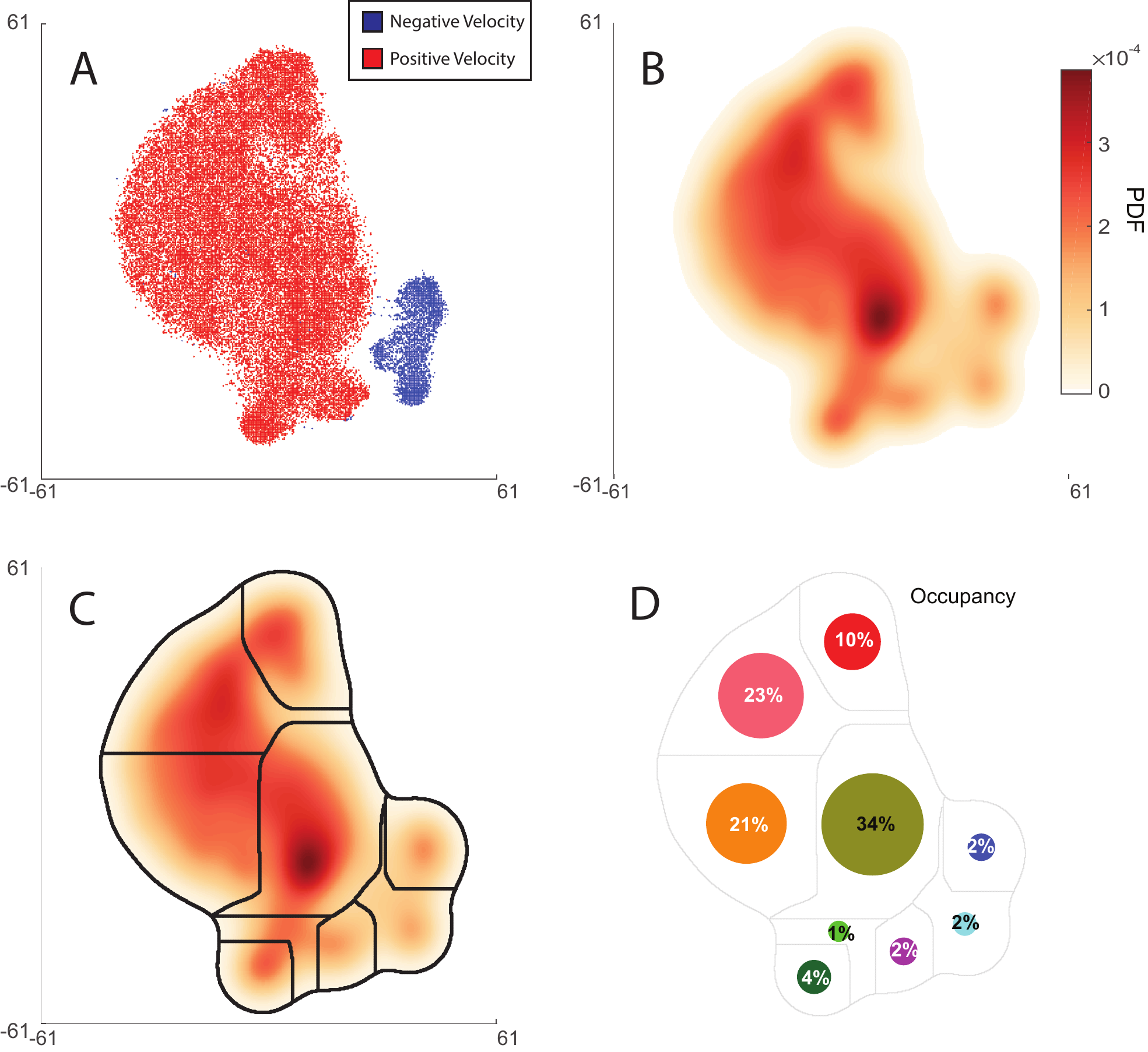}}{ Behavior maps were generated from 2,284 animal-hours of behavior recorded from  P\textit{mec-4::Chrimson} worms during optogenetic stimulation and control conditions.  a)  The sign of the animal's velocity is shown for  55,000  time-points uniformily selected from the recordings.  Distinct  regions in the map  correspond to forward or backward locomotion.    b) Probability density plot shows likelihood that the animal exhibits different behaviors. Peaks in the probability density correspond to stereotyped behaviors.  c) Natural boundaries that separate  stereotyped worm behaviors are found using watershedding. Same as in \autoref{fig:introBehavior}. These regions define  distinct behavior states. d) The probability of occupying a given behavior is shown for animals in an unstimulated condition. Area of circle scales with occupancy probability.  } \label{figs:suppmaps}


\figsupp{Videos of randomly selected animals performing each of the 9 behaviors.}{\includegraphics[width=.8\textwidth]{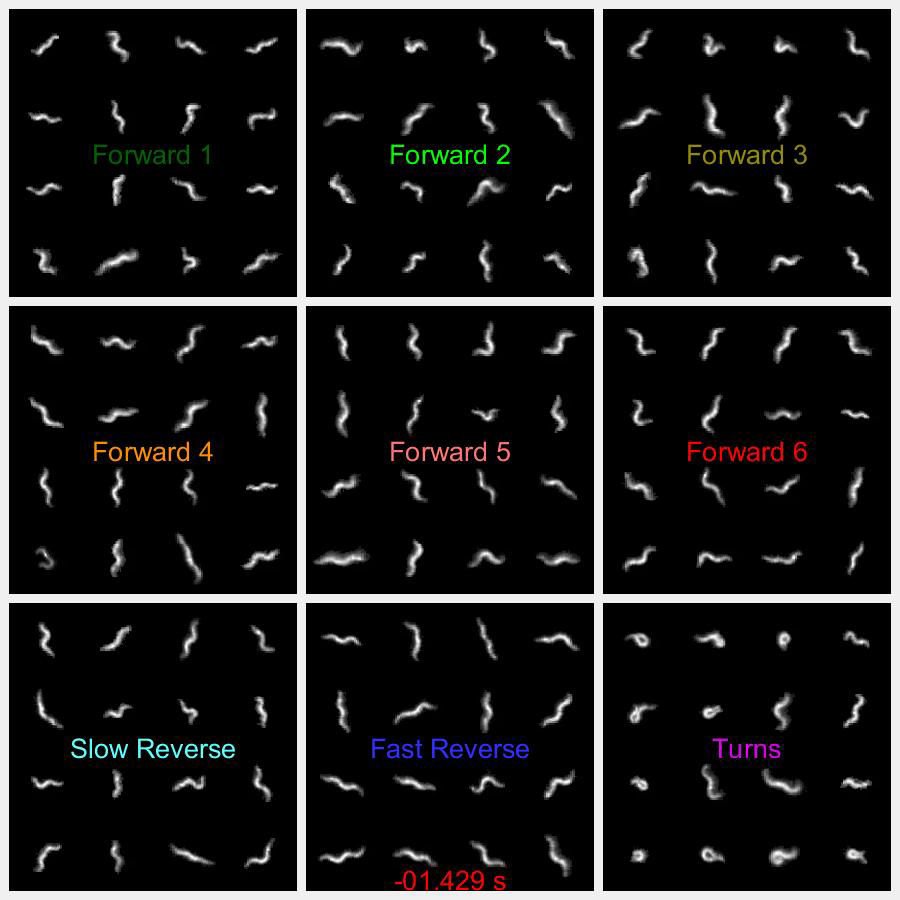}}{ Randomly selected 3-second long examples of animals performing each of the 9 behaviors.  Link to video (\url{https://vimeo.com/259479020})} \label{video:brady}

\figsupp{Video showing path of an animal through behavior space.}{\includegraphics[width=.8\textwidth]{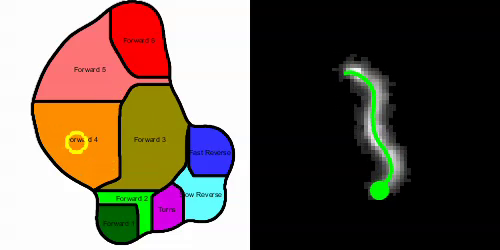}}{ \textbf{right.)} Video of an example animal is shown. Detected centerline (green) is overlaid. Dot denotes head. The animal is kept centered in the video, even though it is moving. \textbf{left.)} Animal's instantaneous behavior is shown (yellow ring) on the behavioral map. Link to video (\url{https://vimeo.com/259479010})} \label{video:trace}
\end{figure}

\subsection{Mechanosensation evokes a range of behavioral responses}
We first  investigated the animal's response to plate tap, a spatially non-localized mechanosensory stimulus generated by tapping the dish containing the animals. Plate taps had  previously been reported to evoke reverse locomotion \citep{rankin_caenorhabditis_1990} and rarely forward accelerations \citep{wicks_integration_1995}.   A solenoid repeatedly delivered a tap stimulus every  60 seconds for  30 minutes to a plate of many Wild-Type (N2) worms,  repeated across 22 plates, resulting in 40,409 total animal-tap presentations. The inter-stimulus interval was chosen to minimize effects of habituation \citep{rankin_mutations_2000}. The animal's behavior was continuously measured and classified using a behavior mapping technique similar to \citep{berman_mapping_2014}. Briefly, statistical inference was performed on all of the animal's posture dynamics to generate a single behavior map. Stereotyped posture dynamics that emerged from this map were defined as behaviors.  Each individual animal's posture dynamics were  projected into this map at each point in time and automatically classified into one of 9 behavior states  that were assigned labels such as Turn. See \autoref{fig:introBehavior}, and \autoref{fig:introBehavior} - Figure Supplements \ref*{figs:pipeline} and \ref*{figs:suppmaps} and methods for a complete description of the behavior mapping. Also see example videos of behavioral mapping in \autoref{fig:introBehavior} - Figure Supplements \ref*{video:brady} and \ref*{video:trace}.

Consistent with previous reports, we observe that taps most dramatically evoke the animal to transition to the Fast Reverse state. Tap stimulus induced a 14-fold increase in the fraction of animals exhibiting  Fast Reverse  immediately post stimuli,  see \autoref{fig:rangeOfBehav}a and \ref*{fig:rangeOfBehav} - Figure Supplement \ref{figs:transitionrates}. 
Additionally,  animals that continued in forward locomotion  exhibited  an overall slowing down, transitioning from fast locomotion states to slower locomotion states, which to our knowledge had not previously been reported.  We also observed a 4.5 fold increase in the fraction of animals exhibiting Turn behavior approximately 5 seconds post stimulus. The fraction of animals exhibiting Slow Reverse also increased slightly upon stimulation.   These measurements suggest that  plate tap evokes a wide-range of behavioral responses in the animal.


\subsection{Optogenetic stimulation mimics tap}
\begin{figure}
\includegraphics[width=\textwidth]{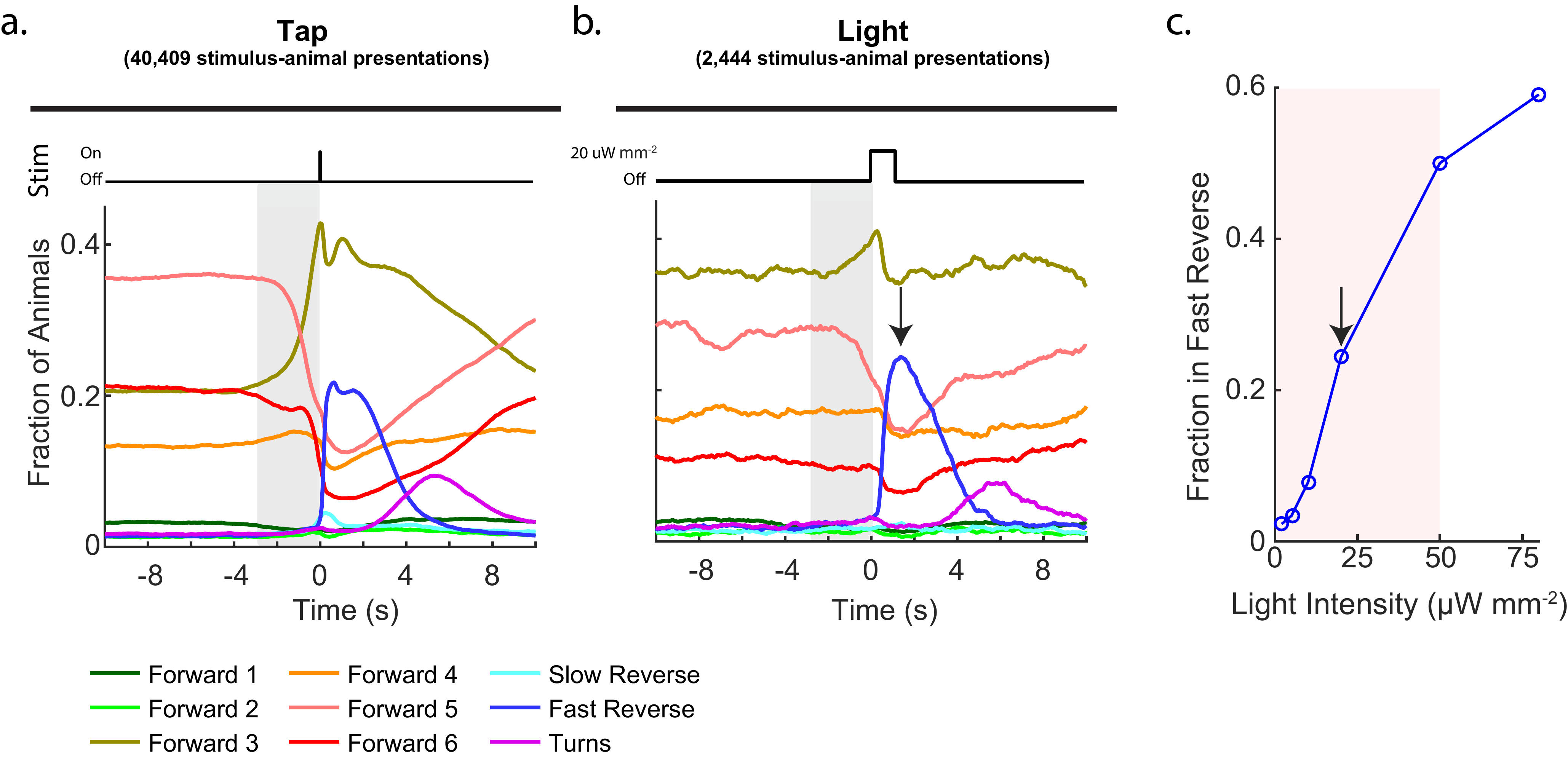}
\caption{ Stimulation evokes a diverse range of behavior responses. \textbf{a.)} Fractions of animals occupying each  behavior state  are shown in response to      a plate tap  (40,409 stimulus-animal presentations) and \textbf{b.)} in response to a 1 second optogenetic light stimulation of the six soft touch mechanosensory neurons (2,444 stimulus-animal presentations, 20 \textmu W mm$^{-2}$). Note the similarity in the behavior responses for light and tap. Gray shaded window indicates inherent  temporal uncertainty in behavior classification. See methods. \textbf{c.)}   Response to optogenetic stimulation depends on light intensity. Peak fraction of animals in the Fast Reverse state in a 6 second window post stimulus are shown for different intensity light pulses.  More than 2,000 stimulus-animal presentations were recorded for each point plotted.  Arrow indicates light intensity used in \textbf{b}. Pink shaded region indicates light range used for subsequent continuous light stimulation experiments, as in \autoref{fig:kernel}. }
\label{fig:rangeOfBehav}
\figsupp{Diagram of high-throughput stimulation and behavior assay.}{\includegraphics[width=.5\textwidth]{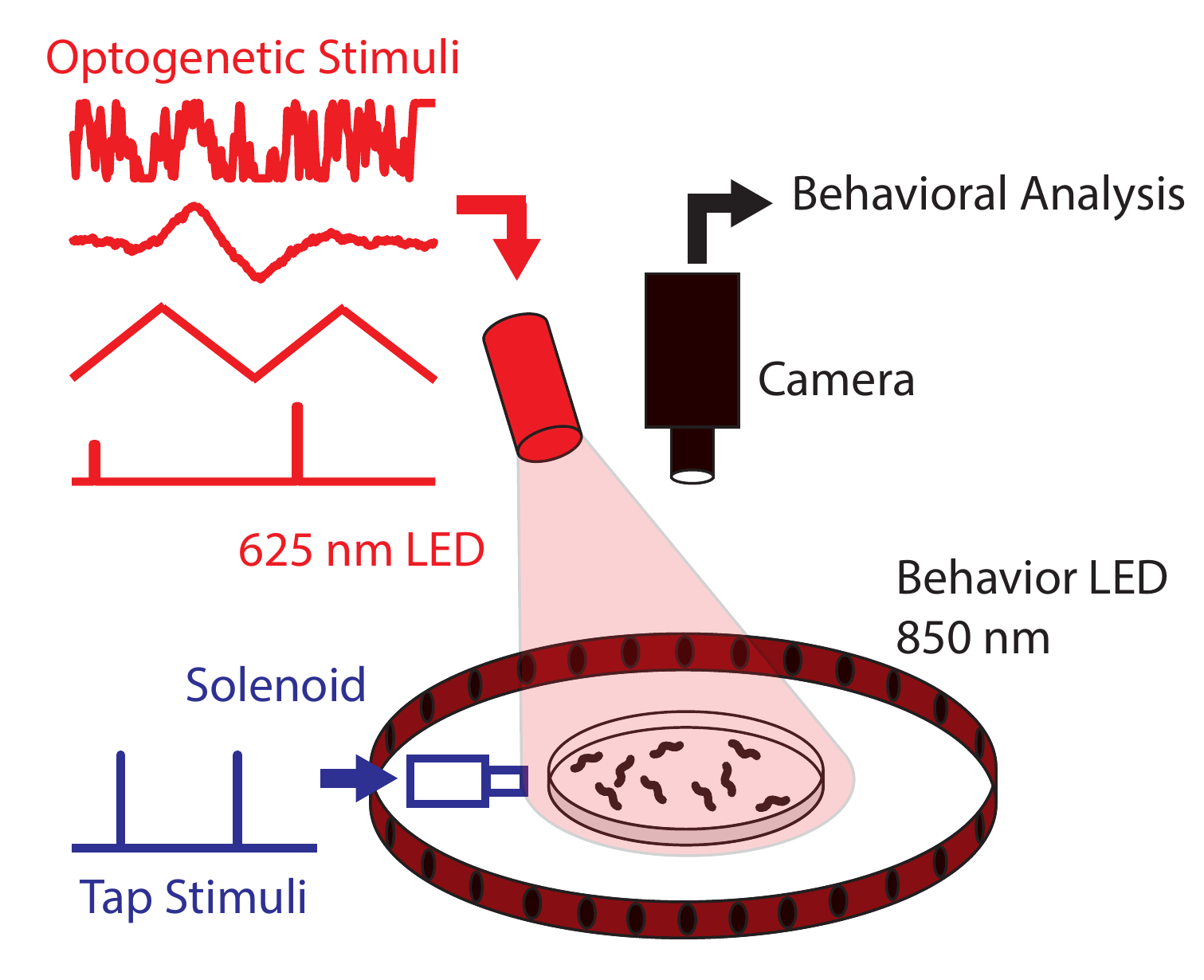}}{Worm behavior is recorded while delivering optogenetic or tap stimulation to a plate containing $63±40$ animals  (mean$\pm$ standard deviation). Optogenetic stimulation is delivered by modulating the light intensity of three 625nm LEDs (only one is shown in the diagram). Taps are delivered to the  plate via a computer controlled solenoid. Recordings last 30 minutes per plate, and each experimental series consists of many plates, see \autoref{tab:experimentsTable}.  } \label{figs:expdesign}
\figsupp{Transition rates for tap and light stimulation.}{\includegraphics[width=.9\textwidth]{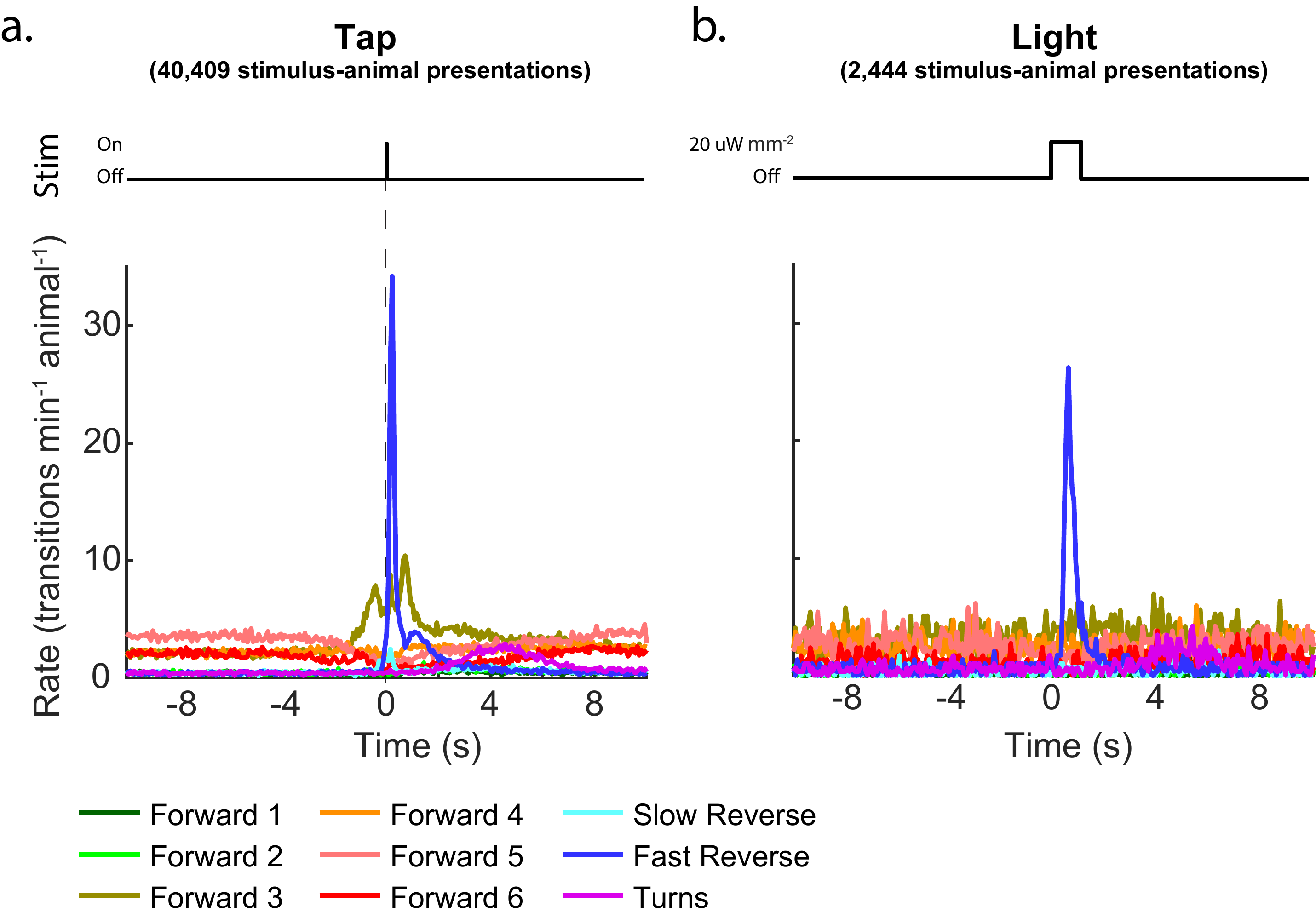}}{ The rate of transitions into each behavior is shown aligned to a tap or 1 s optogenetic light stimulus. Panels \textbf{a} and \textbf{b} correspond to the occupancy plots in \autoref{fig:rangeOfBehav}a and b.} \label{figs:transitionrates}
\figsupp{Control animals grown without ATR are light insensitive.}{\includegraphics[width=.9\textwidth]{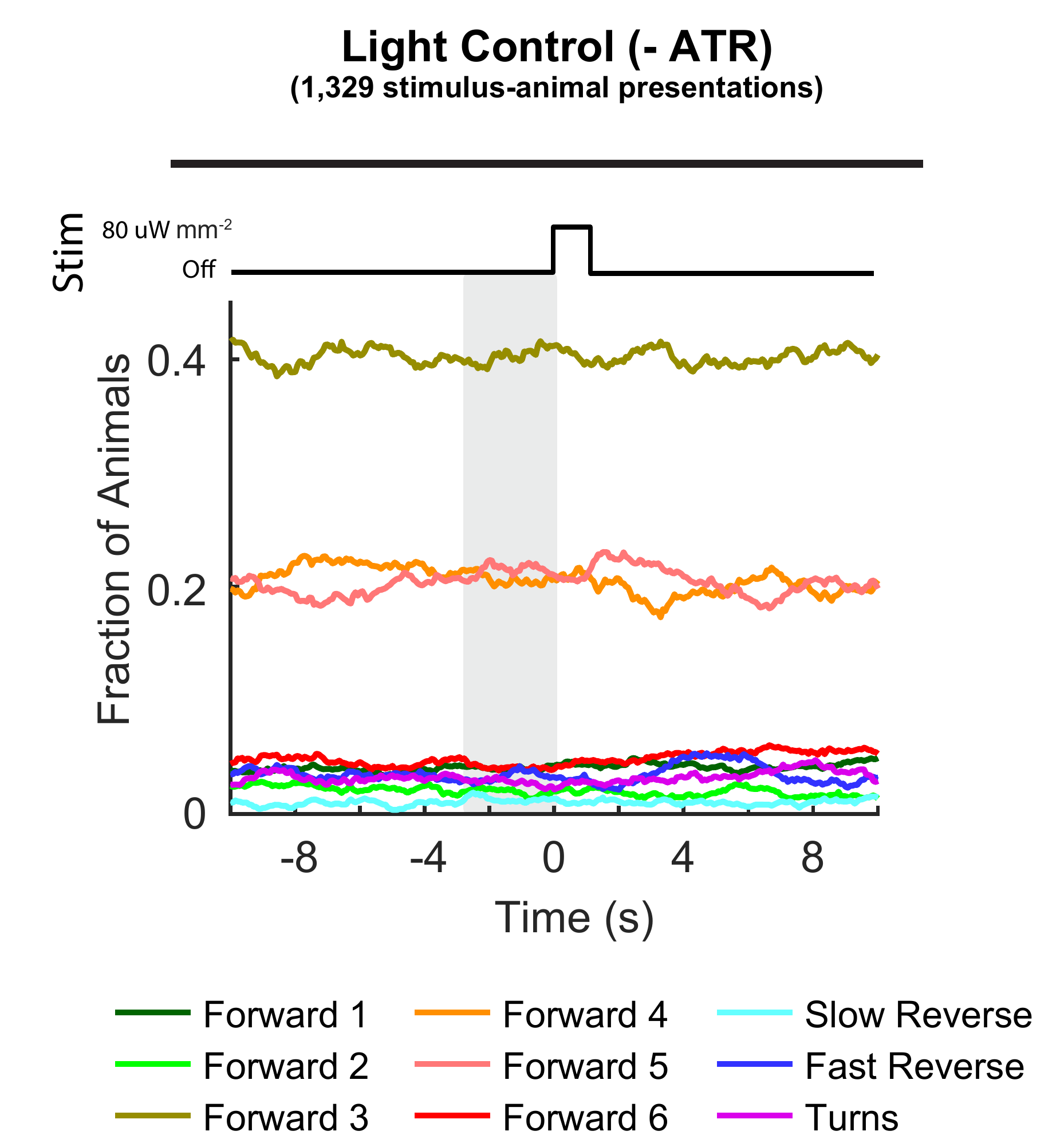}}{ Fractions of control AML67 animals occupying each  behavior state are shown in response to a light pulse stimulus. Animals grown without all-trans-retinal (- ATR) do not respond to light even at a light intensity level of 80 \textmu W mm$^{-2}$. } \label{figs:lightpulsecontrol}
\figsupp{Tap sensitivity of transgenic animals is reduced compared to wildtype.}{\includegraphics[width=.9\textwidth]{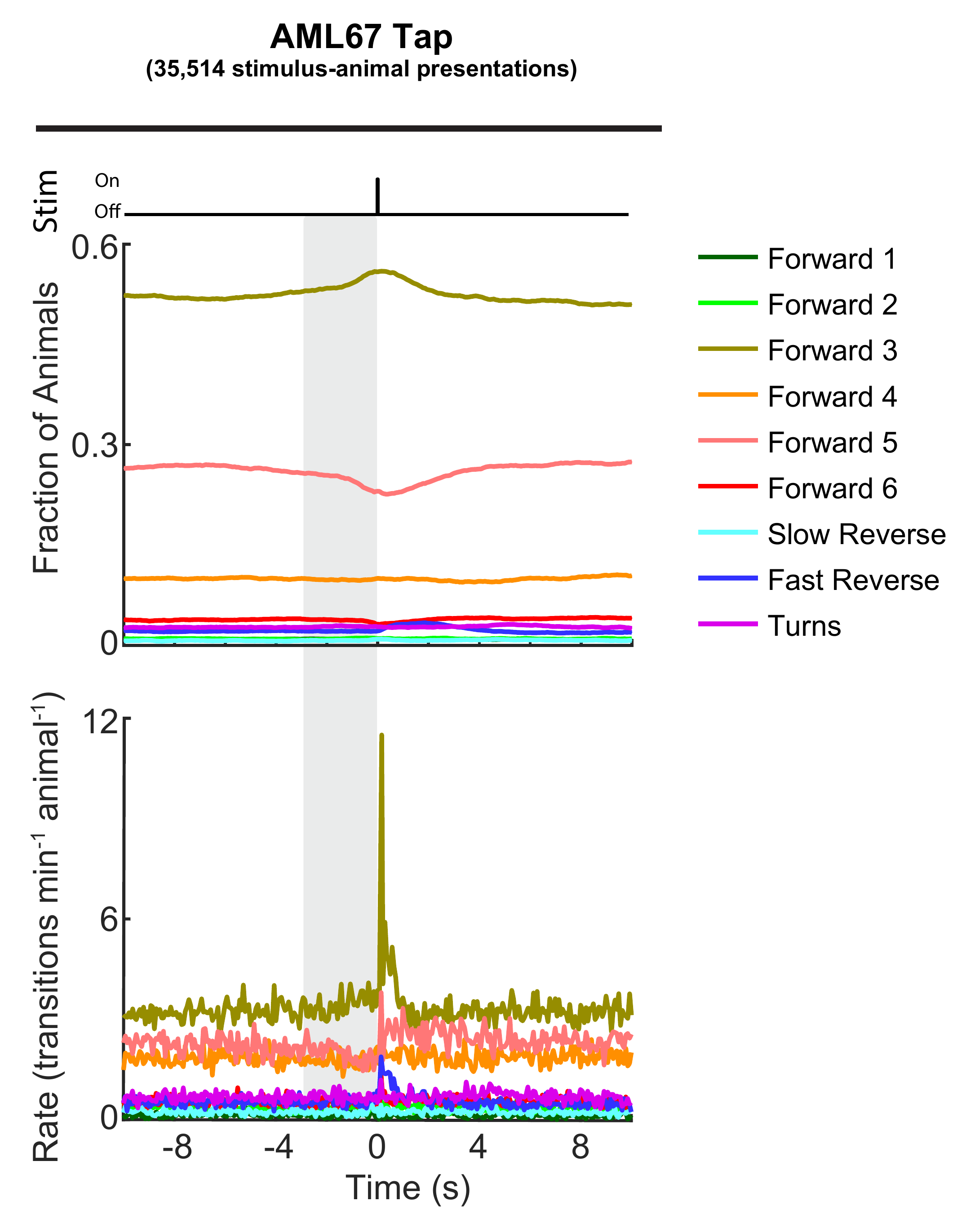}}{ Fractions of AML67 animals occupying each  behavior state and transition rates into each behavior are shown in response to a mechanical tap stimulus.  Recordings from both ATR+ and ATR- conditions are pooled together.  AML67 animals show decreased responsiveness to tap stimulation compared to wild-type presumably because the exogenous mec-4 promoter sequences deplete the endogenous mec-4 transcription factor.} \label{figs:AML67tap}
\end{figure}

We sought to activate the mechanosensory circuit optogenetically because optogenetic stimulation is more amenable to modulation and control. Optogenetic stimulation of the six mechanosensory neurons had previously been shown to evoke reversals and accelerations, similar to tap \citep{nagel_light_2005}. We wondered whether the details of the behavior response to tap that we observed are also present in response to optogenetic activation.  Animals expressing the light gated ion channel Chrimson in the soft touch mechanosensory neurons  (strain AML67 [P\textit{mec-4}::\allowbreak\textit{Chrimson4.2}::\allowbreak\textit{SL2}::\allowbreak\textit{mCherry}::\allowbreak\textit{unc-54}]) were illuminated with red light  for 1 second with a 60 second inter-stimulus interval (2,444 stimulus-animal presentations, 20 \textmu W/mm$^2$,  selected to be in a region of high behavior sensitivity, see \autoref{fig:rangeOfBehav}c).  Consistent with previous reports, light stimulation evoked a behavior response that was quantitatively similar to that of the plate tap, see \autoref{fig:rangeOfBehav}b and required the cofactor all-trans retinal (ATR) \ref*{fig:rangeOfBehav}-Figure supplement \ref{figs:lightpulsecontrol}. For both light and tap, the most salient response was a dramatic increase in animals in Fast Reverse. Both light and tap also evoked an increase in occupancy in Forward 3 and both evoked similar decreases in Forward 4, 5 and 6. Both light and tap also evoked an increase in Turns  that peaked  5 seconds post-stimulus. Hence,  optogenetic stimulation of mechanosensory neurons   evoke  detailed behavior responses similar   to that of a mechanical stimulus. This suggests that our optogenetic stimulation generates physiologically reasonable signals in the mechanosensory neurons and we therefore proceeded to explore the animal's response to optogenetic stimulation.

\begin{figure}
\includegraphics[width=.75\textwidth]{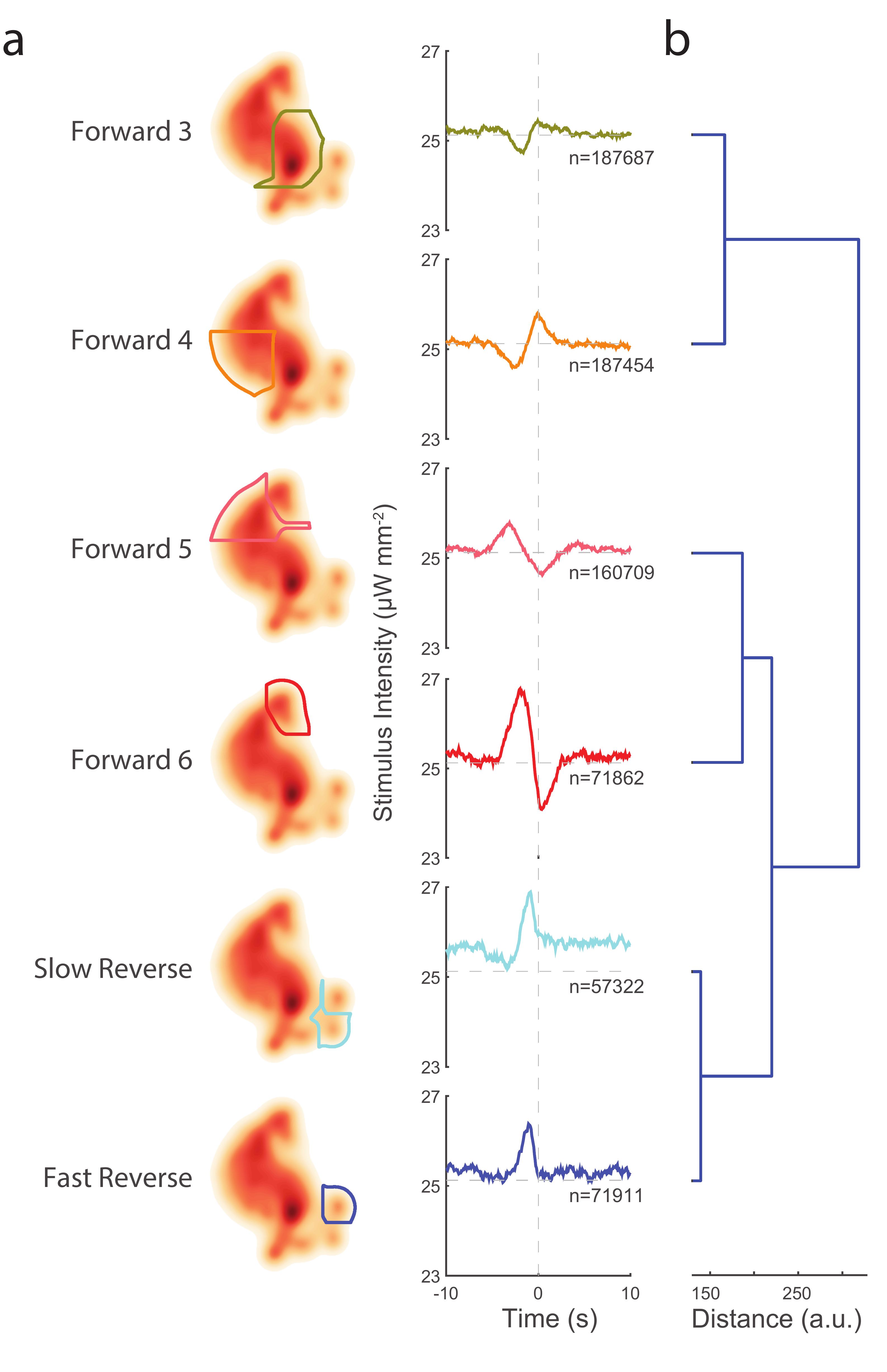}
\caption{Transitions into behavior states 
are tuned to 
higher order temporal features of the stimulus like the derivative.  \textbf{a.)} Random noise time-varying light stimulus is delivered to a population of animals.   Behavior-triggered averages (also referred to as kernels) are calculated for transitions into each behavior state from 1,784 animal-hours of recordings. Each behavior triggered average describes features of the stimulus that correlate with that behavior transition. Only those behavior triggered averages  that pass a significance test compared to a shuffled stimuli are shown.  The shape of the behavior triggered average depends on the behavior. Note that some behaviors have gaussian like shapes, while others have biphasic shapes that act like derivatives.  The number of observed transitions, $n$, into each behavior is listed.  \textbf{b.)} Similar behaviors have similar behavior-triggered averages. Hierarchical clustering was performed on the euclidian distance of  the scaled behavior-triggered averages. Dendrogram is shown. The two reversal states, for example, form a cluster. }
\label{fig:kernel}

\figsupp{Change in behavioral occupancy evoked by random noise stimulation.}{\includegraphics[width=\textwidth]{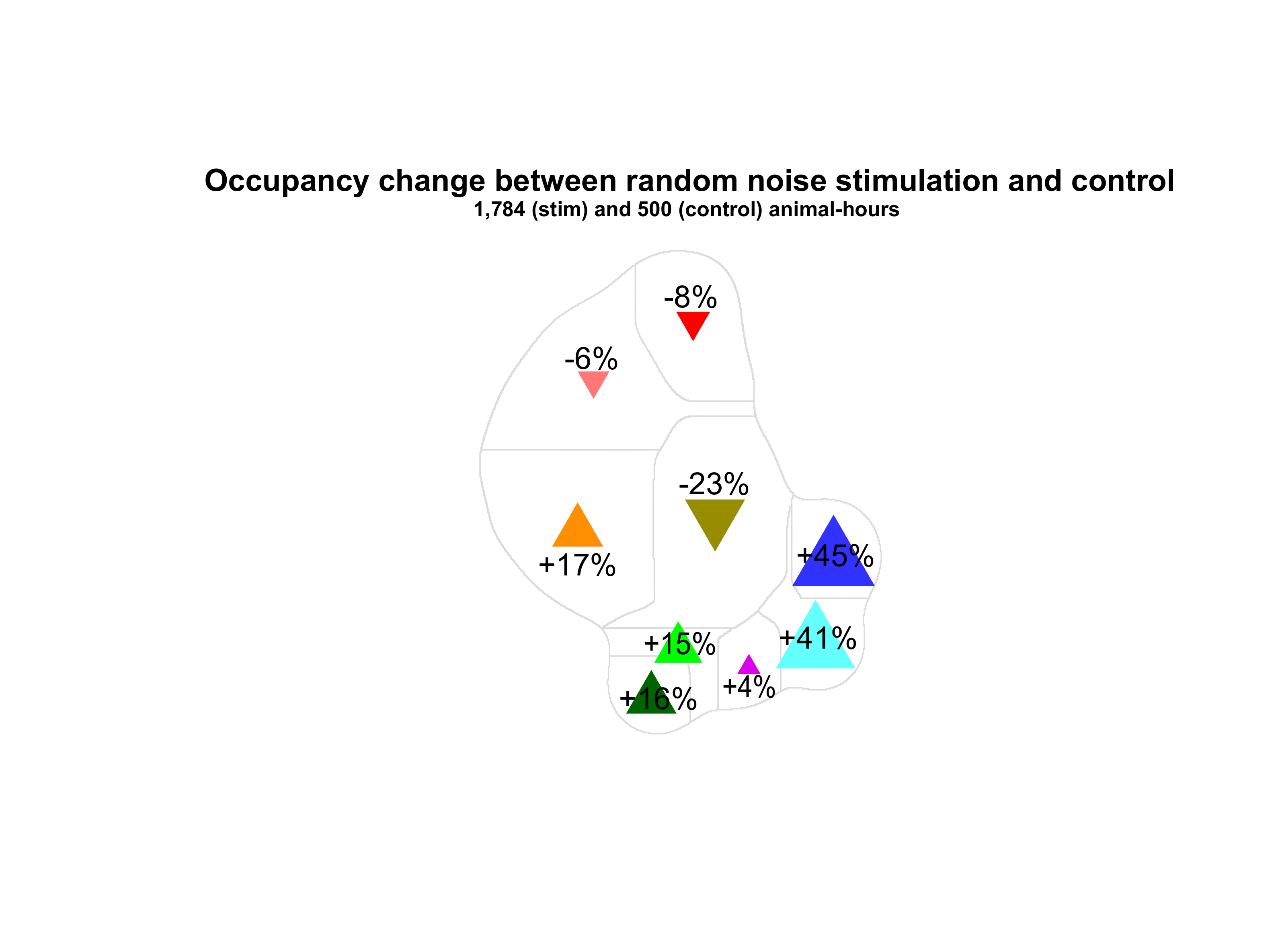}} { The change in occupancy during random noise optogenetic  light stimulation  (1,784 animal hours)  compared to  no-retinal control (500 animal-hours) is shown. Baseline occupancy during no-retinal control is shown in \autoref{fig:introBehavior} - Figure Supplement \ref*{figs:suppmaps}d.   } \label{figs:ATRdiff}
\figsupp{Behavior triggered averages and non-linearities for all behaviors.}{\includegraphics[width=.8\textwidth]{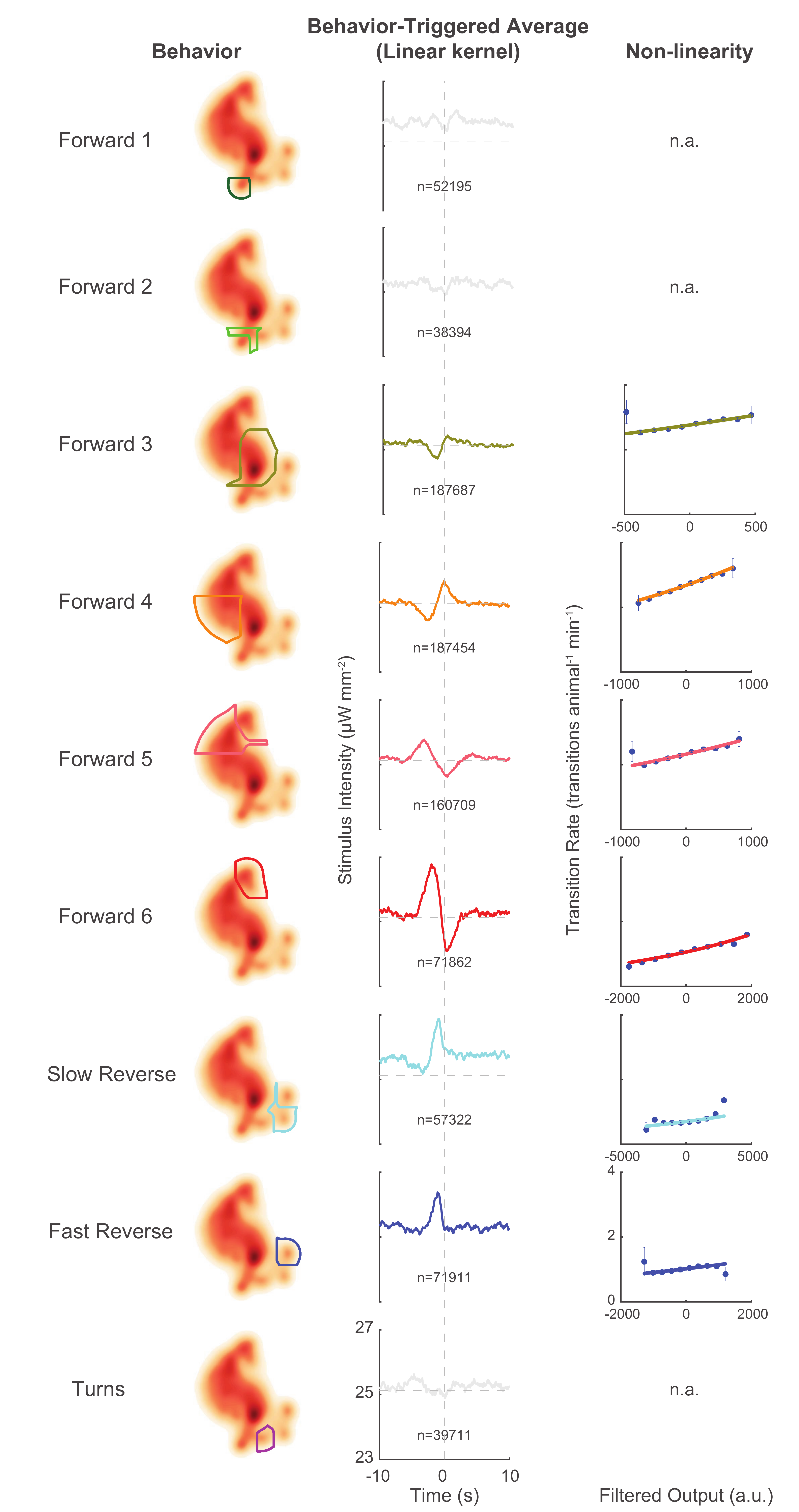}} { Behavior triggered averages and associated non-linearities are shown for transitions into all 9 behavior states. Those behavior triggered averages that fail to pass a shuffled significance threshold are shown in light gray. Non-linearities are only calculated for behaviors whose behavior-triggered averages pass our shuffled significance test.  Note that non-linearites are mostly well-approximated by a line, consistent with the observation in \autoref{fig:rangeOfBehav}c that the animal responds roughly linearly in our stimulus regime.} \label{figs:nonlinearity}
\figsupp{Behavior triggered averages for control animals grown without ATR.}{\includegraphics[width=.5\textwidth]{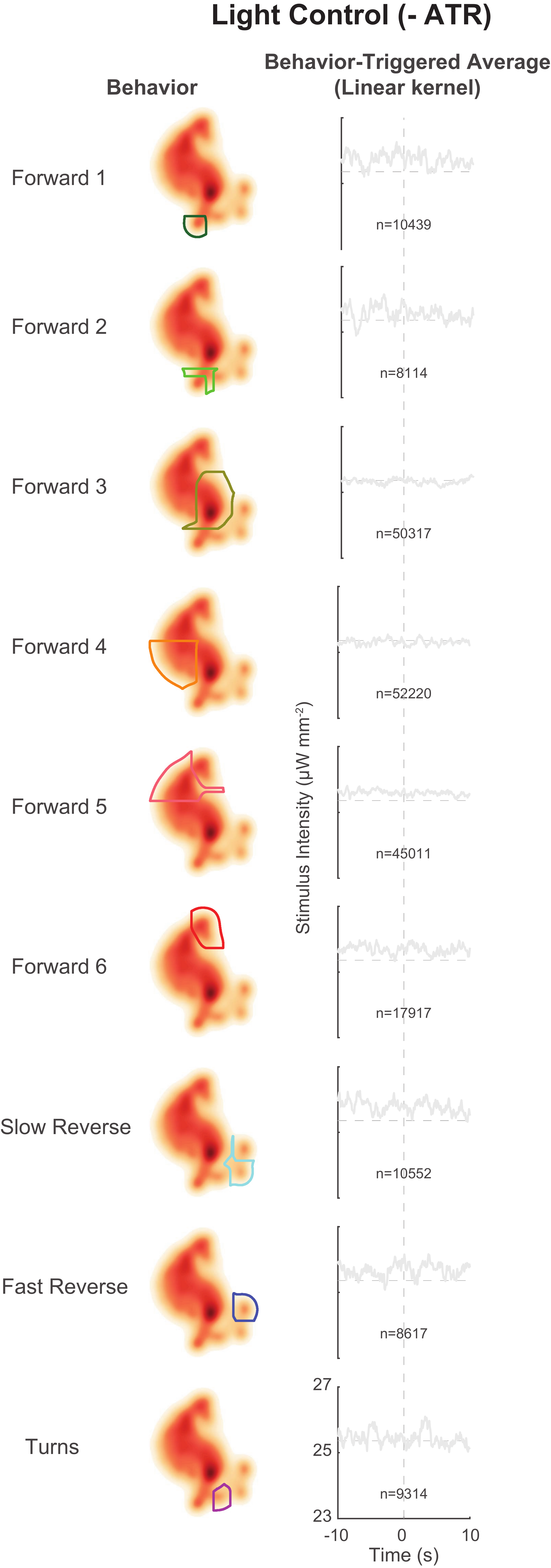}} { Behavior triggered averages  are shown for transitions into all 9 behavior states for control animals grown without the required cofactor all-trans retinal (ATR). As expected, none of the kernels pass a shuffled significance threshold.  Consequently  non-linearities were not calculated.} \label{figs:noretkernels}
\figsupp{Power spectra of a single instantiation of the random noise stimulus.}{\includegraphics[width=.8\textwidth]{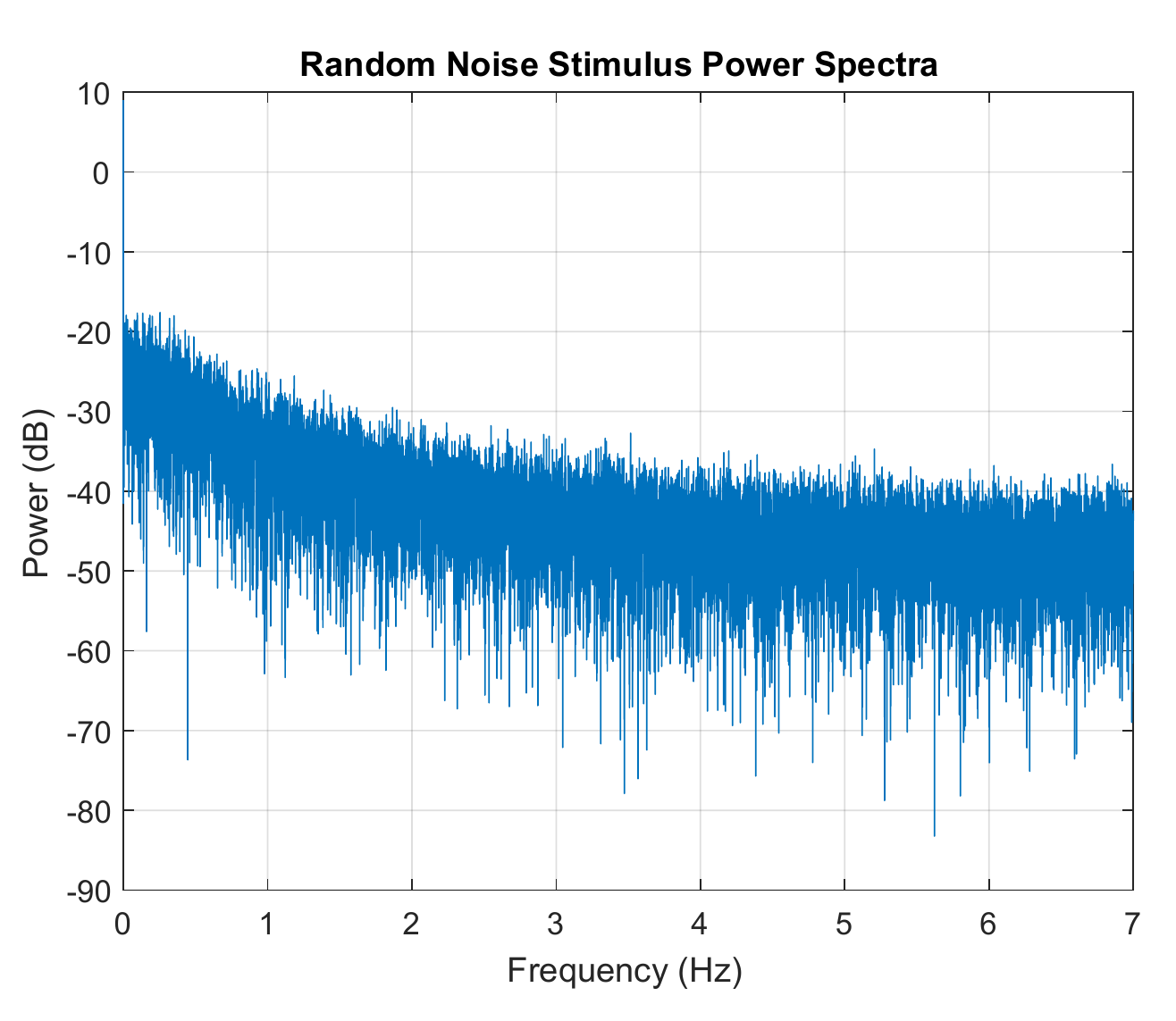}} {The MATLAB periodogram function is used to generate the power spectra of the random noise stimulus for a 30 minute experiment.} \label{figs:powerspectra}

\end{figure}

\subsection{Behavioral responses are correlated to temporal features like the derivative}
When the animal explores its natural environment, crawls through crevices, and interacts with other organisms, it likely experiences time varying mechanical stimuli. 
Therefore we sought to investigate the animal's response to random temporally varying optogenetic stimulation. We find  that the animal's specific behavioral response correlates with higher order temporal features of the stimulus, not merely the amplitude.

To deliver rich temporally varying stimuli, we continuously presented a plate of transgenic animals with light  modulated by broad frequency noise (7 Hz nyquist limit, 0.5 s correlation time, 25 \textmu W mm$^{-2}$ average intensity, min 0, max 50 \textmu W mm$^{-2}$, see power spectra \autoref{fig:kernel} - Figure Supplement \ref*{figs:powerspectra}
). Noise stimulation evoked a wide range of behavioral responses, see \autoref{fig:kernel} - Figure Supplement \ref*{figs:ATRdiff}. We used  reverse correlation to identify the salient features of the  stimulus that correlates with transitions into each behavior.  Reverse correlation yields kernels  that describe how a behavior is tuned to a stimulus.  Kernels are particularly powerful in the context of the linear non-linear (LN) model, a ubiquitous model in neuroscience that can be used to predict a nervous systems' stimulus response \citep{ringach_reverse_2004,schwartz_spike-triggered_2006, coen_dynamic_2014, gepner_computations_2015,hernandez-nunez_reverse-correlation_2015, calhoun_quantifying_2017, clemens_use_2017}. See in particular \citep{gepner_computations_2015}. Briefly, the LN model treats the response to a stimulus as a stochastic process  involving two steps: first the stimulus timeseries $s(t)$ is convolved with a kernel $A$ (linear operation),  and then it is transformed into a response probability $P$ via a non-linear look-up function $f$ (non-linear operation), such that,  
\begin{equation}
P[\textrm{behavior}](t)=f[(A*s)(t)]; \quad (A*s)=\int_0^\infty A(\tau)s(t-\tau)d\tau.
\end{equation}
The shape of the kernel and non-linearity describes how a behavior response  is tuned to the stimulus. 

Kernels can be estimated by finding the behavior triggered average. Briefly,  the stimulus in a time window centered on a behavior transition is averaged across all such behavior transitions. The mean subtracted and time-reversed behavior triggered average is an estimate of the kernel  and so henceforth we use the terms behavior triggered average and kernel interchangeably.  Once the kernels $A$ are calculated, it is straightforward to estimate the non-linearities $f$ from the observed behavior responses (see methods). Kernels and associated non-linearities were computed for transitions into each of the 9 behavior states from over 50,000 behavior transition events per behavior, see \autoref{fig:kernel} and \autoref{fig:kernel} - Figure Supplement \ref*{figs:nonlinearity}.   Kernels for 6 of the 9 behaviors were found to be significant compared to a shuffled stimuli (see methods).  
In contrast, kernels computed from control animals grown without  all-trans retinal all failed to pass our significance threshold, see \autoref{fig:kernel} - Supplementary Figure \ref*{figs:noretkernels}.  
Non-linearities calculated for the 6 behaviors were found to be mostly linear, suggesting that in our case the kernels themselves capture most of the information about how the nervous system responds to our stimulus.  

Our prior understanding of the mechanosensory circuit makes strong predictions about the shape of the kernels that we should observe. If behavior  depends only on which neurons are activated, then all  kernels should have the same shape, scaled linearly, because we are always activating the same set of neurons. (This assumes all six neurons are activated in a linear regime, which seems reasonable  given the approximately linear response observed in \autoref{fig:rangeOfBehav}c). Moreover, if the probability of response depends only on  instantaneous stimulus amplitude, then we further expect all kernels to be narrow gaussians. In contrast to these predictions,  we see a wide diversity of kernels.  Forward locomotion kernels have biphasic waveforms, not at all like gaussians.  Forward 6, for example has the shape of a differentiator suggesting  that the transitions into Forward 6  correlate with decreasing stimuli on a 7 second timescale.    Kernels for Slow Reverse and Fast Reverse, on the other hand, do look like gaussians, consistent with the interpretation that  \textit{reversals} do depend on the stimulus amplitude. Interestingly,  the gaussians are wide, which suggests that the animal may integrate sensory signal over three to four seconds in determining to reverse.

Taken together, we  conclude that the animal's behavior response is not merely correlated with which neurons are stimulated and the stimulus amplitude.  Instead  different behaviors  correlate with different temporal features of signals in the mechanosensory neurons, even though the same six neurons were always activated. The behavioral response correlates with  properties of the stimulus like the derivative or the integral, not just the amplitude.

 \subsection{Similar behavioral responses are tuned to similar stimuli}
We wondered about the organization of the behavioral responses with respect to the stimuli to which they are tuned. Ethologically, one might expect  animals  to have evolved their behavioral response so that similar behaviors are tuned to similar stimuli. 
Indeed, we find that similar behaviors have quantitatively similar kernels. Hierarchical clustering was performed on the euclidian distance of the scaled kernels, see \autoref{fig:kernel}. The two reverse locomotion states have similar kernels and were clustered together.  Forward velocity states fell into two clusters based on speed: Forward 3 and Forward 4 are slower and clustered together, while  Forward 5 and Forward 6 are faster and clustered together. 
That  similarities in the kernels reflect  similarities of their associated behaviors, provides additional confidence in our reverse correlation analysis.

\begin{figure}
\begin{fullwidth}
\begin{center}
\includegraphics[width=.87\textwidth]{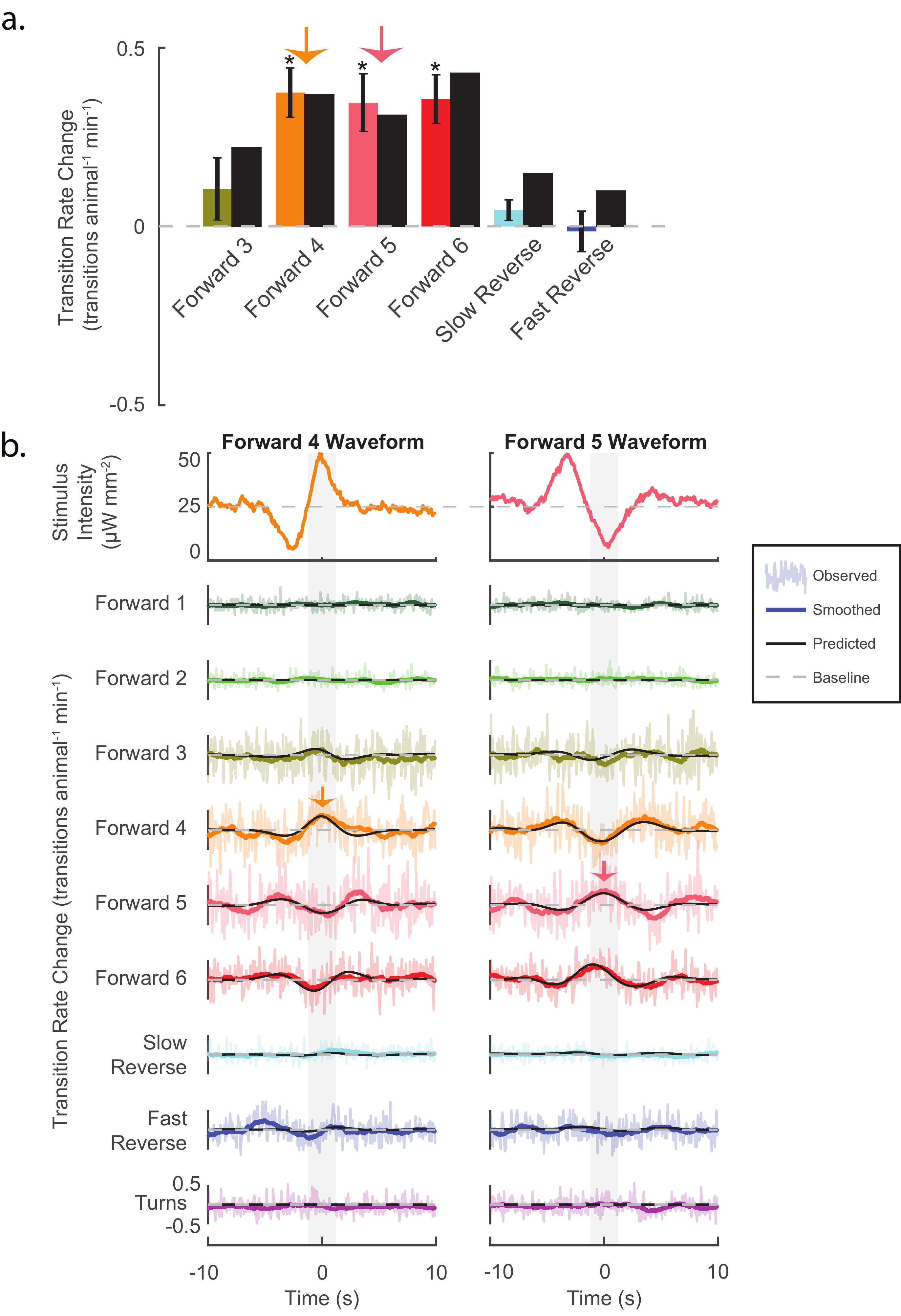}
\end{center}
\caption{Stimuli can be tailored to elicit specific behavioral responses, and the LN model predicts such responses. \textbf{a.)}  Animals are presented with stimuli shaped like the kernels in \autoref{fig:kernel}. Predicted (black bar) and observed (color bar)  changes in transition rate are shown for transitions into each kernel-shaped stimulus' corresponding behavior. For example, a Forward 3 shaped  stimulus  increases transitions into Forward 3 (mustard bar).
 For five of six behaviors, stimulation evoked increased transitions into their corresponding behaviors, as predicted. 
Transition rate changes are measured with respect to baseline (see methods).
Significance is estimated via a t-test and error bars show standard error of the mean. The number of stimulus-animal presentations, from left to right, were \{14,238, 13,612
, 14,699, 14,424, 14,194, 13,708.\}. Of these, the number of timely transitions observed were  \{1,400,  1,428, 1,692, 944, 191, 513.\}. The p-values were \{ \num{2.2e-1},  \num{5.6e-6}, \num{1e-4}, \num{3.4e-5}, \num{7.5e-2}, \num{9.5e-1}.\}.  \textbf{b.)} The LN model predicts details of the animal's behavioral response.  For each point in time,  the LN model predicts the change from baseline of transition rates for all nine behaviors in response to a stimulus. Detailed responses to Forward 4- and Forward 5-kernel-shaped stimuli are shown (see \autoref{fig:playback} - Figure Supplement \ref*{figs:allplayback} for the rest).  Raw transitions rates (light colored shading), smoothed transition rates (colored line) and LN prediction (solid black line) are shown. For stimuli shaped like Forward 4, the LN model correctly predicts not only that transitions into Forward 4 increase, but also that transitions into Forward 5 and 6 decrease. Light gray shading indicates the 2 s time window  used to calculate transition rates for the transitions shown in  \textbf{a} (orange and pink arrows).   Of 13,612 and 14,699 presentations for  Forward 4- and 5-kernel shaped stimuli, respectively, the following number of transitions were observed in the 20 second window, by row for Forward 4-shaped: \{1,265, 1,330, 12,312, 11,962, 13,436, 6,861, 1,735, 4,934, 2,864\}  and for Forward 5-shaped \{1,198, 1,437, 13,657, 13,538, 14,656, 7,295, 1,673, 5,506, 3,118\}.}
\label{fig:playback}

\figsupp{Behavioral responses to all kernel-shaped stimuli.}{\includegraphics[width=\textwidth]{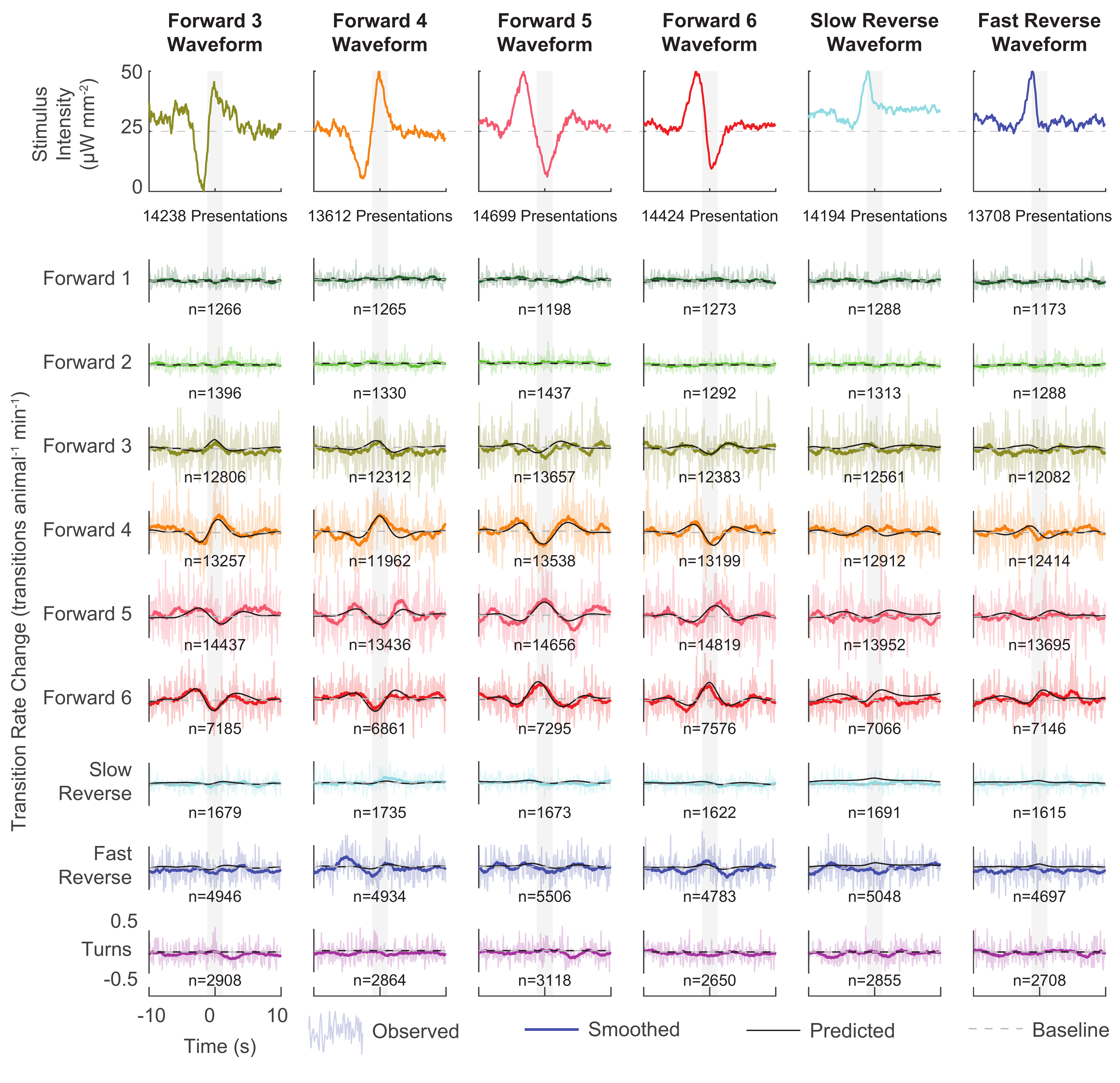}} { The LN model predicts the transition rate change from baseline for the nine behaviors in response to six stimuli constructed from statistically significant behavior triggered averages. $n$ refers to the number of transitions of the corresponding behavior observed during the 20 second window. Presentation numbers refer to stimulus-animal presentations.} \label{figs:allplayback}

\figsupp{Control animals grown without ATR do not respond to kernel-shaped stimulus.}{\includegraphics[width=\textwidth]{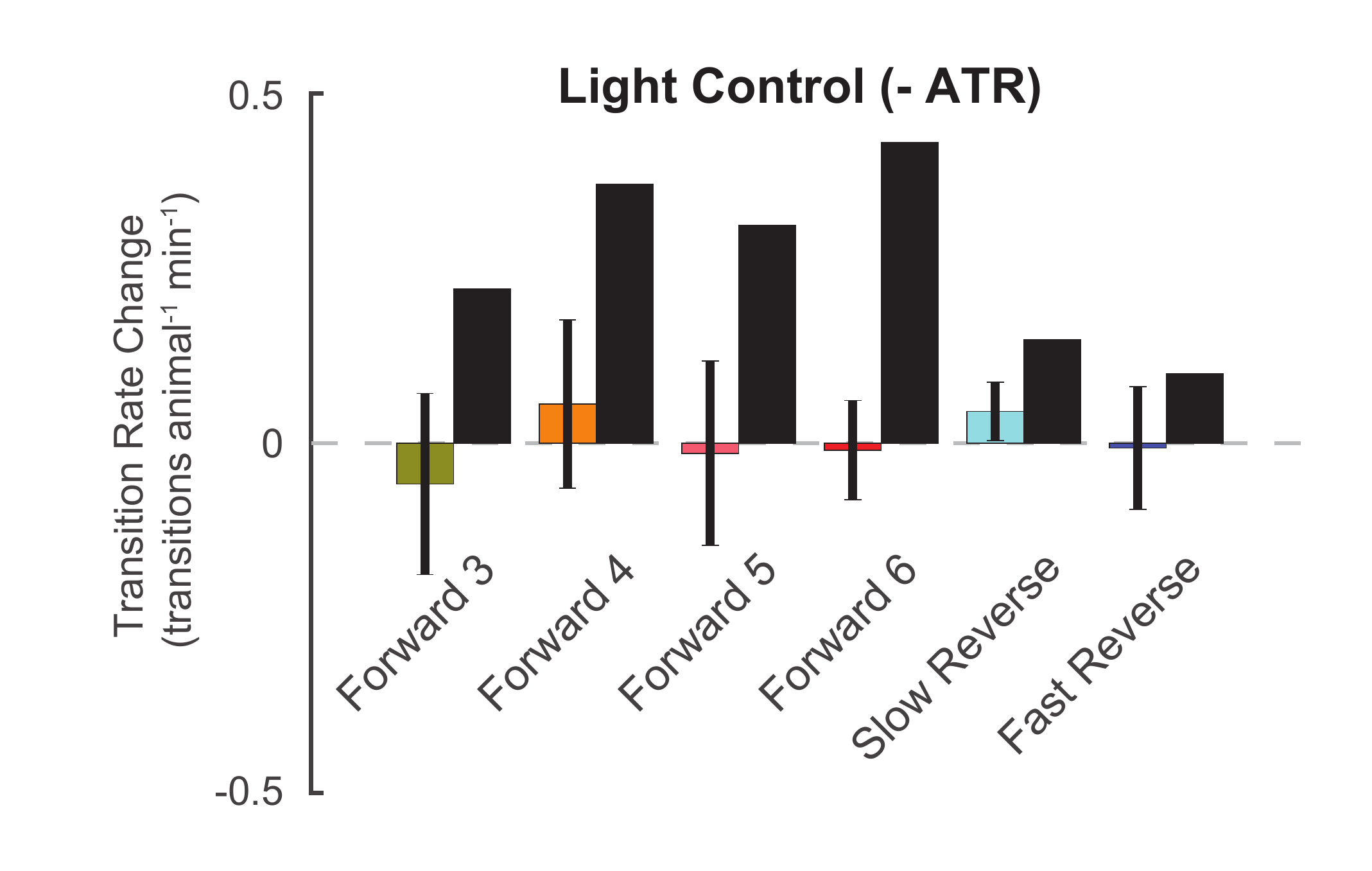}} { Kernel-shaped stimuli are delivered to AML67 control animals grown without the required co-factor ATR. The change in transition rate into the kernel-shaped stimuli's corresponding behavior is shown, as in  \autoref{fig:playback}a. Control animals do not exhibit a significant increase in transitions into the expected behavior states.  Error bars show standard error of the mean. Black bars show LN predictions for light sensitive animals. The number of stimulus-animal presentations, from left to right, were \{3,474, 3,880, 3,988, 3,639, 4,351, 3,534.\}. Of these, the number of timely transitions were  \{355,  225, 396, 162, 68, 95.\}. The p-values from a t-test were \{0.63,  0.97, 0.92, 0.87, 0.33, 0.68.\} } \label{figs:playbackcontrol}

\end{fullwidth}
\end{figure}

\subsection{Stimuli can be tailored to generate specific behavioral responses}
To causally test whether specific signals in the mechanosensory neurons can bias the animal towards specific behaviors as predicted, we generated stimuli that were tailored to elicit specific behavioral responses. The kernels found in \autoref{fig:kernel} purport to describe how each behavioral response is tuned to stimuli. Therefore, stimuli shaped like  one of the kernels should drive an \textit{increase} in transitions into its  respective behavior. If, however,  the behavioral response is tuned differently,  then the kernel-shaped stimulus may evoke \textit{decreases} in transitions to that behavior. (We already know that the animal  can respond to some stimuli by decreasing transitions to certain behaviors  because we saw this with tap and Forward 6, for example, see \autoref{fig:rangeOfBehav} - Figure Supplement \ref*{figs:transitionrates} ).  

We tested whether stimuli  shaped like the kernels in \autoref{fig:kernel} increased transitions  into its associated behaviors.  Kernel waveforms were presented to a plate of animals in a randomized order (6 kernels, $>$13,500 animal-stimulus presentations per kernel; 40 s inter-stimulus interval). Five of six kernels elicited increased transitions to their respective behaviors as predicted, three of the six significantly so,  see \autoref{fig:playback}a. None significantly decreased transitions to their respective behaviors.    We therefore conclude that the kernels correctly depict tuning of the behavioral responses. Consequently   we conclude that mechanosensory signals (even in the same neurons) can be tailored to evoke specific behaviors just by altering the stimulus waveform. 

\subsection{LN model predicts behavioral response, including to novel stimuli}
The LN model provides an analytical framework to predict how an animal responds to a stimulus. The LN model correctly predicted that kernel shaped waveforms should increase transitions into each kernel's associated behavior state, see \autoref{fig:playback}a. The kernel-shaped waveforms  also evoked other behavioral responses. For example, stimuli shaped like  the Forward 4 kernel increased transitions to both  Forward 4 and  Forward 3; but decreased transitions to  Forward 5 and 6, see \autoref{fig:playback}b.    How well, we wondered, does the LN model predict those responses? We compared the observed behavioral responses (colored lines) to detailed time-dependent predictions made by the  LN model (black lines).  To the resolution with which we could observe, we were reassured to find that the LN model correctly predicted the sign and temporal profile of changes in transition rates for all nine behavior states in response to each of the six kernel stimuli \autoref{fig:playback}b and \autoref{fig:playback} - Figure Supplement \ref*{figs:allplayback}, suggesting that the LN model captures myriad details of the animal's behavioral response.

\begin{figure}
\begin{center}
\includegraphics[width=.6\textwidth]{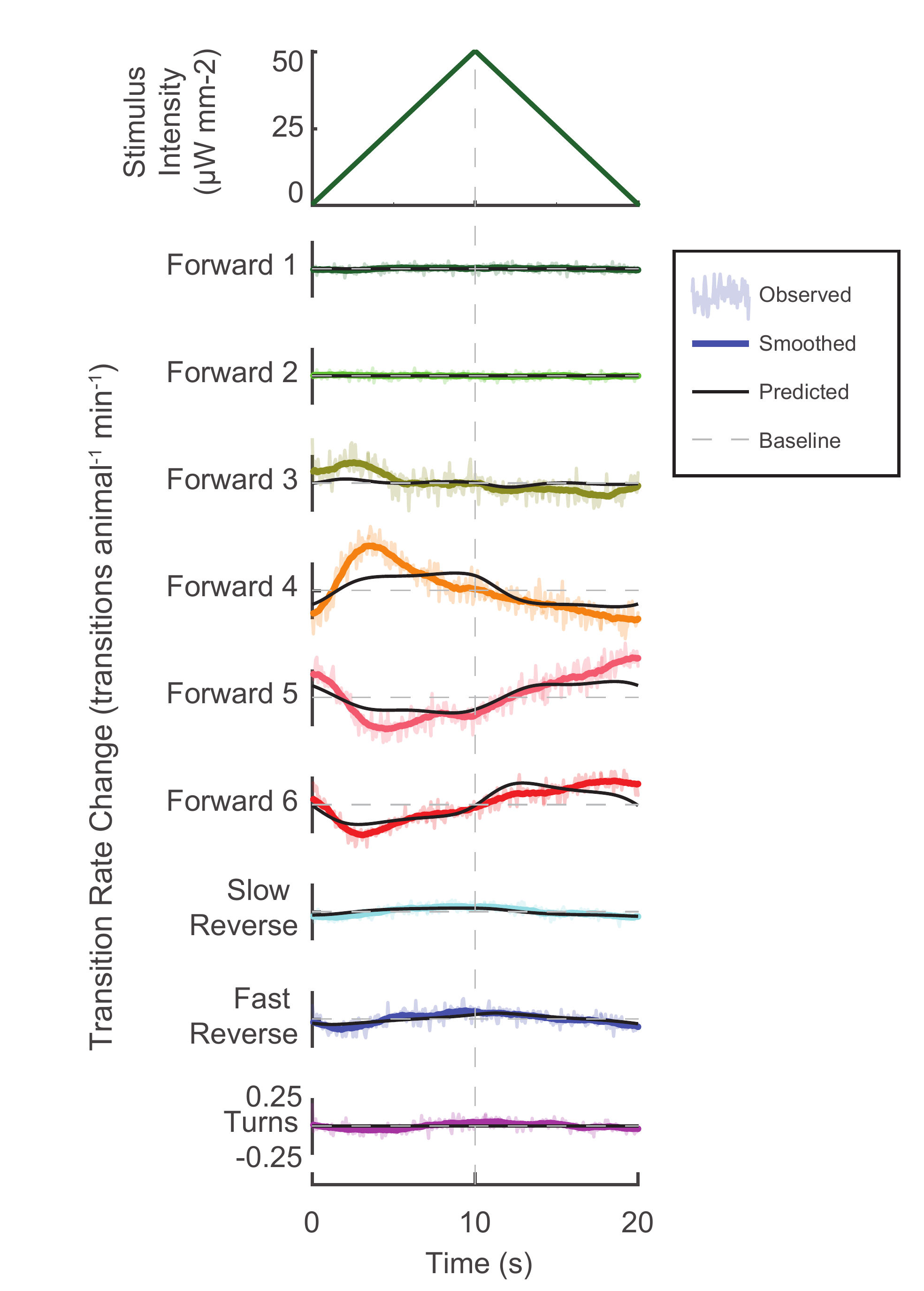}
\end{center}
\caption{Novel stimuli can be constructed to  enrich specific mechanosensory  responses. A novel triangle-wave optogenetic light stimulus was repeatedly presented to animals. Change in transition rates are shown for transitions into each behavior (raw, light color shaded; smoothed,  solid color line). Changes to transition rate as predicted by the LN model are also shown (black line). Increasing light intensity increases transitions into Forward 3 and 4, while decreasing light intensity increases transitions into Forward 5 and 6. Transitions into Slow and Fast Reverse are highest during highest stimulus intensity. The LN model  predicts these trends (though not all the details) despite  never previously experiencing  this particular stimulus. Of 340,757 animal-stimulus presentations the following number of transitions were observed (by row, from top to bottom): \{33,315, 31,243, 298,400, 343,474, 327,509, 160,332, 43,909, 106,743, 57,439\}.} 
\label{fig:trianglewave}

\figsupp{Control animals grown without ATR do not respond to triangle wave stimuli.}{\includegraphics[width=\textwidth]{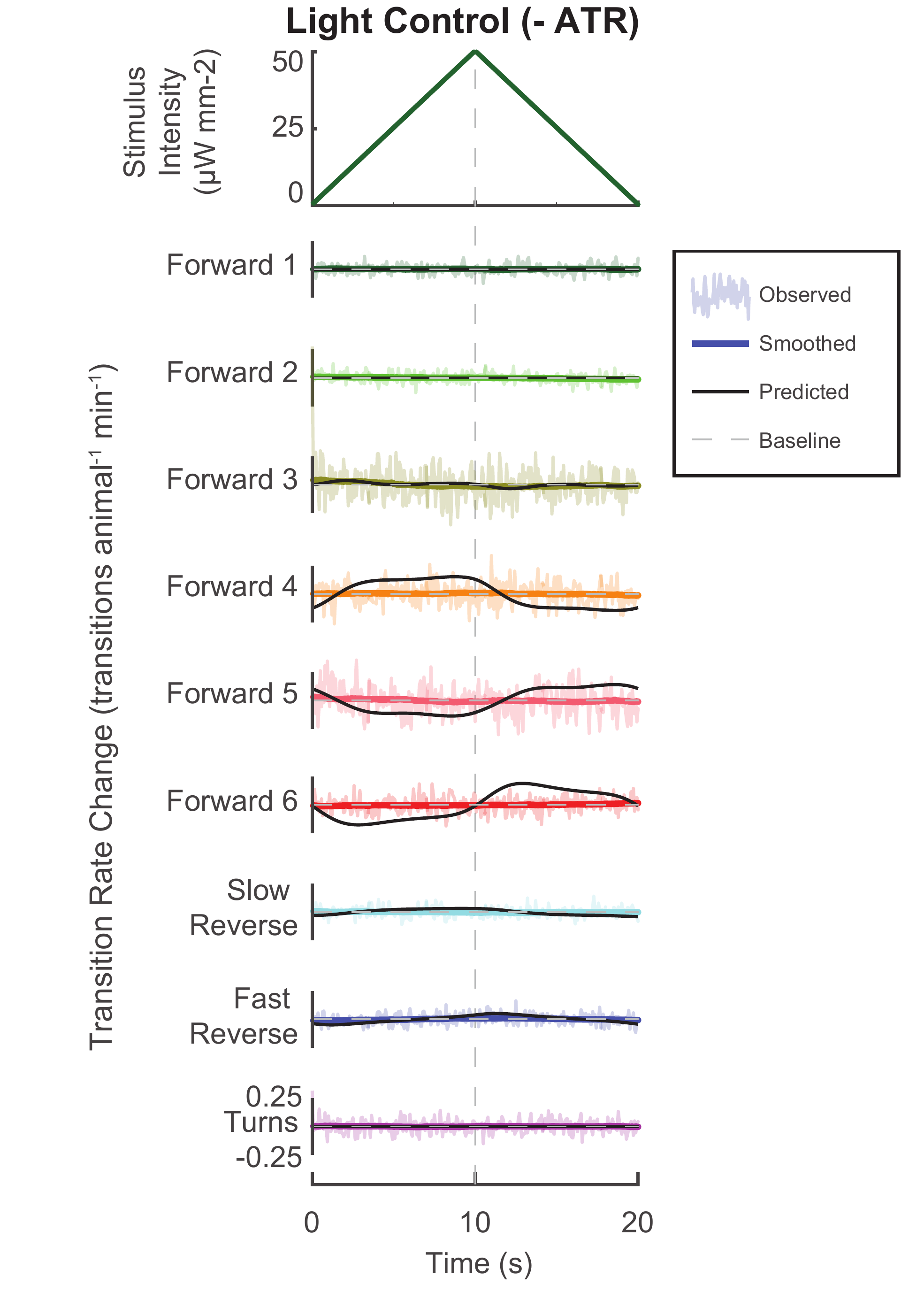}} { A novel traingle-wave optogenetic light stimulus, as in  \autoref{fig:trianglewave} , was repeatedly presented to control animals grown without the required co-factor retinal. Control animals do not respond to the stimulus. Observed response is shown as well as LN predicted response  for light-sensitive animals. Of 142,461 animal-stimulus presentations the following number of transitions were observed (by row, from top to bottom): \{14,575, 15,886, 146,149, 82,216, 131,060, 42,080, 16,871, 18,581, 28,318\}.} \label{figs:trianglecontrol}

\end{figure}

We further challenged our understanding of the animal's behavioral response to stimulus by presenting an entirely novel stimulus, a triangle-wave (340,757 stimulus-animal presentations), see \autoref{fig:trianglewave} and \autoref{fig:trianglewave} - Figure supplement 1. How well does the LN model predict the animal's behavior response to this novel stimulus?    The LN model captured the sign and general
trend (though not all features)  of the time-dependent change in the transition rate to all nine behaviors in response to the triangle wave. Moreover, the LN model provides a framework for understanding the animal's response by inspecting features of the kernel waveform. For example, the Fast Reverse kernel is symmetric in time and its mean-subtracted integral is positive. Therefore the shape of the Fast Reverse  kernel suggests that Fast Reverse should be tuned to the overall stimulus intensity but not its derivative. Indeed we observe a very slight increase in the rate of transitions to Fast Reverse during peak stimulus intensity.  Conversely the Forward 6 kernel is asymmetric in time and its biphasic waveform resembles that of the negative derivative of a gaussian. Therefore Forward 6 should be tuned to  decreases in stimulus intensity, as we observe.   

Taken together,  our experiments show that the animal can be driven to transition into different specific behavior states  by modulating the temporal profile of signals in the same mechanosensory neurons, and that the  LN model predicts the animal's response.

\subsection{Sensory processing is context dependent}
\begin{figure}
\begin{center}
\includegraphics[width=.6\textwidth]{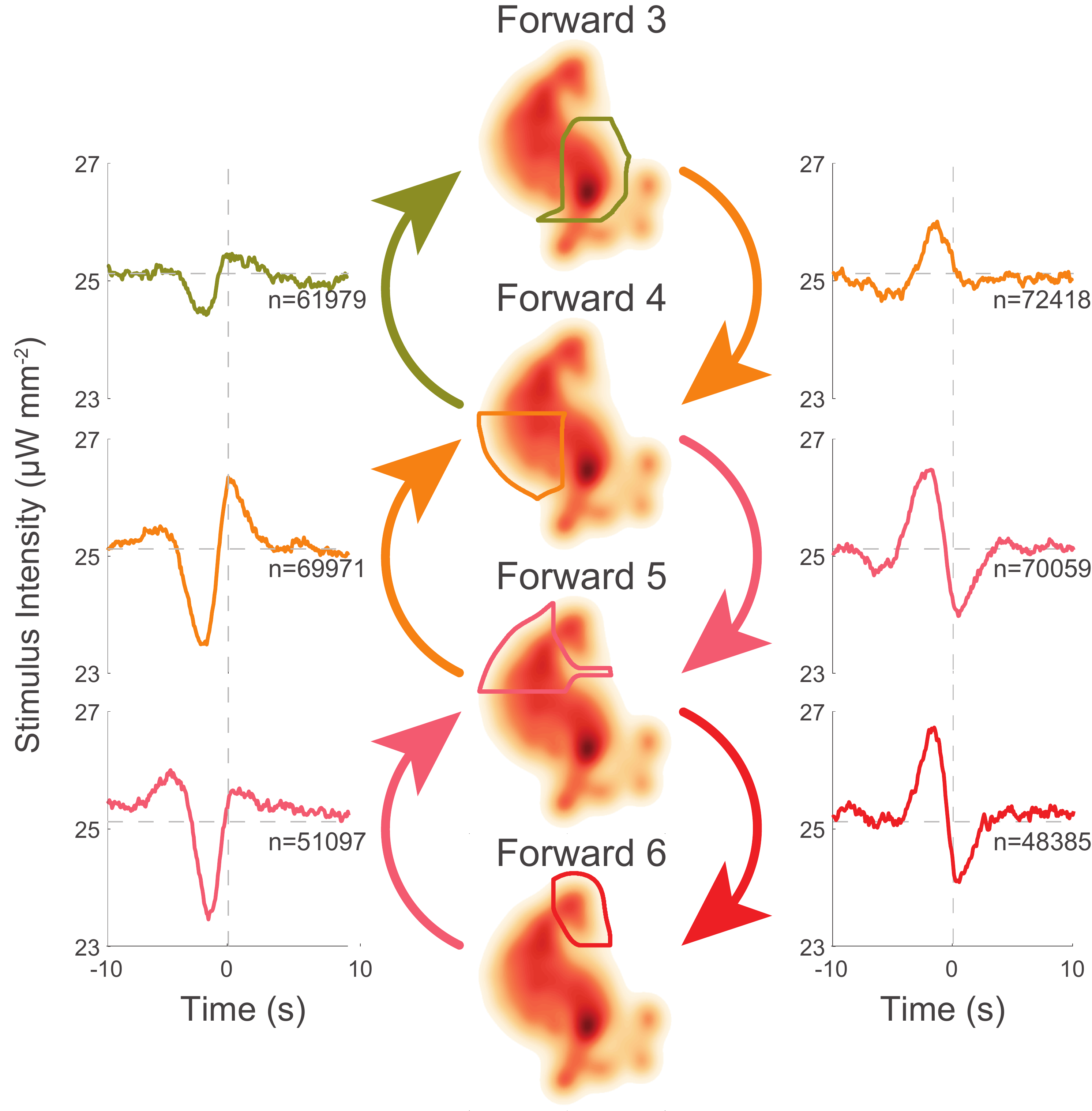}
\end{center}
\caption{Behavior transitions that involve slowing down and speeding  up have stereotyped tuning.  Selected context-dependent kernels are shown for transitions amongst forward locomotory states, where higher numbered states have higher velocities.  Kernels for slowing transitions (left column) are all similar, while kernels for speeding up transitions (right column) are also similar. Slowing and speeding-up kernels resemble horizontal reflections of one another.} 
\label{fig:velocitycontext}
\figsupp{All 72 pairwise context-dependent behavior triggered averages.}{\includegraphics[width=\textwidth]{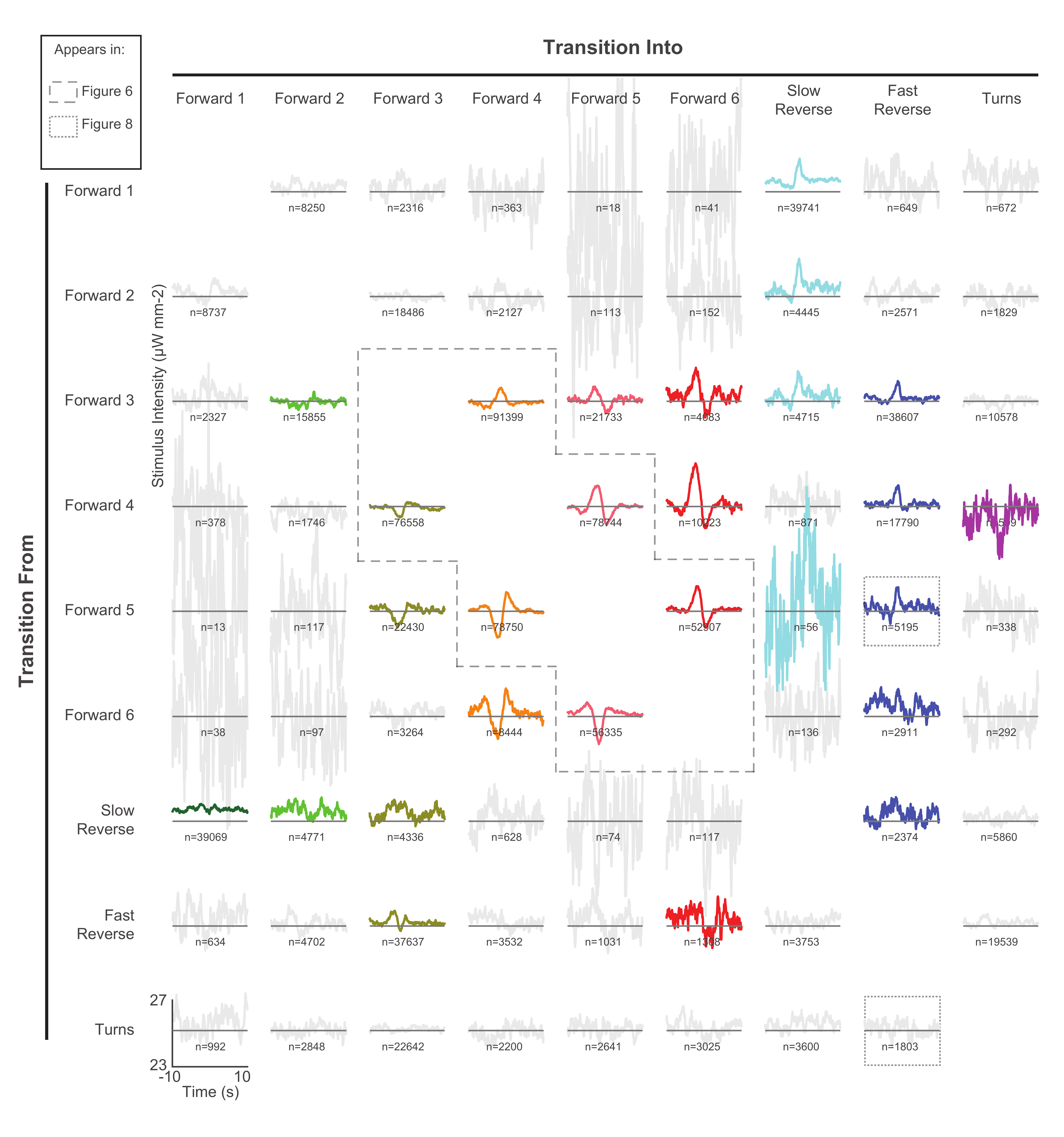}} { Pairwise behavior triggered averages (also referred to as kernels) are shown for transitions from one specified behavior to another. Kernels are calculated from 1,784 animal-hours of continuous random noise stimulation, same as in \autoref{fig:kernel} . For transitions into a given behavior (column), the kernel waveforms differ depending on the behavior that the animal originated in (rows). This suggests that the animal's behavioral response to stimulus depends on the animal's current behavior state.  For each kernel, the vertical axis spans 23 to 27 \textmu W mm$^{-2}$ and horizontal axis spans -10 to 10 seconds. $n$ indicates number of transitions observed. Kernels that fail to pass a shuffled significance threshold are grayed out (see methods). } \label{figs:allkernels}
\figsupp{All 72 pairwise context-dependent behavior triggered averages for control animals grown without ATR.}{\includegraphics[width=\textwidth]{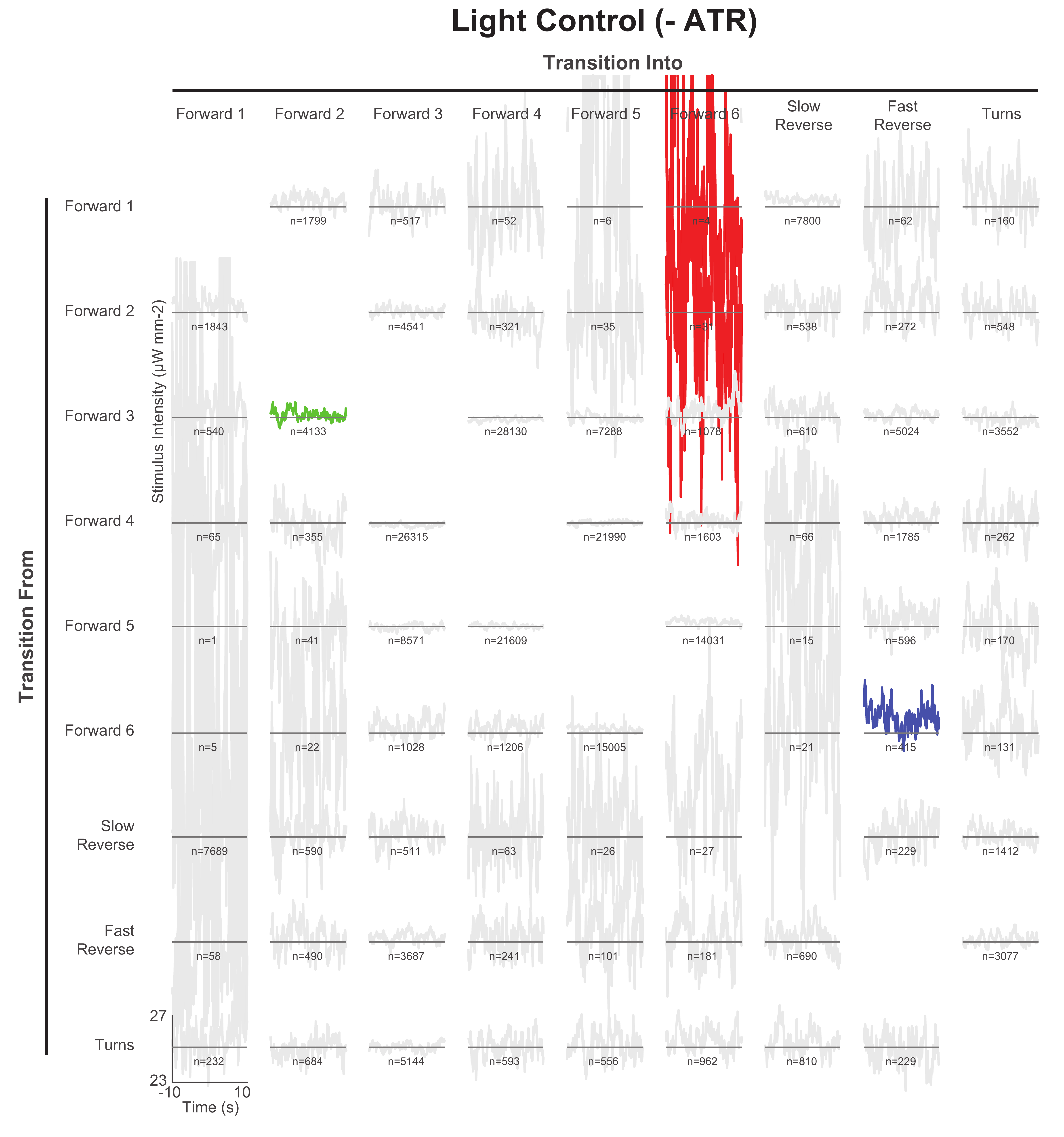}} { Pairwise behavior triggered averages (also referred to as kernels) are shown for transitions from one specified behavior to another. For each kernel, the vertical axis spans 23 to 27 \textmu W mm$^{-2}$ and horizontal axis spans -10 to 10 seconds. $n$ indicates number of transitions observed. Kernels that fail to pass a shuffled ignificance threshold are grayed out (see methods). } \label{figs:noretallkernels}
\end{figure}

\textit{C. elegans} is known to respond differently to the  same stimuli when it is in different long-lived behavior states like hunger \citep{ghosh_neural_2016}, or quiescence \citep{schwarz_reduced_2011, nagy_homeostasis_2014} and arousal \citep{cho_multilevel_2014}.   We wondered whether  mechanosensory processing might additionally be influenced by short-lived behavior states, like the Turn, Reverse or Forward locomotory states measured here. 
To investigate tuning of the animal's behavioral response conditional on its current behavior state, we calculated context-dependent kernels, one for each pairwise transition, see \autoref{fig:velocitycontext} - Figure supplement \ref*{figs:allkernels}.
Of 72 possible pairwise transitions, 27 had kernels that passed our shuffled significance threshold (compared to only 4 for our off-retinal control, see \autoref{fig:velocitycontext} - Figure Supplement \ref*{figs:noretallkernels}). Transitions to some behavior states, like Forward 4,  had kernels that changed dramatically depending on which behavior the animal originated from,  see  columns in \autoref{fig:velocitycontext} - Figure Supplement \ref*{figs:allkernels}. The  pairwise-specific kernels  provided evidence of two types of context-dependent sensory processing in \textit{C. elegans} that occur on short-time scales. In both cases the animal appears to respond to the same stimuli differently depending on its current behavior.  In the first,   the animal responds to certain mechanosensory signals by speeding up or slowing down.  In the second type, the animal suppresses its response to mechanosensory stimuli during turning behavior. These two types of context-dependency are described below.

\subsection{There are mechanosensory signals for speeding up or slowing down}
Behavior transitions that involve slowing down have similar tuning. For example, the Forward 5$\rightarrow$4 kernel has a similar shape to the  Forward 4$\rightarrow$3 kernel, see \autoref{fig:velocitycontext}, left column. Likewise, transitions involving speeding up also have similar kernels. For example, Forward 3$\rightarrow$4 and Forward 4$\rightarrow$5 have similar kernels, see \autoref{fig:velocitycontext}, right column. Moreover,  the two classes of kernels appear to be reflections of one another about the line of mean stimulus intensity.   
The stereotypy of the speed up and slow down kernels suggests that the
animal has evolved to respond to certain  stimuli  by slowing down or speeding up in a relative way instead of transitioning to a stimulus-defined velocity.  This is of interest because it implies a form of context dependency: it suggests that the same stimulus will drive the animal into  forward locomotory states of different speeds depending on the animal's current state.


\begin{figure}
\begin{center}
\includegraphics[width=\textwidth]{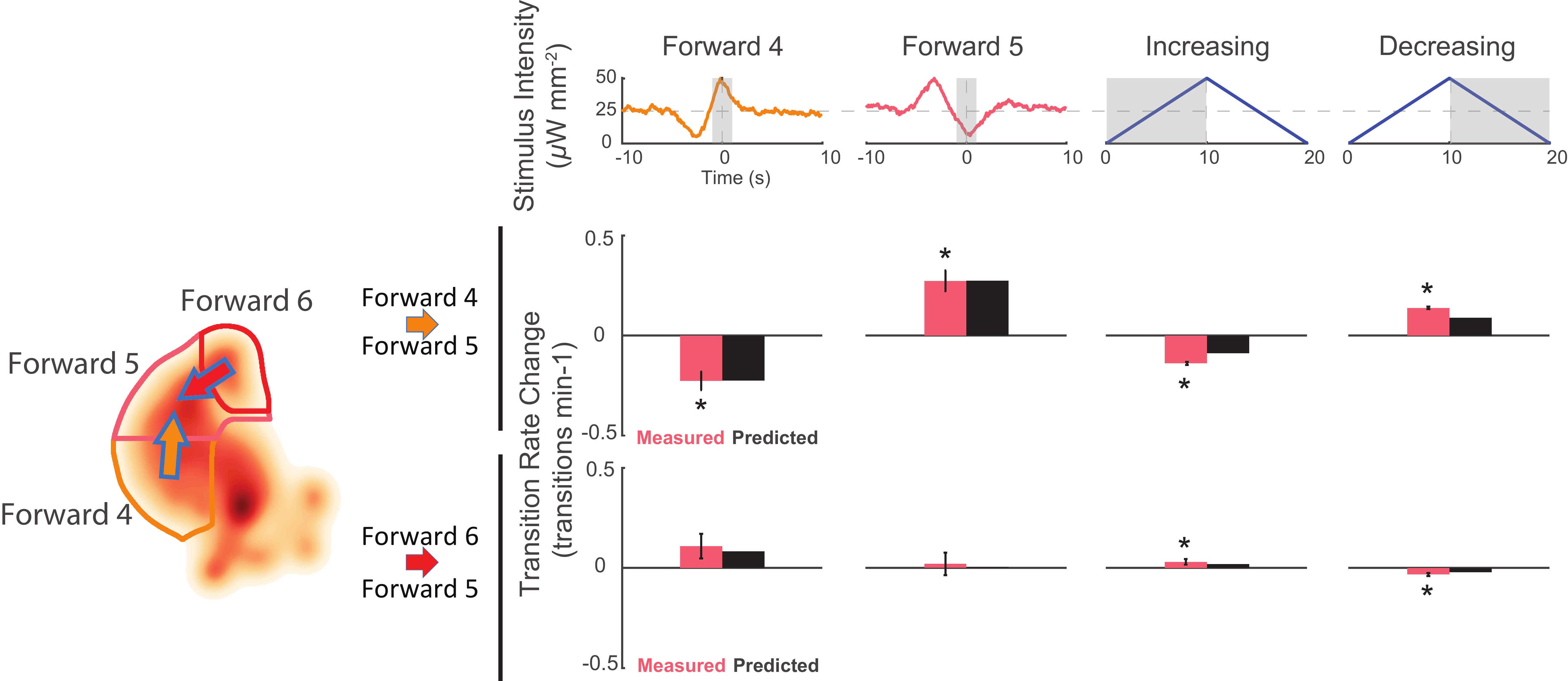}
\end{center}
\caption{Animals respond to the same stimuli differently depending on their current behavior state.
The change  in transition rate from baseline is shown for transitions into Forward 5 from either Forward 4 (middle row) or Forward 6 (bottom row) in response to four different stimuli (columns).  Observed transition rates (colored bars) are compared to LN model predictions (black bars).  
The stimulus effects the rate of transitions into Forward 5 differently depending on whether the animal was in Forward 4 or Forward 6 at the time of stimulus. For example, consistent with the animal responding to a slowing down signal,  the Forward 4-shaped stimulus decreases  Forward 4$\rightarrow$5 transitions, but increases Forward 6$\rightarrow$5 transitions.
Star indicates significant change in transition rate from baseline. 
Gray shaded region indicates the time window over which the transition rate is calculated. Baseline is defined slightly differently for the kernel-shaped stimuli compared to the triangle waves, see methods.
Of 13,612 and 14,699 stimulus-animal presentations for Forward 4 and Forward 5 kernel-shaped stimuli, and 340,757 stimulus-animal presentations for the triangle wave, the following number of  transitions were observed: {26, 24, 2,604, 2,634} for Forward 4$\rightarrow$5  (top row) and  {6, 7, 713, 791} for Forward 6$\rightarrow$5 (bottom row). To test signifiant change from baseline a t-test was used and the following p-values were observed: \{\num{5.5e-5}, \num{6.9e-6}, \num{1.8e-19}, \num{5.1e-48}\} for Forward 4$\rightarrow$5  (top row) and  \{ \num{6.3e-2}, \num{8.7e-1}, \num{3.9e-2}, \num{6.7e-5}\} for Forward 6$\rightarrow$5 (bottom row). Error bars show the standard error of the mean.  } 
\label{fig:samestimDiffresponse}

\end{figure}

To  validate whether  the stereotyped speed-up or slow-down stimulus indeed causes the animal to speed up or slow-down,  we again inspected the animal's response to the kernel-shaped stimuli or the triangle-wave stimulus. Indeed, we find that the same stimulus drives the animal into a different forward locomotory state depending on the animal's current state, see \autoref{fig:samestimDiffresponse}.
 For example, animals in the slower Forward 4 state responded to a Forward 4 kernel-shaped stimulus by  \textit{decreasing} their transitions to  Forward 5.  In contrast animals in the faster Forward 6 state responded to the same stimulus by \textit{increasing} their transitions into Forward 5.  This was one of  multiple instances where we observed the animal  responding to the same stimuli with opposite responses depending on its current behavior. During triangle wave stimulation, for example, an increasing ramp causes slowing down, while  a decreasing ramp causes speeding up,  see \autoref{fig:samestimDiffresponse}.  We therefore conclude that  stereotyped mechanosensory signals drive the animal to speed up or slow down.

 
\begin{figure}
\begin{center}
\includegraphics[width=\textwidth]{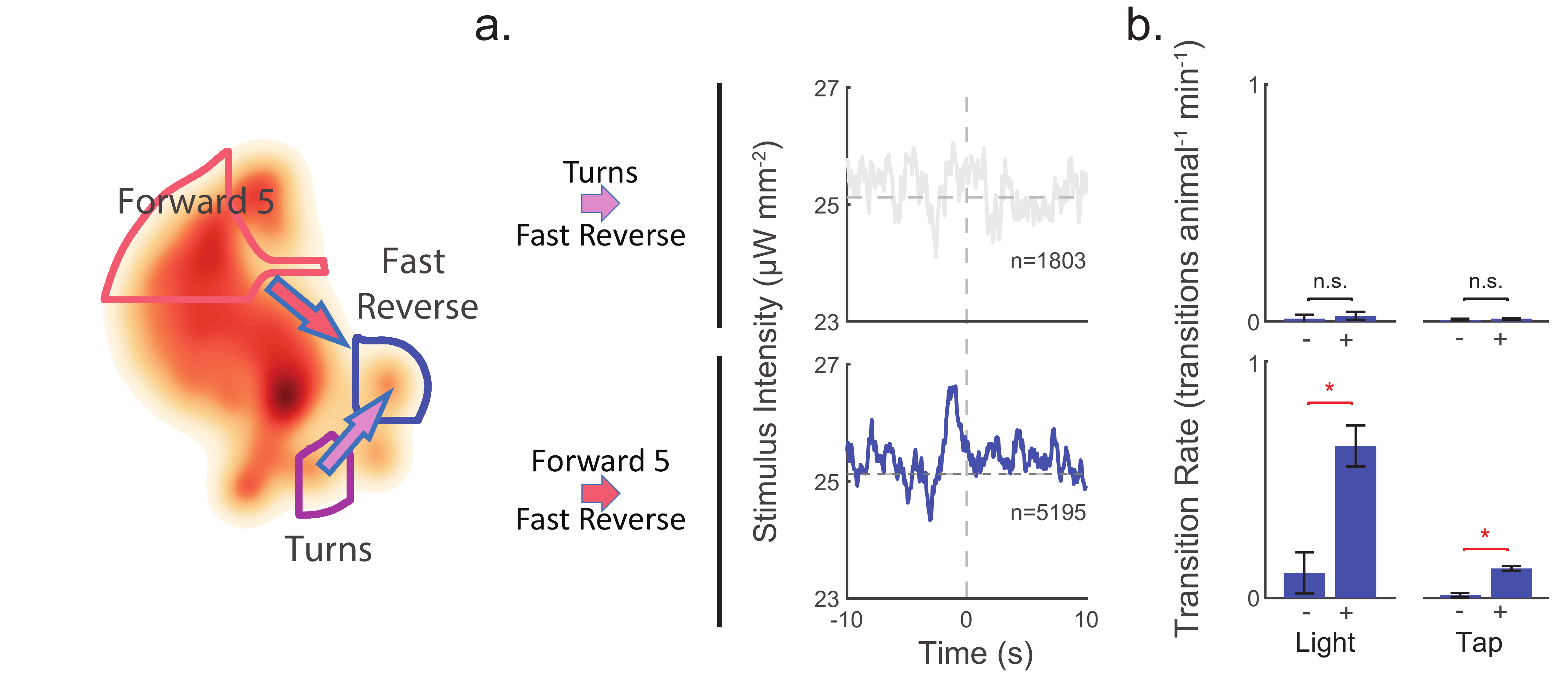}
\end{center}
\caption{Attention to mechanosensory signals depends on behavior. When the animal is in the turn state it ignores mechanosensory stimuli. 
\textbf{a.)}  Kernels are shown for two context-dependent transitions into Fast Reverse. Transitions into Fast Reverse originating from Forward 5 are correlated with stimulus and have a significant kernel while those originating from Turns are not and fail our shuffled significance threshold, see methods. Kernels shown are same as in \autoref{fig:velocitycontext} - Figure Supplement \ref*{figs:allkernels} \textbf{b.)}  Transition rate in response to light and tap are shown. Animals in the turn state show no significant change in transition rates in response to light or tap, while animals in other states, like Forward 5 do. The 2 second post-stimulus mean transition rate into Fast Reverse is shown in response to a 1 s  light stimulation (+), mechanical tap (+) or a mock control (-).  Star indicates significance, calculated using an E-test, see methods.  Error bars show  standard error of the mean. 2,487 and 37,000 stimulus-animal presentations were analyzed for light(+) and tap(+) respectively, and 2,427 and 40,012 mock controls(-) for light and tap. The following number of transitions were observed  \{1, 2, 11, 15\} for Turns$\rightarrow$Fast Reverse (top row) and \{9, 55, 18, 160\} for Forward 5$\rightarrow$Fast Reverse (bottom row).  P-values for the E-test are \{0.68, 0.34\} for Turns$\rightarrow$Fast Reverse (top row) and \{\num{1.96e-9},  0\} for Forward 5$\rightarrow$Fast Reverse (bottom row)}
\label{fig:attention}

\figsupp{Transition rates in response to light pulse for all  pairwise transitions.}{\includegraphics[width=\textwidth]{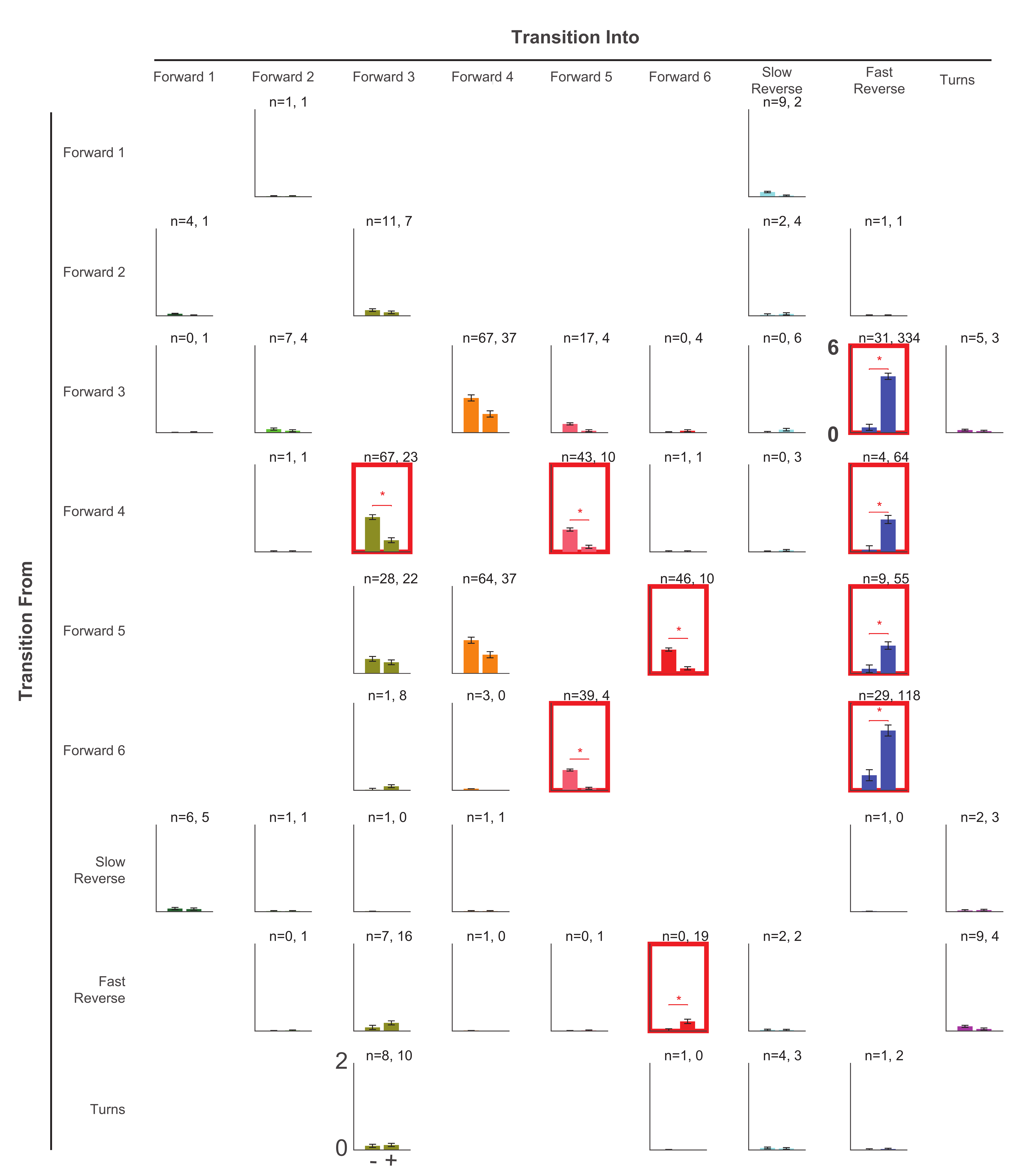}}{Transition rate  is shown for all observed pairwise transitions in response to 1 s light pulse (+, right bar, 2,487 stimulus-animal presentations) and mock control (-, left bar, 2,427 stimulus-animal presentations). Number of observed transitions $n$ for each bar is listed. No bars are shown for pairwise transitions that were not observed. Red square and star indicate significance, calculated by  a multiple hypothesis corrected E-test, see methods.  P-values are listed in \autoref{fig:attention} - Figure supplement \ref*{figs:optotappval}. Note that $y$ axis range is 0 to 2 transitions animal$^{-1}$ min$^{-1}$ for all cases except for Forward 3$\rightarrow$Fast Reverse, where it is 0 to 6 transitions animal$^{-1}$ min$^{-1}$.} 
\label{figs:optotap}

\figsupp{P-values for  transition rates in response to light pulse for all  pairwise transitions.}{
\tiny
\begin{tabular}{|l*{9}{|S[table-format=1.2e-2, table-figures-decimal = 2]}|} \hline
             & {Forward 1}   & {Forward 2}   & {Forward 3}   & {Forward 4}   & {Forward 5}   & {Forward 6}   & {Slow Reverse} & {Fast Reverse} & {Turn }       \\ \hline
Forward 1    & 	&	8.65E-01	&		&		&		&		&	3.17E-02	&		&	 \\ \hline
Forward 2    & 1.85E-01	&		&	3.30E-01	&		&		&		&	4.77E-01	&	8.65E-01	&	 \\ \hline
Forward 3    & 4.82E-01	&	3.73E-01	&		&	2.17E-03	&	3.24E-03	&	4.13E-02	&	9.66E-03	&	0	&	4.91E-01 \\ \hline
Forward 4    & 	&	8.65E-01	&	1.42E-06	&		&	1.97E-06	&	8.65E-01	&	8.84E-02	&	0	&	 \\ \hline
Forward 5    & 	&		&	3.56E-01	&	4.89E-03	&		&	4.23E-07	&		&	1.96E-09	&	 \\ \hline
Forward 6    & 	&		&	1.85E-02	&	5.94E-02	&	9.53E-09	&		&		&	0	&	 \\ \hline
Slow Reverse & 7.52E-01	&	8.65E-01	&	4.82E-01	&	8.65E-01	&		&		&		&	4.82E-01	&	7.37E-01 \\ \hline
Fast Reverse & 	&	4.82E-01	&	7.17E-02	&	4.82E-01	&	4.82E-01	&	1.37E-06	&	9.08E-01	&		&	1.65E-01 \\ \hline
Turns        & 	&		&	6.75E-01	&		&		&	4.82E-01	&	7.27E-01	&	6.81E-01	&	 \\ \hline
\end{tabular}
\normalsize

}{P-values are listed for the multiple hypothesis corrected E-tests performed in \autoref{fig:attention} - Figure Supplement \ref{figs:optotap}. Row specifies ``transition from'' and column specifies ``transition into.''} \label{figs:optotappval}

\figsupp{Transition rates in response to tap for all  pairwise transitions.}{\includegraphics[width=\textwidth]{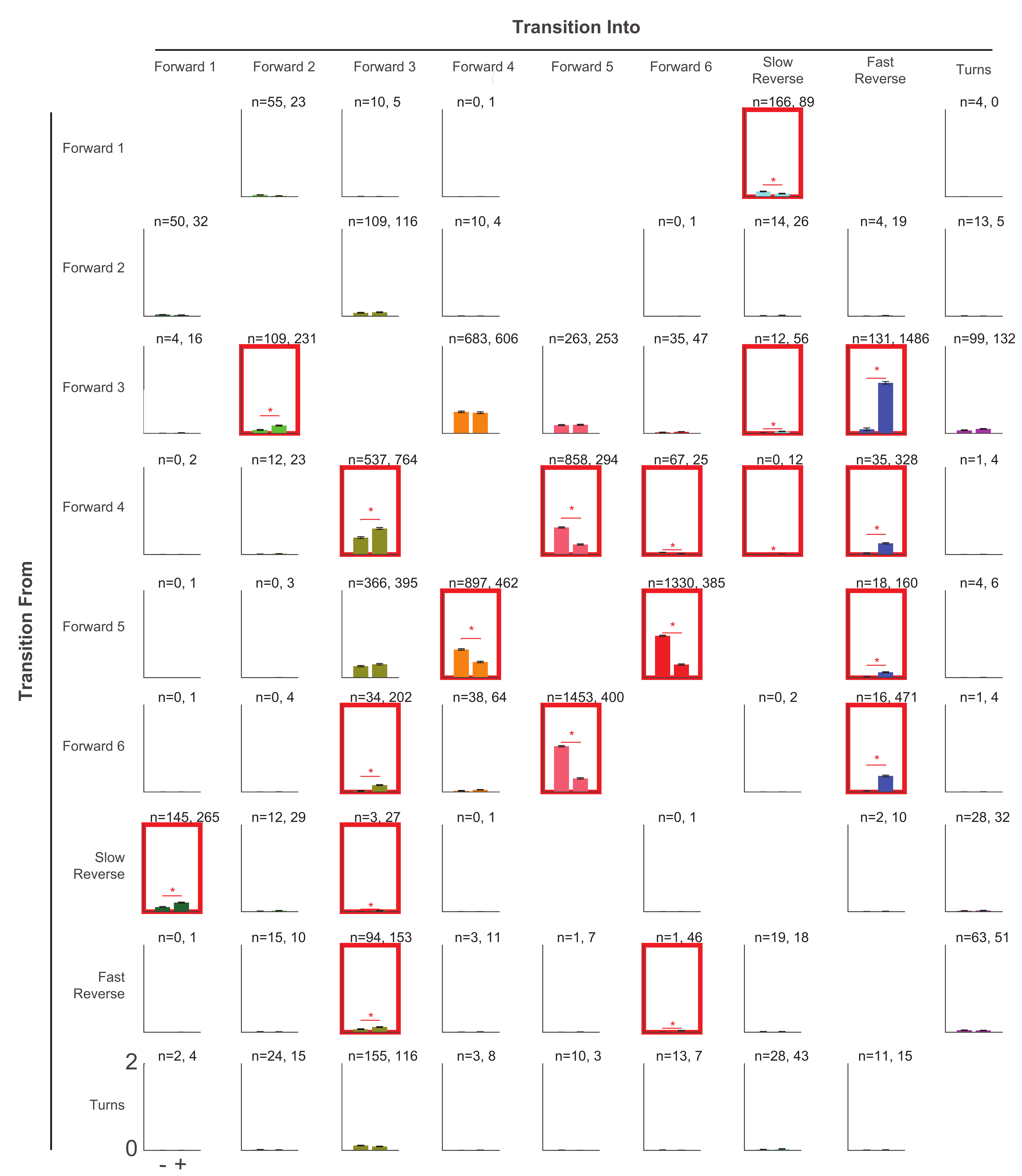}}{Transition rate  is shown for all observed pairwise transitions in response to tap (+, right bar, 37,000 stimulus-animal presentations) and mock control (-, left bar, 40,012 stimulus-animal presentations). Number of observed transitions $n$ for each bar is listed. No bars are shown for pairwise transitions that were not observed. Red square and star indicate significance, calculated by  a multiple hypothesis corrected E-test, see methods. P-values are listed in \autoref{fig:attention} - Figure supplement \ref*{figs:platetappval}. All $y$ axis range is 0 to 2 transitions animal$^{-1}$ min$^{-1}$. } \label{figs:platetap}

\figsupp{P-values for  transition rates in response to tap for all  pairwise transitions.}{
\tiny
\begin{tabular}{|l*{9}{|S[table-format=1.2e-2, table-figures-decimal = 2]}|} \hline
             & {Forward 1}   & {Forward 2}   & {Forward 3}   & {Forward 4}   & {Forward 5}   & {Forward 6}   & {Slow Reverse} & {Fast Reverse} & {Turn }       \\ \hline
Forward 1  &   & 7.98e-04 & 2.60e-01 & 4.82e-01 &          &          & 1.99e-05 &          & 4.18e-02        \\ \hline
Forward 2    & 9.97e-02 &          & 2.94e-01 & 1.47e-01 &  & 4.82e-01 & 3.48e-02 & 8.43e-04 & 8.27e-02 \\ \hline
Forward 3    & 4.20e-03 & 0 &          & 4.58e-01 & 6.54e-01 & 9.61e-02 & 9.59e-09 & 0 & 5.95e-03 \\ \hline
Forward 4    & 1.25e-01 & 3.92e-02 & 0 &          & 0 & 3.84e-05 & 1.02e-04 & 0 & 2.03e-01 \\ \hline
Forward 5    & 4.82e-01 & 5.78e-02 & 3.36e-02 & 0 &          & 0 &          & 0 & 4.75e-01 \\ \hline
Forward 6    & 4.82e-01 & 4.18e-02 & 0 & 3.17e-03 & 0 &          & 1.25e-01 & 0 & 2.03e-01 \\ \hline
Slow Reverse & 4.61e-12 & 3.83e-03 & 2.60e-06 & 4.82e-01 &          & 4.82e-01 &          & 1.62e-02 & 4.14e-01 \\ \hline
Fast Reverse & 4.82e-01 & 4.09e-01 & 1.37e-05 & 2.32e-02 & 2.98e-02 & 6.17e-14 & 9.38e-01 &          & 4.70e-01 \\ \hline
Turns        & 4.20e-01 & 2.28e-01 & 8.31e-02 & 1.13e-01 & 6.33e-02 & 2.46e-01 & 3.57e-02 & 3.40e-01 &      \\ \hline
\end{tabular}
\normalsize

}{P-values are listed for the multiple hypothesis corrected E-tests performed in \autoref{fig:attention} - Figure Supplement \ref{figs:platetap}. Row specifies ``transition from'' and column specifies ``transition into.'' } \label{figs:platetappval}

\end{figure}

\subsection{Attention to mechanosensory signals depends on the animal's current behavior}
When the animal turns it ignores all tested mechanosensory signals. This surprising observation is predicted by reverse-correlation analysis and confirmed by optogenetic and tap stimulation.  
Transitions out of Turn are uncorrelated with stimulus, and kernels for those transitions all fail to pass our shuffled significance threshold, see  bottom row in \autoref{fig:velocitycontext} - Figure Supplement \ref*{figs:allkernels}. Consequently, the kernels predict that the animal should ignore mechanosensory stimuli during turns. In contrast, for every other behavior state there is always at least one (and often many)   transitions  exiting out of the state whose kernels  pass our significance  threshold (all rows other than Turn have at least one significant kernel).  

To further test whether the animal indeed ignores stimuli during turns, we investigated the animal's context-dependent  response to light pulses or tap. 
When the animal was in the Turn state, neither light pulse nor tap evoked a significant  change in the rate of transitions into any other  behavior, see bottom row \autoref{fig:attention} - Figure Supplement \ref*{figs:optotap} and \ref*{figs:platetap} (multiple-hypothesis corrected E-test, see methods). In contrast, when the animal was in  other states like  Forward 5, both tap and light pulses evoked significant changes in the transition rate into other behaviors. In fact, every other behavior state except for Forward 2 had at least one behavior transition exiting the state whose transition rate was significantly affected by either light or tap.  The Turn behavior state alone  was unique in that none of the kernels for transitions originating in Turn were significant, and no transition rates changed significantly in response to either light or tap. We therefore conclude that in Turn, but not other states, the animal ignores mechanosensory stimuli.

Transitions into Fast Reverse provide an illustrative example, see \autoref{fig:attention}.  When the animal is in the Turn state, there is no significant difference in the rate of Turn$\rightarrow$Fast Reverse transition  between shuffled and  stimuli.  But when the animal is in  Forward 5, light and taps caused a significant increase in Forward 5$\rightarrow$Fast Reverse.  Taken together, we conclude that the animal  attends to mechanosensory signals during most  behavior states, like Forward 5, but ignores it during  Turns.   In the discussion we explore why this may be ethologically relevant.

\section{Discussion}
This work revises a number of implicit assumptions about \textit{C. elegans} sensory processing.  First, the animal's behavioral response  is not merely determined by the amplitude of signals in the touch neurons. Instead, 
it is also tuned to  temporal properties of these signals, like the derivative, that extend over many seconds in time.   Moreover, by adjusting a stimulus waveform, mechanosensory signals in the same neurons can be tailored to elicit different behavioral responses. Second, mechanosensory signals influence a  broader set of behaviors than previously reported. Mechanosensation not only drives reversals and  accelerations  but  it can also evoke  the animal to slow down. Third, even short term behavior states can influence the animal's sensory processing. Earlier work has emphasized context-dependent sensory processing for behaviors with timescales of minutes to hours, like hunger-satiety \citep{ghosh_neural_2016},  quiescence \citep{schwarz_reduced_2011, nagy_homeostasis_2014} and arousal \citep{cho_multilevel_2014}.  
Here we show that even seconds-long timescale behaviors can dramatically alter how the animal responds to a stimulus. 
Most dramatically, when the animal turns it appears to ignore mechanosensory signals completely.

In more complex sensory systems like the retina, we have come to expect that the nervous system is carefully tuned to temporal properties of sensory signals \citep{meister_neural_1999}. Recently it was shown that in drosophila,  temporal processing is important for behavioral responses to odors, light and sound \citep{behnia_processing_2014, coen_dynamic_2014, gepner_computations_2015,hernandez-nunez_reverse-correlation_2015}. And in the much simpler \textit{C. elegans}, temporal processing on order second timescale has been observed in thermosensation \citep{clark_afd_2006,clark_temporal_2007}, and  chemosensation \citep{kato_temporal_2014} where it is known to be crucial for guiding thermotaxis or chemotaxis. 
In the \textit{C. elegans} mechanosensory circuit, it had been known that  temporal processing occurs  at the receptor level to convert applied forces into evoked  currents  on the tens of milliseconds timescale  \cite{eastwood_tissue_2015}, but it had remained unclear whether  the nervous system used temporal information downstream for determining the animal's behavioral response. 
In this work we now see evidence of temporal processing on behavior-relevant timescales that guides the animal's behavioral response. That such behavior-relevant temporal processing is observed even in the simple mechanosensory circuit raises the possibility that  temporal processing may be ubiquitous across sensory systems for driving behavior. 
Why might it be beneficial for the \textit{C. elegans} nervous system to have evolved to tune its behavioral response to temporal properties of mechanosensory signals, like the derivative, over seconds?   We speculate that perhaps it is important for the worm to react differently if mechanosensory signals are increasing or decreasing, rather than merely base its decision on the overall strength.

That the animal ignores mechanosensory inputs during turning is striking and surprising. Why might the animal have evolved to ignore such signals during  turns? The turn is part of the \textit{C. elegans} escape response, an avoidance behavior with similarities to escape responses in  other organisms like crayfish, mullusks or goldfish \citep{pirri_neuroethology_2012}.   \textit{C. elegans} escape consists of  reverse locomotion, followed by a turn and then forward locomotion in a new direction. The turn allows the animal to reorient and navigate away from a predator, and defects in this circuit have been shown to decrease survivability \citep{maguire_c._2011}. Failing to complete the turn could inadvertently cause the animal to retrace its steps and return to danger. 

Ultimately we see evidence of two kinds of internal processes that govern how the animal interprets sensory signals. First the animal integrates mechanosensory information over seconds timescale. Second, the animal interprets these signals differently depending on the animal's behavior state.  An exciting future direction will be to identify the neural circuit mechanisms which allows the worm's nervous system to integrate mechanosensory signals over time; and to rapidly alter its response to mechanosensory signals depending on behavior state. This could shed insight into how internal brain states rapidly modulate sensory processing in a simple model system.


\section{Methods and Materials}

\subsection{Strains}
The two strains used in this study were wild-type N2 Bristol animals; and AML67 (wtfIs46[p\textit{mec-4}::\allowbreak\textit{Chrimson}::\allowbreak\textit{SL2}::\allowbreak\textit{mCherry}::\allowbreak\textit{unc-5}]), a transgenic strain that expresses the light-gated ion channel Chrimson and a fluorescent protein mCherry in mechanosensory neurons. To generate AML67, 40 ng of plasmid (pAL::p\textit{mec-4}::\textit{Chrimson}::\textit{SL2}::\textit{mCherry}::\textit{unc-54}) were injected into N2 animals,  integrated via UV irradiation \citep{evans_transformation_2006}, and outcrossed with N2 six times. AML67 has been deposited in the public Caenorhabditis Genome Center repository. Plasmid pAL::p\textit{mec-4}::\textit{Chrimson}::\textit{SL2}::\textit{mCherry}::\textit{unc-54}  (\href{https://www.addgene.org/107745/}{https://www.addgene.org/107745/}) was engineered using  HiFi Cloning Kit (NEB). Chrimson was a kind gift of Boyden Lab. mCherry and backbone was amplified from pJIM20, gift from John Murray, U Penn. Promoter sequence (mec-4), splicing sequence (SL2) and 3$^\prime$-utr sequence (\textit{unc-54}) were amplified using primers as listed in \autoref{tab:primerTable}. Construct was sequenced confirmed before injection.

\begin{table}[]
\centering
\caption{Forward and reverse primer sequences used to to generate pAL::p\textit{mec-4}::\textit{Chrimson}::\textit{SL2}::\textit{mCherry}::\textit{unc-54}.\label{tab:primerTable}}
\begin{tabular}{ll}
Primer & Sequence \\
\toprule
mec-4\_fwd                   & AAGCTTCAATACAAGCTC            \\
mec-4\_rev                   & TAACTTGATAGCGATAAAAAAAATAG    \\
CHRIMSON\_fwd                & ATGGCTGAGCTTATTTCATC          \\
CHRIMSON\_rev                & AACAGTATCTTCATCTTCC           \\
SL2\_fwd                     & GGTACCGCTGTCTCATCC            \\
SL2\_rev                     & GATGCGTTGAAGCAGTTTC           \\
mCherry\_fwd                 & ATGGTCTCAAAGGGTGAAG           \\
mCherry\_rev                 & TTATACAATTCATCCATGCC          \\
U54\_fwd                     & GCGCCGGTCGCTACCATTAC          \\
U54\_rev                     & AAGGGCCCGTACGGCCGA           
\end{tabular}
\end{table}

Transgenic animals exhibited reduced sensitivity to tap or touch compared to wild-type, presumably because Chrimson competes with endogenous MEC-4 protein for transcription, see \autoref{fig:rangeOfBehav} - Figure Supplement \ref*{figs:AML67tap}.   From the alleles we had generated, we selected AML67 for use in this study because it was the most sensitive to tap and touch, despite being reduced compared to wild-type.

\subsection{Nematode handling}
Strains were  maintained on 9 cm NGM agar plates seeded with OP50 \textit{E. coli}  food, at \SI{20}{\celsius}.  Worms were bleached 3 days prior to experiment to provide 1 day old adults. For optogenetic experiments, bleached worms were placed on plates seeded with 1mL of 0.5 mM all-trans-retinal (ATR) mixed with OP50. Control plate lacked ATR. To avoid inadvertent optogenetic activation, plates were wrapped in aluminum foil,  handled in the dark, and viewed  under dissection microscopes using dim blue light.

To harvest worms for high-throughput experiments, roughly 100 to 200 worms were cut from agar and washed and then spun-down in a 1.5 mL microcentrifuge tube. Worms at the bottom of the tube were placed on an unseeded 9 cm NGM agar plate via micropipette. Excess liquid on the plate was carefully wicked away using tissue paper. Worms were  allowed to adapt to their new environment for 25 minutes before recordings or stimulation. 

\subsection{High-throughput imaging}
Experiments were conducted in a custom-built high-throughput imaging rig (\autoref{fig:rangeOfBehav} - Figure Supplement \ref*{figs:expdesign}). Plates of  animals were  recorded while undergoing 30 minutes of optogenetic or tap stimulation. Imaging was performed as follows: The agar plate was illuminated by a ring of 850 nm infrared LEDs (irrf850-5050-60-reel, environmentallights.com). A 2,592 x 1,944 pixel CMOS camera (ACA2500-14um, Basler) recorded worm movements at 14 frames per second and a magnification of 20 \textmu m per pixel so as to provide sufficient spatiotemporal resolution to capture posture dynamics.  The field of view of the camera was centered on the plate and included approximately 50\% of plate surface.  Custom LabVIEW software acquired images from the camera and controlled stimulus delivery as described below.

\subsection{Tap delivery}
Taps were delivered to the side of 9 cm plates containing the animals by means of a solenoid, similar to  \citep{swierczek_high-throughput_2011}.  An electric solenoid tapper (Small Push-Pull Solenoid, Adafruit) was driven with a 70 ms, 24 V, DC pulse under Labview control via a LabJack DAQ and a solid-state relay. During tap experiments, taps were delivered to the plate once per minute for 30 minutes, see \autoref{tab:experimentsTable}. The 1 minute inter-stimulus interval was chosen to minimize habituation \citep{timbers_intensity_2013}.

\subsection{Optogenetic stimulation}
Experiments involving optogenetic stimulation are summarized in \autoref{tab:experimentsTable}. Optogenetic stimulation was delivered by three 625 nm LEDs  (M625L3, Thorlabs) positioned such that their light approximately tiles the agar plate visible in the camera's field of view. LED's were driven by a diode driver  (L2C210C, Thorlabs) under the control of  LabVIEW via an analog signal from a LabJack DAQ (Model U3-HV with LJTick-DAC).  The range of the light intensity for optogenetic stimulation averaged at the plate spanned from 0 to 80 \textmu W mm$^{-2}$. Small spatial inhomogeneities  in light intensity were characterized and accounted for in software so as to calculate the precise light intensity delivered to each animal. A infared long pass filter (FEL0800, Thorlabs) in front of the camera blocked light from the stimulus LEDs  and only permitted light from the infrared behavior LEDs. 

\subsubsection{Optogenetic pulse stimulus}
For optogenetic pulse experiments, as in \autoref{fig:rangeOfBehav}, a 1 second light pulses was delivered  once per minute for 30 minutes. Initial experiments measured the behavioral responses to pulses of different light intensities. In those experiments,  shown in \autoref{fig:rangeOfBehav}c, the light intensity of the pulse was randomly shuffled such that five pulses each of  2, 5, 10, 50, and 80  \textmu W mm$^{-2}$ were delivered during the 30 minute recording.

\subsubsection{Random noise optogenetic stimulus}
Experiments involving reverse correlation all used a light stimulus with intensity modulated by random broad-spectrum noise.   The random noise stimulus was generated according to,
\begin{equation}
s(t+1) = As(t)+Bn_{\textrm{rand}}+C,
\end{equation}
where $A \equiv \exp{-(\tau_{\textrm{period}}/\tau_c)}$ and  $B \equiv \sigma_{\textrm{rms}} \sqrt{1-A^2}$.
Here \(s(t+1)\) is the stimulus intensity at the next time-point, \(A\) is the weighting of the previous stimulus \(s(t)\), \(B\) is the weighting of a random number, \(n_{\textrm{rand}}\), drawn from a Gaussian distribution with standard deviation given by  \(\sigma_{\textrm{rms}}\), and \(C\) is a constant offset that sets the average stimulus intensity. The weighting $A$ is related to  correlation time \(\tau_c\) and the duration of our time step  \(\tau_{\textrm{period}}\).  Because in our setup the stimulus is updated with each image acquisition, the time step \(\tau_{\textrm{period}}\) is the inverse of the image acquisition rate, or approximately  0.07 s for 14 Hz.

Both \(C\) and \(\sigma_{\textrm{rms}}\) were chosen to be 25 \textmu W mm$^{-2}$ so that the the function generated intensities that mostly fell in the intensity range of 0 to 50  \textmu W mm$^{-2}$, a regime that appeared to be most sensitive to behavior response (see \autoref{fig:rangeOfBehav}c; ).  \(\tau_c\) was chosen to be 0.5 s as this roughly matched our intuition about the timescale of temporally varying mechanical stimuli that the animal might encounter while navigating its natural environment. Finally, the stimulus was clipped and forced to stay in the range of 0 to 50 \textmu W mm$^{-2}$. Frequency spectra of our stimuli is shown in \autoref{fig:kernel} - Figure Supplement \ref*{figs:powerspectra}.

\subsubsection{Triangle wave optogenetic stimulus}
Triangle wave  stimuli  were also generated.  Triangle waves were linearly increasing ramps of light intensity  from 0 \textmu W mm$^{-2}$ to 50 \textmu W mm$^{-2}$ for 10 seconds followed by linearly decreasing ramps  50 \textmu W mm$^{-2}$ to 0 \textmu W mm$^{-2}$ for 10 seconds, repeated continuously for 30 minutes.

\subsubsection{Kernel-shaped (tailored) stimulus}
In the tailored stimulation experiments,  stimuli were generated from the behavior triggered averages found using reverse correlation. The six behavior triggered averages from  \autoref{fig:kernel} were scaled in intensity until  either their minimum was at 0 \textmu W mm$^{-2}$ or the maximum was at 50 \textmu W mm$^{-2}$. These were then shuffled and played back one per minute such that each behavior triggered averages was delivered 5 times per 30 minute experiment.  25 \textmu W mm$^{-2}$ of constant light intensity was delivered between stimulus presentation.

\begin{landscape}
\begin{table}[bt]
\caption{\label{tab:experimentsTable} Summary of experimental conditions. Each experimental series consisted of recordings of multiple plates usually spread across multiple days, as indicated. Recordings were all 30 mins in duration per plate. Note two methods were used to tally the number of stimulus-animal presentations (see methods). Here a stimulus presentation is counted even if the track was interrupted mid-presentation.}

\begin{tabular}{llllllllllll}
\toprule
\begin{tabular}[x]{@{}c@{}}Experiment\\Series\end{tabular} &Strain &Stim& ATR & \begin{tabular}[x]{@{}c@{}}\# of\\Plates\end{tabular} & \begin{tabular}[x]{@{}c@{}}\# of\\Days\end{tabular}  & \begin{tabular}[x]{@{}c@{}}Interstim\\Interval\\(s)\end{tabular} & \begin{tabular}[x]{@{}c@{}}Stimulus\\Duration\\(s)\end{tabular} & \begin{tabular}[x]{@{}c@{}}Total\\Animal-\\Stimulus\\Presentations\end{tabular} & \begin{tabular}[x]{@{}c@{}}Cumulative\\Recording\\Length\\(animal-hours)\end{tabular} & \begin{tabular}[c]{@{}c@{}}Animals\\per frame\\(Mean$\pm$Stdev)\end{tabular} & Figures \\
\toprule
\multirow{2}{*}[-1em]{\begin{tabular}[x]{@{}c@{}}Random\\Noise\end{tabular}} & \multirow{2}{*}[-1em]{AML67} & \multirow{2}{*}[-1em]{Light}& + & 58 &3  & n/a & n/a & n/a & 1,784 & 62$\pm$34 & \begin{tabular}[c]{@{}l@{}}\autoref{fig:introBehavior}, \autoref{fig:introBehavior}-Supp \ref*{figs:suppmaps},\\ 
\autoref{fig:introBehavior}-Supp \ref*{video:brady},\\ \autoref{fig:introBehavior}-Supp \ref*{video:trace}, \autoref{fig:kernel},\\ \autoref{fig:kernel}-Supp \ref*{figs:ATRdiff}, \autoref{fig:kernel}-Supp \ref*{figs:nonlinearity},\\ \autoref{fig:kernel}-Supp \ref*{figs:powerspectra}, \autoref{fig:velocitycontext},\\ \autoref{fig:velocitycontext}-Supp \ref*{figs:allkernels}, \autoref{fig:attention}\end{tabular} 
\\ \cline{4-12}
&  &  & - & 20 &3 & n/a & n/a & n/a & 500 & 50$\pm$23 & \begin{tabular}[c]{@{}l@{}}\autoref{fig:introBehavior},\autoref{fig:introBehavior}-Supp \ref*{figs:suppmaps},\\ \autoref{fig:introBehavior}-Supp \ref*{video:brady}, \autoref{fig:kernel}-Supp \ref*{figs:ATRdiff}, \\ \autoref{fig:kernel}-Supp \ref*{figs:noretkernels}, \autoref{fig:velocitycontext}-Supp \ref*{figs:noretallkernels}\end{tabular} \\
\toprule
\multirow{2}{*}{\begin{tabular}[x]{@{}c@{}}Triangle\\Wave\end{tabular}} &\multirow{2}{*}{AML67} & \multirow{2}{*}{Light} & + & 62 &3  & 0 & 20 & 340,757 & 1,912 & 62$\pm$42 & \autoref{fig:trianglewave}, \autoref{fig:samestimDiffresponse} \\ \cline{4-12}
&  &  & - & 20 &3  & 0 & 20 & 142,461 & 800 & 80$\pm$55 & \autoref{fig:trianglewave}-Supp \ref*{figs:trianglecontrol} \\
\toprule
\multirow{2}{*}[.8em]{\begin{tabular}[x]{@{}c@{}}Kernel-\\Shaped\\Stimuli\end{tabular}} &\multirow{2}{*}{AML67} & \multirow{2}{*}{Light} & + & 44 &3  & 40 & 20 & 84,875 & 1,453 & 66$\pm$40 & \begin{tabular}[c]{@{}l@{}}\autoref{fig:playback}, \autoref{fig:playback}-Supp \ref*{figs:allplayback}\\ \autoref{fig:samestimDiffresponse}\end{tabular} \\ \cline{4-12}
& & & - & 12 &3 & 40 & 20 & 22,866 & 392 & 65$\pm$33 & \autoref{fig:playback}-Supp \ref*{figs:playbackcontrol} \\
\toprule
\multirow{2}{*}{\begin{tabular}[x]{@{}c@{}}Light Pulse\end{tabular}}& \multirow{2}{*}{AML67} & \multirow{2}{*}{Light}& + & 12 &1 & 59 & 1 & 15,128 & 260 & 43$\pm$24 & \autoref{fig:rangeOfBehav}, \autoref{fig:attention} \\ \cline{4-12}
& &  & - & 6 &1 & 59 & 1 & 8,107 & 139 & 46$\pm$18 & \autoref{fig:rangeOfBehav}-Supp \ref*{figs:lightpulsecontrol} \\
\toprule
\multirow{3}{*}{\begin{tabular}[x]{@{}c@{}}Plate Tap\end{tabular}} &\multirow{2}{*}{AML67} & \multirow{2}{*}{Tap} & + & 7 &2  & 60 & Impulse & 21,117 & 366 & 105$\pm$60 &\multirow{2}{*}{ \autoref{fig:rangeOfBehav}-Supp \ref*{figs:AML67tap} }\\ \cline{4-11}
&  &  & - & 8 &2 & 60 & Impulse & 14,646 & 254 & 64$\pm$46 & \\ \cline{2-12}
& N2 & Tap & - & 22 &3 & 60 & Impulse & 40,409 & 695 & 63$\pm$25 & \begin{tabular}[c]{@{}l@{}}\autoref{fig:rangeOfBehav}, \autoref{fig:attention},\\ \autoref{fig:attention}-Supp \ref*{figs:platetap}\end{tabular} \\
\bottomrule
Total & &  & & 271 & & & & &8,554 & 63$\pm$40 & 
\end{tabular}
\end{table}
\end{landscape}

\subsection{Measuring animal behavior}
The unsupervised behavior mapping approach used in this work is adapted from the fly \citep{berman_mapping_2014} and is similar in spirit to work in rodents  \citep{wiltschko_mapping_2015}. It also builds upon decades of methodological advances quantifying \textit{C. elegans}  behavior \citep{croll_behavoural_1975, croll_components_1975, stephens_dimensionality_2008, ramot_parallel_2008, brown_dictionary_2013, yemini_database_2013, gyenes_deriving_2016, gomez-marin_hierarchical_2016}.  

Animal behavior was measured and classified using an analysis pipeline, summarized in \autoref{fig:introBehavior} - Figure Supplement \ref*{figs:pipeline}. First, worms were located and tracked, then their posture was extracted, and finally their posture dynamics were clustered and classified. Details of the pipeline are described below. The pipeline was written in MATLAB and run on the  Princeton University's high performance parallel computing cluster. Source code is available at (\href{https://github.com/leiferlab/liu-temporal-processing}{https://github.com/leiferlab/liu-temporal-processing}). 

\subsubsection{Animal location tracking}
To first identify animals and track their location, raw video of animals on plates  were analyzed using a modified version of the Parallel Worm Tracker \citep{ramot_parallel_2008}. Animal's were found via binary thresholding and centroid tracking
.

\subsubsection{Animal posture extraction}
The animal's posture was found  by extracting the animal's centerline from the video using custom MATLAB scripts.   Videos of each individual worm were first generated  by cropping a 70x70 pixel region around the worm's centroid at every frame. A centerline with 20 points was fitted to the image at each frame using an active contour model similar to \citep{nguyen_automatically_2017}, inspired by the one described by \citep{deng_efficient_2013}. The algorithm for fitting the centerline was specifically optimized to measure posture of the worm in a variety of conditions, including when the animal crossed over itself during turns. The active contour model fits the centerline by relaxing contiguous points along a gradient defined by four forces: (1) an image force that fit the contour to the image of the worm; (2) a tip force that guides the beginning and end of the contour to the worm's presumptive head and tail; (3) a spring force that guides the contour to be similar lengths; (4) and a repel force that makes sure that the contour does not stick to itself. To ensure continuity in time, the active contour of the following frame is initialized by the relaxed contour of the previous frame. The head and the tail of the worm was determined by assuming the worm moves forward the majority of the time. A quality score was calculated to estimate how well the centerline fit the image and how much it displaced from the previous centerline. On the rare occasion when the quality score of a frame fell below  threshold, that  frame was dropped, and the track was split into two.


\subsubsection{Posture dimensionality reduction}
To more efficiently interpret thhe animal's posture, the dimensionality of the animal's centerline was reduced from 20 position $(x,y)$ coordinate to five posture coefficients using principle component analysis (PCA), following \citep{stephens_dimensionality_2008}. Principle components of posture were extracted from recordings of approximately 2 million animal-frames of freely behaving N2 worms. Centerlines were converted into a series of angles oriented such that the mean angle is 0. The first five principle components explain  >98\% of the posture variance.  The animal's posture dynamics were thus represented as a time-series of five coefficients, one for each of the five principle posture modes.

\subsubsection{Generating spectrograms of posture dynamics}
To characterize posture dynamics,  a spectrogram was generated for each of the posture mode coefficients, as in \citep{berman_mapping_2014}. A Morlet continuous wavelet transform was performed on each of the 5 coefficient time series at 25 frequencies dyadically spaced between 0.3 Hz and 7 Hz. The low frequency bound was chosen to reflect our intuition regarding the timescale of \textit{C. elegans} behavior and the high frequency bound was set by the Nyquist sampling frequency of our image acquisition. The spectrogram  provides information about the frequency spectra of the animal's posture dynamics but it lacks information about the phase of the animal's posture, which is important for discerning forward from backward locomotion. To preserve forward and backward information, we  introduced  a binary ``directionality'' vector that is 2 when the worm centroid is moving forward, and 1 when the worm centroid is backwards. Directionality was calculated by taking  the sign of the dot product of the head vector with a tangent vector of the animal's centroid trajectory. Together, the five spectrograms and directionality vector provide a 126 dimensional feature vector that describe the animal's behavior at each time point.  It is this feature vector that is clustered, as described below.

\subsubsection{Defining the behavioral map and behavior states}
To classify behavior into discrete stereotyped behavior states that emerge naturally from our recordings, we  followed a behavior-mapping strategy described in \citep{berman_mapping_2014}. A single behavior map was generated so that behaviors were defined consistently across all experiments. To generate the behavior map, 50,000 animal-time points were uniformly sampled from the 2,284 animal-hours of behavior recordings made during random-noise optogenetic stimulation. Each animal-time point contributes a 126-dimensional feature vector describing the animal's instantaneous behavior.  We generated a 2D map of these feature vectors by embedding the 126 dimensional space in a plane using non-linear dimensionality reduction technique called t-distributed stochastic neighbor embedding (t-SNE) \citep{maaten_visualizing_2008}. Under t-SNE, each feature vector is embedded such that the local distance between feature vectors is conserved but  long distance scales are distorted (\autoref{fig:rangeOfBehav} - Figure Supplement \ref*{figs:suppmaps}a).  

We then generated a probability density histogram of behavior by projecting all $10^8$ behavior time points from the 2,284 animal-hours of random noise optogenetic stimulation  (\autoref{fig:introBehavior} - Figure Supplement \ref*{figs:suppmaps}b) into the 2D map. Clusters of high probability in this density map corresponded to a distinct stereotyped behavior.  Stereotype behaviors were defined by water-shedding the probability density map  (\autoref{fig:introBehavior} - Figure Supplement \ref*{figs:suppmaps}c)   and each region was assigned a name like ``Forward 3''  (\autoref{fig:introBehavior}b).  Examples of worms in each region are shown in \autoref{fig:introBehavior} - Figure Supplement \ref*{video:brady}.  Time points from subsequent recordings were similarly projected into this map for the purposes of classifying animal behavior.

\subsubsection{Identifying behavioral transitions}
At each time point, the worm belongs to a point in the 2D behavior map described above (see \autoref{fig:introBehavior} - Figure Supplement \ref*{video:trace}).  Animals that dwelled in one behavior region for at least 0.5 seconds were classified as exhibiting that behavior during all contiguous time points   in that behavior region. Animal's inhabiting a behavior region for less then 0.5 seconds were classified as ``in transition.''  

A  ``transition into behavior $X$''  is defined to occur on  the first time point that the animal is classified as in $X$. Transitions from  behavior $W\rightarrow X$ were defined to occur on the first time point the animal is classified as in  $X$  provided that:  the animal  transitioned directly from $W$ to $X$; or the animal had previously been classified as in $W$, was then  classified as ``in transition'' and then was classified as in state $X$. Cases where the animal was in  $X$ , then ``in transition'' and then returned to   $X$, were ignored.

\subsubsection{Ambiguities in temporal definition of behavior}
The wavelet spectrogram introduces an inherent uncertainty in  the precise timing of a behavior transition.  This ultimately arises from the uncertainty principle: behavior dynamics that have low frequency components provide less temporal resolution than higher frequency dynamics.  An equivalent view  is that the spectrogram feature vector at any given moment is influenced by temporally adjacent postural dynamics in the past and future, and this influence is stronger at lower frequencies than higher ones. 

This temporal uncertainty or ``bleeding over'' of future behavior,  causes the animal occasionally to appear to (but not actually to) respond to a stimulus prior to its delivery. In the worst case, the  time-scale of this leakage is set by our choice of the lowest frequency wavelet, which is 0.3 Hz (i.e. 2.7 seconds). Behaviors with strong higher-frequency components have shorter timescale uncertainties. We take large time windows of 20 seconds to define our kernels, in part, so that a few second time-shift does not result in any loss of information.

\subsection{Reverse correlation}
Reverse correlation was used to find a linear kernel and non-linearity that describe the relationship between the animal's behavior transitions and an applied stimulus.

\subsubsection{Calculating kernels}
Linear kernels for each behavior were estimated by computing the behavior-triggered-average of the stimulus,
\begin{equation}
\hat{A} = \dfrac{1}{N} \sum_{n=1}^N \vec{s}(t_n),
\end{equation}
where \(t_n\) is the time of \textit{n}th behavioral transition, \(\vec{s}(t_n)\) is a vector representing the stimuli presented during a 20 second temporal window around \(t_n\) and \(N\) is the total number of behavioral transitions \citep{schwartz_spike-triggered_2006}. The linear kernel was estimated to be the mean-subtracted, time-reversed behavior-triggered average.

\subsubsection{Kernel significance}
Behavior triggered averages (also referred to as kernels) were deemed significant if their magnitude (L2 norm) exceeded the top 1 percent of a distribution of random kernels found by shuffling the stimulus in time.   Shuffling was performed in such a way as to preserve the temporal properties of the transition train while completely decorrelating it from the stimulus.   Specifically, shuffling was performed by circle-shifting the transition timings within every track by a randomly selected integer  between 1 and the number of time points in the track.  Shuffled kernel distributions for each behavior were generated by recalculating the behavior triggered-average 100 times, each with different circle-shifted timings.

\subsubsection{Estimating the non-linearity}
The non-linearity $f$ allows the probability of a behavior transition to be estimated from the filtered signal namely the stimulus convolved with the linear kernel \citep{gepner_computations_2015}.  Non-linearities were estimated  from the ratio of two histograms: the first is a histogram of time-point counts versus filtered signal given  a behavioral transition at  that time-point, and the second is a histogram of time-point counts versus filtered signal for all time-points \citep{schwartz_spike-triggered_2006}. Histograms were tabulated with 10 equally spaced bins spanning the  range of the filtered signal.  Bin-wise division of the two histograms   yielded 10 points relating probability of behavior to filtered signals (\autoref{fig:kernel} - Figure Supplement \ref*{figs:nonlinearity} ). For each point, we calculate a propagated error, $E$, assuming Poisson counting statistics,
\begin{equation}
E = \sqrt{\dfrac{T-1}{F^2} + \dfrac{T^2(F-1)}{F^4}},
\end{equation}
where $T$ is the number of behavioral transitions in that bin, and $F$ is the number of filtered signal time-points in that bin. We then fitted a 2 parameter exponential to the 10 points, weighing each point by the inverse of the error in order to reduce the influence of noise. This fitted exponential function is our estimate of the non-linearity.

\subsection{Calculating transition rates}
When presented as a timeseries of rates, as in \autoref{fig:rangeOfBehav} - Figure Supplement \ref*{figs:transitionrates},  transition rates were calculated according to the following:  Behavior timeseries from all recordings were cropped in a time window around each stimulus, commingled, and then time aligned to the stimulus. The fraction of all animals undergoing a transition was calculated at each time step. The fractions of animal were directly converted into a rate of transitions per animal per minute, yielding the timeseries of rates.

\subsubsection{Calculating transition rate changes}
Transition rate change, as in \autoref{fig:playback}b, \autoref{fig:playback} - Figure Supplement \ref*{figs:allplayback}, \autoref{fig:trianglewave}, and \autoref{fig:samestimDiffresponse}, were calculated as follows: an average transition rate was found in a time window during a stimulus (as described above), and then a baseline was subtracted off.  For kernel-shaped stimuli experiments   (\autoref{fig:playback}a and \autoref{fig:playback} - Figure Supplement \ref*{figs:allplayback}) the baseline is defined as the average transition rate in a 20 second time window prior to each stimulus. For the triangle wave in \autoref{fig:trianglewave}, the baseline was defined to be the overall mean transition rate throughout the recording.

In cases where a bar is shown  (\autoref{fig:playback}a, \autoref{fig:samestimDiffresponse}), a  change in transition rate was calculated by averaging the timeseries of rates over a time window (indicated in figures with gray shading). 

\subsubsection{Measuring transition rates for  tap or light induced context dependent transitions}
Transition rates were calculated slightly differently in \autoref{fig:attention} and \autoref{fig:attention} - Figure Supplement \ref*{figs:optotap} and \ref*{figs:platetap} to facilitate significance testing via the E-test \cite{krishnamoorthy_more_2004}. The transition rate in a two second time window immediately following light pulse or tap (+) was compared to a transition rate in a two second window immediately following a mock control (-). Mock controls were set to occur at the mid point between consecutive stimuli.

Instead of calculating the transition rates at each time bin and then averaging across time, as described previously, we instead calculated a single transition rate for the entire two second time window by comingling transitions from all time bins, as follows. We 1) selected tracks that were uninterrupted for the 2 seconds, (2) counted the total number of  given transition in the 2 seconds after stimulus onset across all of our experiments, (3) divided  by the total number of tracked time points, and (4) converted the value to transitions per animal per minute. The number of stimulus-animal presentations differs slightly from those in \autoref{fig:rangeOfBehav} because now tracks are required to be contiguous for 2 seconds after stimulus presentation, which was not a requirement previously. 

P-values were attained using an E-test \citep{krishnamoorthy_more_2004}. To account for testing 72  behavior transitions concurrently, we use the Bonferroni multiple-hypothesis correction. Only p-values less than  $\alpha = 0.05/72 = 7\cdot10^{-1}$ are considered significant.

In our analysis of light pulse response, we grouped all stimulation light intensities together. 

\section{Data}

Behavioral analysis and stimulation data for all tracked animals in all experiments in \autoref{tab:experimentsTable} are available at \url{https://doi.org/10.6084/m9.figshare.5956348}. See dataset README for details. All recorded data, including raw images  (>1 TB), are being made available at  \url{http://dx.doi.org/10.21227/H27944}. 

\section{Acknowledgments}
We thank  Marc Gershow (NYU)  and Gordon Berman  (Emory) for productive discussions and troubleshooting. We also thank 
Mala Murthy and Jonathan Pillow, both of Princeton University, for productive discussions and feedback.  Alicia Castillo Bahena and Tayla Duarte contributed to preliminary studies of this work. 

Wild type N2-type strains were procured from Caenorhabditis Genetic Center (CGC), Minnesota.

This work was supported by grants from the Simons Foundation (SCGB \#324285, and SCGB \#543003, A.M.L.). This work was also supported by Princeton University's Dean for Research Innovation Fund to A.M.L.~ and J.W.S.  Research reported in this publication was supported by the National Human Genome Research Institute of the National Institutes of Health under Award Number T32HG003284. The content is solely the responsibility of the authors and does not necessarily represent the official views of the National Institutes of Health. 

\section{Author Contributions}
M.L.~conducted all experiments and analysis. A.K.S.~performed all transgenics. A.M.L., J.W.S. and M.L.~conceived all experiments.  All authors reviewed and edited the manuscript.   



\begin{thebibliography}{55}
\providecommand{\natexlab}[1]{#1}
\providecommand{\urlprefix}{}
\providecommand{\doiprefix}{doi: }

\bibitem[{Ardiel et~al.(2017)Ardiel, Evan L. and Yu, Alex J. and Giles, Andrew
  C. and Rankin, Catharine H.}]{ardiel_habituation_2017}
\textbf{\color{eLifeMediumGrey} Ardiel EL}, Yu AJ, Giles AC, Rankin CH.
\newblock Habituation as an adaptive shift in response strategy mediated by
  neuropeptides.
\newblock npj Science of Learning.  2017 Aug; 2(1):9.
\newblock \urlprefix\url{https://www.nature.com/articles/s41539-017-0011-8},
  \href{10.1038/s41539-017-0011-8}{\doiprefix 10.1038/s41539-017-0011-8}.

\bibitem[{Behnia et~al.(2014)Behnia, Rudy and Clark, Damon A. and Carter, Adam
  G. and Clandinin, Thomas R. and Desplan, Claude}]{behnia_processing_2014}
\textbf{\color{eLifeMediumGrey} Behnia R}, Clark DA, Carter AG, Clandinin TR,
  Desplan C.
\newblock Processing properties of {ON} and {OFF} pathways for
  \textit{{Drosophila}} motion detection.
\newblock Nature.  2014 Aug; 512(7515):427--430.
\newblock \urlprefix\url{https://www.nature.com/articles/nature13427},
  \href{10.1038/nature13427}{\doiprefix 10.1038/nature13427}.

\bibitem[{Berman et~al.(2014)Berman, Gordon J. and Choi, Daniel M. and Bialek,
  William and Shaevitz, Joshua W.}]{berman_mapping_2014}
\textbf{\color{eLifeMediumGrey} Berman GJ}, Choi DM, Bialek W, Shaevitz JW.
\newblock Mapping the stereotyped behaviour of freely moving fruit flies.
\newblock Journal of the Royal Society, Interface / the Royal Society.  2014
  Oct; 11(99).
\newblock \href{10.1098/rsif.2014.0672}{\doiprefix 10.1098/rsif.2014.0672}.

\bibitem[{Brown et~al.(2013)Brown, André E. X. and Yemini, Eviatar I. and
  Grundy, Laura J. and Jucikas, Tadas and Schafer, William
  R.}]{brown_dictionary_2013}
\textbf{\color{eLifeMediumGrey} Brown AEX}, Yemini EI, Grundy LJ, Jucikas T,
  Schafer WR.
\newblock A dictionary of behavioral motifs reveals clusters of genes affecting
  {Caenorhabditis} elegans locomotion.
\newblock Proceedings of the National Academy of Sciences of the United States
  of America.  2013 Jan; 110(2):791--796.
\newblock \href{10.1073/pnas.1211447110}{\doiprefix 10.1073/pnas.1211447110}.

\bibitem[{Calhoun and Murthy(2017)Calhoun, Adam J and Murthy,
  Mala}]{calhoun_quantifying_2017}
\textbf{\color{eLifeMediumGrey} Calhoun AJ}, Murthy M.
\newblock Quantifying behavior to solve sensorimotor transformations: advances
  from worms and flies.
\newblock Current Opinion in Neurobiology.  2017 Oct; 46:90--98.
\newblock
  \urlprefix\url{http://www.sciencedirect.com/science/article/pii/S095943881730123X},
  \href{10.1016/j.conb.2017.08.006}{\doiprefix 10.1016/j.conb.2017.08.006}.

\bibitem[{Chalfie et~al.(1985)Chalfie, M. and Sulston, J. E. and White, J. G.
  and Southgate, E. and Thomson, J. N. and Brenner, S.}]{chalfie_neural_1985}
\textbf{\color{eLifeMediumGrey} Chalfie M}, Sulston JE, White JG, Southgate E,
  Thomson JN, Brenner S.
\newblock The neural circuit for touch sensitivity in {Caenorhabditis} elegans.
\newblock The Journal of Neuroscience: The Official Journal of the Society for
  Neuroscience.  1985 Apr; 5(4):956--964.

\bibitem[{Chalfie and Sulston(1981)Chalfie, Martin and Sulston,
  John}]{chalfie_developmental_1981}
\textbf{\color{eLifeMediumGrey} Chalfie M}, Sulston J.
\newblock Developmental genetics of the mechanosensory neurons of
  {Caenorhabditis} elegans.
\newblock Developmental Biology.  1981 Mar; 82(2):358--370.
\newblock
  \urlprefix\url{http://www.sciencedirect.com/science/article/pii/0012160681904590},
  \href{10.1016/0012-1606(81)90459-0}{\doiprefix 10.1016/0012-1606(81)90459-0}.

\bibitem[{Chiba and Rankin(1990)Chiba, C. M. and Rankin, C.
  H.}]{chiba_developmental_1990}
\textbf{\color{eLifeMediumGrey} Chiba CM}, Rankin CH.
\newblock A developmental analysis of spontaneous and reflexive reversals in
  the nematode {Caenorhabditis} elegans.
\newblock Journal of Neurobiology.  1990 Jun; 21(4):543--554.
\newblock \href{10.1002/neu.480210403}{\doiprefix 10.1002/neu.480210403}.

\bibitem[{Cho and Sternberg(2014)Cho, Julie Y. and Sternberg,
  Paul W.}]{cho_multilevel_2014}
\textbf{\color{eLifeMediumGrey} Cho J}, Sternberg P.
\newblock Multilevel {Modulation} of a {Sensory} {Motor} {Circuit} during
  {C}. elegans {Sleep} and {Arousal}.
\newblock Cell.  2014 Jan; 156(1):249--260.
\newblock
  \urlprefix\url{http://www.sciencedirect.com/science/article/pii/S0092867413015201},
  \href{10.1016/j.cell.2013.11.036}{\doiprefix 10.1016/j.cell.2013.11.036}.

\bibitem[{Clark et~al.(2006)Clark, Damon A and Biron, David and Sengupta, Piali
  and Samuel, Aravinthan D T}]{clark_afd_2006}
\textbf{\color{eLifeMediumGrey} Clark DA}, Biron D, Sengupta P, Samuel ADT.
\newblock The {AFD} sensory neurons encode multiple functions underlying
  thermotactic behavior in {Caenorhabditis} elegans.
\newblock The Journal of Neuroscience: The Official Journal of the Society for
  Neuroscience.  2006 Jul; 26(28):7444--7451.
\newblock \urlprefix\url{http://www.ncbi.nlm.nih.gov/pubmed/16837592},
  \href{10.1523/JNEUROSCI.1137-06.2006}{\doiprefix
  10.1523/JNEUROSCI.1137-06.2006}.

\bibitem[{Clark et~al.(2007)Clark, Damon A. and Gabel, Christopher V. and
  Gabel, Harrison and Samuel, Aravinthan D. T.}]{clark_temporal_2007}
\textbf{\color{eLifeMediumGrey} Clark DA}, Gabel CV, Gabel H, Samuel ADT.
\newblock Temporal {Activity} {Patterns} in {Thermosensory} {Neurons} of
  {Freely} {Moving} {Caenorhabditis} elegans {Encode} {Spatial} {Thermal}
  {Gradients}.
\newblock J Neurosci.  2007 Jun; 27(23):6083--6090.
\newblock
  \urlprefix\url{http://www.jneurosci.org/cgi/content/abstract/27/23/6083},
  \href{10.1523/JNEUROSCI.1032-07.2007}{\doiprefix
  10.1523/JNEUROSCI.1032-07.2007}.

\bibitem[{Clemens and Murthy(2017)Clemens, Jan and Murthy,
  Mala}]{clemens_use_2017}
\textbf{\color{eLifeMediumGrey} Clemens J}, Murthy M.
\newblock The {Use} of {Computational} {Modeling} to {Link} {Sensory}
  {Processing} with {Behavior} in {\textless}{Emphasis}
  {Type}="{Italic}"{\textgreater}{Drosophila}{\textless}/{Emphasis}{\textgreater}.
\newblock In: \emph{Decoding {Neural} {Circuit} {Structure} and {Function}}
  Springer, Cham; 2017.p. 241--260.
\newblock
  \urlprefix\url{https://link.springer.com/chapter/10.1007/978-3-319-57363-2_9},
  dOI: 10.1007/978-3-319-57363-2\_9.

\bibitem[{Coen et~al.(2014)Coen, Philip and Clemens, Jan and Weinstein, Andrew
  J. and Pacheco, Diego A. and Deng, Yi and Murthy, Mala}]{coen_dynamic_2014}
\textbf{\color{eLifeMediumGrey} Coen P}, Clemens J, Weinstein AJ, Pacheco DA,
  Deng Y, Murthy M.
\newblock Dynamic sensory cues shape song structure in \textit{{Drosophila} }.
\newblock Nature.  2014 Mar; 507(7491):233--237.
\newblock \urlprefix\url{https://www.nature.com/articles/nature13131},
  \href{10.1038/nature13131}{\doiprefix 10.1038/nature13131}.

\bibitem[{Croll(1975{\natexlab{a}})Croll, NA}]{croll_behavoural_1975}
\textbf{\color{eLifeMediumGrey} Croll N}.
\newblock Behavoural analysis of nematode movement.
\newblock Advances in Parasitology.  1975; 13:71--122.

\bibitem[{Croll(1975{\natexlab{b}})Croll, NA}]{croll_components_1975}
\textbf{\color{eLifeMediumGrey} Croll N}.
\newblock Components and patterns in the behavior of the nematode
  {Caenorhabditis} elegans.
\newblock Journal of Zoology.  1975; 176:159--176.
\newblock
  \urlprefix\url{http://www.wormbase.org/db/misc/biblio?name=BEHAV%2FMOVEMENT&category=&class=Keyword&abstract=WBPaper00000075#WBPaper00000075}.

\bibitem[{Deng et~al.(2013)Deng, Yi and Coen, Philip and Sun, Mingzhai and
  Shaevitz, Joshua W.}]{deng_efficient_2013}
\textbf{\color{eLifeMediumGrey} Deng Y}, Coen P, Sun M, Shaevitz JW.
\newblock Efficient {Multiple} {Object} {Tracking} {Using} {Mutually}
  {Repulsive} {Active} {Membranes}.
\newblock PLoS ONE.  2013 Jun; 8(6):e65769.
\newblock \urlprefix\url{http://dx.doi.org/10.1371/journal.pone.0065769},
  \href{10.1371/journal.pone.0065769}{\doiprefix 10.1371/journal.pone.0065769}.

\bibitem[{Driscoll and Kaplan(1997)Driscoll, Monica and Kaplan,
  Joshua}]{driscoll_mechanotransduction_1997}
\textbf{\color{eLifeMediumGrey} Driscoll M}, Kaplan J.
\newblock Mechanotransduction.
\newblock In: Riddle DL, Blumenthal T, Meyer BJ, Priess JR, editors. \emph{C.
  elegans {II}}, 2nd ed. Cold Spring Harbor (NY): Cold Spring Harbor Laboratory
  Press; 1997.\urlprefix\url{http://www.ncbi.nlm.nih.gov/books/NBK20177/}.

\bibitem[{Eastwood et~al.(2015)Eastwood, Amy L. and Sanzeni, Alessandro and
  Petzold, Bryan C. and Park, Sung-Jin and Vergassola, Massimo and Pruitt, Beth
  L. and Goodman, Miriam B.}]{eastwood_tissue_2015}
\textbf{\color{eLifeMediumGrey} Eastwood AL}, Sanzeni A, Petzold BC, Park SJ,
  Vergassola M, Pruitt BL, Goodman MB.
\newblock Tissue mechanics govern the rapidly adapting and symmetrical response
  to touch.
\newblock Proceedings of the National Academy of Sciences.  2015 Dec;
  112(50):E6955--E6963.
\newblock \urlprefix\url{http://www.pnas.org/content/112/50/E6955},
  \href{10.1073/pnas.1514138112}{\doiprefix 10.1073/pnas.1514138112}.


\bibitem[{Evans(2006)Evans, Thomas}]{evans_transformation_2006}
\textbf{\color{eLifeMediumGrey} Evans T}.
\newblock Transformation and microinjection.
\newblock WormBook.  2006;
  \urlprefix\url{http://www.wormbook.org/chapters/www_transformationmicroinjection/transformationmicroinjection.html},
  \href{10.1895/wormbook.1.108.1}{\doiprefix 10.1895/wormbook.1.108.1}.


\bibitem[{Gepner et~al.(2015)Gepner, Ruben and Skanata, Mirna Mihovilovic and
  Bernat, Natalie M. and Kaplow, Margarita and Gershow,
  Marc}]{gepner_computations_2015}
\textbf{\color{eLifeMediumGrey} Gepner R}, Skanata MM, Bernat NM, Kaplow M,
  Gershow M.
\newblock Computations underlying {Drosophila} photo-taxis, odor-taxis, and
  multi-sensory integration.
\newblock eLife.  2015 May; 4:e06229.
\newblock \urlprefix\url{https://elifesciences.org/content/4/e06229v2},
  \href{10.7554/eLife.06229}{\doiprefix 10.7554/eLife.06229}.

\bibitem[{Ghosh et~al.(2016)Ghosh, D. Dipon and Sanders, Tom and Hong, Soonwook
  and McCurdy, Li Yan and Chase, Daniel L. and Cohen, Netta and Koelle, Michael
  R. and Nitabach, Michael N.}]{ghosh_neural_2016}
\textbf{\color{eLifeMediumGrey} Ghosh DD}, Sanders T, Hong S, McCurdy LY, Chase
  DL, Cohen N, Koelle MR, Nitabach MN.
\newblock Neural {Architecture} of {Hunger}-{Dependent} {Multisensory}
  {Decision} {Making} in {C}. elegans.
\newblock Neuron.  2016 Dec; 92(5):1049--1062.
\newblock \href{10.1016/j.neuron.2016.10.030}{\doiprefix
  10.1016/j.neuron.2016.10.030}.

\bibitem[{Gomez-Marin et~al.(2016)Gomez-Marin, Alex and Stephens, Greg J. and
  Brown, André E. X.}]{gomez-marin_hierarchical_2016}
\textbf{\color{eLifeMediumGrey} Gomez-Marin A}, Stephens GJ, Brown AEX.
\newblock Hierarchical compression of {Caenorhabditis} elegans locomotion
  reveals phenotypic differences in the organization of behaviour.
\newblock Journal of The Royal Society Interface.  2016 Aug; 13(121):20160466.
\newblock
  \urlprefix\url{http://rsif.royalsocietypublishing.org/content/13/121/20160466},
  \href{10.1098/rsif.2016.0466}{\doiprefix 10.1098/rsif.2016.0466}.

\bibitem[{Gyenes and Brown(2016)Gyenes, Bertalan and Brown, André E.
  X.}]{gyenes_deriving_2016}
\textbf{\color{eLifeMediumGrey} Gyenes B}, Brown AEX.
\newblock Deriving {Shape}-{Based} {Features} for {C}. elegans {Locomotion}
  {Using} {Dimensionality} {Reduction} {Methods}.
\newblock Frontiers in Behavioral Neuroscience.  2016; 10.
\newblock
  \urlprefix\url{https://www.frontiersin.org/articles/10.3389/fnbeh.2016.00159/full},
  \href{10.3389/fnbeh.2016.00159}{\doiprefix 10.3389/fnbeh.2016.00159}.

\bibitem[{Hernandez-Nunez et~al.(2015)Hernandez-Nunez, Luis and Belina, Jonas
  and Klein, Mason and Si, Guangwei and Claus, Lindsey and Carlson, John R. and
  Samuel, Aravinthan Dt}]{hernandez-nunez_reverse-correlation_2015}
\textbf{\color{eLifeMediumGrey} Hernandez-Nunez L}, Belina J, Klein M, Si G,
  Claus L, Carlson JR, Samuel AD.
\newblock Reverse-correlation analysis of navigation dynamics in {Drosophila}
  larva using optogenetics.
\newblock eLife.  2015 May; 4.
\newblock \href{10.7554/eLife.06225}{\doiprefix 10.7554/eLife.06225}.

\bibitem[{Kato et~al.(2014)Kato, Saul and Xu, Yifan and Cho, Christine E. and
  Abbott, L. F. and Bargmann, Cornelia I.}]{kato_temporal_2014}
\textbf{\color{eLifeMediumGrey} Kato S}, Xu Y, Cho CE, Abbott LF, Bargmann CI.
\newblock Temporal {Responses} of {C}. elegans {Chemosensory} {Neurons} {Are}
  {Preserved} in {Behavioral} {Dynamics}.
\newblock Neuron.  2014 Feb; 81(3):616--628.
\newblock
  \urlprefix\url{http://www.cell.com/neuron/abstract/S0896-6273(13)01088-X},
  \href{10.1016/j.neuron.2013.11.020}{\doiprefix 10.1016/j.neuron.2013.11.020}.

\bibitem[{Kindt et~al.(2007)Kindt, Katie S. and Quast, Kathleen B. and Giles,
  Andrew C. and De, Subhajyoti and Hendrey, Dan and Nicastro, Ian and Rankin,
  Catharine H. and Schafer, William R.}]{kindt_dopamine_2007}
\textbf{\color{eLifeMediumGrey} Kindt KS}, Quast KB, Giles AC, De S, Hendrey D,
  Nicastro I, Rankin C, Schafer WR.
\newblock Dopamine {Mediates} {Context}-{Dependent} {Modulation} of {Sensory}
  {Plasticity} in {C}. elegans.
\newblock Neuron.  2007 Aug; 55(4):662--676.
\newblock
  \urlprefix\url{http://www.sciencedirect.com/science/article/pii/S0896627307005442},
  \href{10.1016/j.neuron.2007.07.023}{\doiprefix 10.1016/j.neuron.2007.07.023}.

\bibitem[{Kitamura et~al.(2001)Kitamura, Kei-ichiro and Amano, Shigetoyo and
  Hosono, Ryuji}]{kitamura_contribution_2001}
\textbf{\color{eLifeMediumGrey} Kitamura Ki}, Amano S, Hosono R.
\newblock Contribution of neurons to habituation to mechanical stimulation in
  {Caenorhabditis} elegans.
\newblock Journal of Neurobiology.  2001 Jan; 46(1):29--40.
\newblock
  \urlprefix\url{http://onlinelibrary.wiley.com/doi/10.1002/1097-4695(200101)46:1<29::AID-NEU3>3.0.CO;2-8/abstract},
  \href{10.1002/1097-4695(200101)46:1<29::AID-NEU3>3.0.CO;2-8}{\doiprefix
  10.1002/1097-4695(200101)46:1<29::AID-NEU3>3.0.CO;2-8}.

\bibitem[{Krishnamoorthy and Thomson(2004)Krishnamoorthy, K. and Thomson,
  Jessica}]{krishnamoorthy_more_2004}
\textbf{\color{eLifeMediumGrey} Krishnamoorthy K}, Thomson J.
\newblock A more powerful test for comparing two {Poisson} means.
\newblock Journal of Statistical Planning and Inference.  2004 Jan;
  119(1):23--35.
\newblock \urlprefix\url{http://www.ucs.louisiana.edu/~kxk4695/JSPI-04.pdf},
  \href{10.1016/S0378-3758(02)00408-1}{\doiprefix
  10.1016/S0378-3758(02)00408-1}.

\bibitem[{Leifer et~al.(2011)Leifer, Andrew M and Fang-Yen, Christopher and
  Gershow, Marc and Alkema, Mark J and Samuel, Aravinthan D
  T}]{leifer_optogenetic_2011}
\textbf{\color{eLifeMediumGrey} Leifer AM}, Fang-Yen C, Gershow M, Alkema MJ,
  Samuel ADT.
\newblock Optogenetic manipulation of neural activity in freely moving
  {Caenorhabditis} elegans.
\newblock Nature Methods.  2011 Feb; 8(2):147--152.
\newblock \urlprefix\url{http://www.ncbi.nlm.nih.gov/pubmed/21240279},
  \href{10.1038/nmeth.1554}{\doiprefix 10.1038/nmeth.1554}.

\bibitem[{Maaten and Hinton(2008)Maaten, Laurens van der and Hinton,
  Geoffrey}]{maaten_visualizing_2008}
\textbf{\color{eLifeMediumGrey} Maaten Lvd}, Hinton G.
\newblock Visualizing {Data} using t-{SNE}.
\newblock Journal of Machine Learning Research.  2008; 9(Nov):2579--2605.
\newblock \urlprefix\url{http://jmlr.org/papers/v9/vandermaaten08a.html}.

\bibitem[{Mazzochette et~al.(2018)Mazzochette, Eileen A. and Nekimken, Adam L.
  and Loizeau, Frederic and Whitworth, John and Huynh, Brian and Goodman,
  Miriam B. and Pruitt, Beth L.}]{mazzochette_tactile_2018}
\textbf{\color{eLifeMediumGrey} Mazzochette EA}, Nekimken AL, Loizeau F,
  Whitworth J, Huynh B, Goodman MB, Pruitt BL.
\newblock The {Tactile} {Receptive} {Fields} of {Freely} {Moving}
  {Caenorhabditis} elegans {Nematodes}.
\newblock bioRxiv.  2018 Mar; p. 259937.
\newblock
  \urlprefix\url{https://www.biorxiv.org/content/early/2018/03/05/259937},
  \href{10.1101/259937}{\doiprefix 10.1101/259937}.

\bibitem[{Maguire et~al.(2011)Maguire, Sean M. and Clark, Christopher M. and
  Nunnari, John and Pirri, Jennifer K. and Alkema, Mark J.}]{maguire_c._2011}
\textbf{\color{eLifeMediumGrey} Maguire S}, Clark C, Nunnari J, Pirri J, Alkema
  M.
\newblock The {C}. elegans {Touch} {Response} {Facilitates} {Escape} from
  {Predacious} {Fungi}.
\newblock Current Biology.  2011 Aug; 21(15):1326--1330.
\newblock
  \urlprefix\url{http://www.sciencedirect.com/science/article/pii/S096098221100769X},
  \href{10.1016/j.cub.2011.06.063}{\doiprefix 10.1016/j.cub.2011.06.063}.

\bibitem[{Meister and Berry(1999)Meister, M. and Berry, M.
  J.}]{meister_neural_1999}
\textbf{\color{eLifeMediumGrey} Meister M}, Berry MJ.
\newblock The neural code of the retina.
\newblock Neuron.  1999 Mar; 22(3):435--450.

\bibitem[{Nagel et~al.(2005)Nagel, Georg and Brauner, Martin and Liewald, Jana
  F. and Adeishvili, Nona and Bamberg, Ernst and Gottschalk,
  Alexander}]{nagel_light_2005}
\textbf{\color{eLifeMediumGrey} Nagel G}, Brauner M, Liewald JF, Adeishvili N,
  Bamberg E, Gottschalk A.
\newblock Light {Activation} of {Channelrhodopsin}-2 in {Excitable} {Cells} of
  {Caenorhabditis} elegans {Triggers} {Rapid} {Behavioral} {Responses}.
\newblock Current Biology.  2005 Dec; 15(24):2279--2284.
\newblock
  \urlprefix\url{http://www.cell.com/current-biology/abstract/S0960-9822(05)01407-7},
  \href{10.1016/j.cub.2005.11.032}{\doiprefix 10.1016/j.cub.2005.11.032}.

\bibitem[{Nagy et~al.(2014{\natexlab{a}})Nagy, Stanislav and Raizen, David M.
  and Biron, David}]{nagy_measurements_2014}
\textbf{\color{eLifeMediumGrey} Nagy S}, Raizen DM, Biron D.
\newblock Measurements of behavioral quiescence in {Caenorhabditis} elegans.
\newblock Methods.  2014 Aug; 68(3):500--507.
\newblock
  \urlprefix\url{http://www.sciencedirect.com/science/article/pii/S1046202314001029},
  \href{10.1016/j.ymeth.2014.03.009}{\doiprefix 10.1016/j.ymeth.2014.03.009}.

\bibitem[{Nagy et~al.(2014{\natexlab{b}})Nagy, Stanislav and Tramm, Nora and
  Sanders, Jarred and Iwanir, Shachar and Shirley, Ian A. and Levine, Erel and
  Biron, David}]{nagy_homeostasis_2014}
\textbf{\color{eLifeMediumGrey} Nagy S}, Tramm N, Sanders J, Iwanir S, Shirley
  IA, Levine E, Biron D.
\newblock Homeostasis in {C}. elegans sleep is characterized by two
  behaviorally and genetically distinct mechanisms.
\newblock eLife.  2014 Dec; 3:e04380.
\newblock \urlprefix\url{https://elifesciences.org/articles/04380},
  \href{10.7554/eLife.04380}{\doiprefix 10.7554/eLife.04380}.

\bibitem[{Nekimken et~al.(2017)Nekimken, Adam L. and Mazzochette, Eileen A. and
  Goodman, Miriam B. and Pruitt, Beth L.}]{nekimken_forces_2017}
\textbf{\color{eLifeMediumGrey} Nekimken AL}, Mazzochette EA, Goodman MB,
  Pruitt BL.
\newblock Forces applied during classical touch assays for {Caenorhabditis}
  elegans.
\newblock PLOS ONE.  2017 May; 12(5):e0178080.
\newblock
  \urlprefix\url{http://journals.plos.org/plosone/article?id=10.1371/journal.pone.0178080},
  \href{10.1371/journal.pone.0178080}{\doiprefix 10.1371/journal.pone.0178080}.

\bibitem[{Nguyen et~al.(2017)Nguyen, Jeffrey P. and Linder, Ashley N. and
  Plummer, George S. and Shaevitz, Joshua W. and Leifer, Andrew
  M.}]{nguyen_automatically_2017}
\textbf{\color{eLifeMediumGrey} Nguyen JP}, Linder AN, Plummer GS, Shaevitz JW,
  Leifer AM.
\newblock Automatically tracking neurons in a moving and deforming brain.
\newblock PLOS Computational Biology.  2017 May; 13(5):e1005517.
\newblock
  \urlprefix\url{http://journals.plos.org/ploscompbiol/article?id=10.1371/journal.pcbi.1005517},
  \href{10.1371/journal.pcbi.1005517}{\doiprefix 10.1371/journal.pcbi.1005517}.

\bibitem[{O'Hagan et~al.(2005)O'Hagan, Robert and Chalfie, Martin and Goodman,
  Miriam B.}]{ohagan_mec-4_2005}
\textbf{\color{eLifeMediumGrey} O'Hagan R}, Chalfie M, Goodman MB.
\newblock The {MEC}-4 {DEG}/{ENaC} channel of {Caenorhabditis} elegans touch
  receptor neurons transduces mechanical signals.
\newblock Nature Neuroscience.  2005 Jan; 8(1):43--50.
\newblock
  \urlprefix\url{http://www.nature.com/neuro/journal/v8/n1/full/nn1362.html},
  \href{10.1038/nn1362}{\doiprefix 10.1038/nn1362}.

\bibitem[{Petzold et~al.(2013)Petzold, Bryan C. and Park, Sung-Jin and
  Mazzochette, Eileen A. and Goodman, Miriam B. and Pruitt, Beth
  L.}]{petzold_mems-based_2013}
\textbf{\color{eLifeMediumGrey} Petzold BC}, Park SJ, Mazzochette EA, Goodman
  MB, Pruitt BL.
\newblock {MEMS}-based force-clamp analysis of the role of body stiffness in
  {C}. elegans touch sensation.
\newblock Integrative Biology.  2013 May; 5(6):853--864.
\newblock
  \urlprefix\url{http://pubs.rsc.org/en/content/articlelanding/2013/ib/c3ib20293c},
  \href{10.1039/C3IB20293C}{\doiprefix 10.1039/C3IB20293C}.

\bibitem[{Pirri and Alkema(2012)Pirri, Jennifer K and Alkema, Mark
  J}]{pirri_neuroethology_2012}
\textbf{\color{eLifeMediumGrey} Pirri JK}, Alkema MJ.
\newblock The neuroethology of {C}. elegans escape.
\newblock Current Opinion in Neurobiology.  2012 Apr; 22(2):187--193.
\newblock
  \urlprefix\url{http://www.sciencedirect.com/science/article/pii/S095943881100225X},
  \href{10.1016/j.conb.2011.12.007}{\doiprefix 10.1016/j.conb.2011.12.007}.

\bibitem[{Porto et~al.(2017)Porto, Daniel A. and Giblin, John and Zhao, Yiran
  and Lu, Hang}]{porto_reverse-correlation_2017}
\textbf{\color{eLifeMediumGrey} Porto DA}, Giblin J, Zhao Y, Lu H.
\newblock Reverse-{Correlation} {Analysis} of {Mechanosensation} {Circuit} in
  {C}. elegans {Reveals} {Temporal} and {Spatial} {Encoding}.
\newblock bioRxiv.  2017 Aug;
  \urlprefix\url{http://biorxiv.org/content/early/2017/08/07/147363.abstract},
  \href{10.1101/147363}{\doiprefix 10.1101/147363}.

\bibitem[{Ramot et~al.(2008)Ramot, Daniel and Johnson, Brandon E. and Berry,
  Jr, Tommie L. and Carnell, Lucinda and Goodman, Miriam
  B.}]{ramot_parallel_2008}
\textbf{\color{eLifeMediumGrey} Ramot D}, Johnson BE, Berry TL Jr, Carnell L,
  Goodman MB.
\newblock The {Parallel} {Worm} {Tracker}: {A} {Platform} for {Measuring}
  {Average} {Speed} and {Drug}-{Induced} {Paralysis} in {Nematodes}.
\newblock PLoS ONE.  2008 May; 3(5):e2208.
\newblock \urlprefix\url{http://dx.plos.org/10.1371/journal.pone.0002208},
  \href{10.1371/journal.pone.0002208}{\doiprefix 10.1371/journal.pone.0002208}.

\bibitem[{Rankin and Wicks(2000)Rankin, Catharine H. and Wicks, Stephen
  R.}]{rankin_mutations_2000}
\textbf{\color{eLifeMediumGrey} Rankin CH}, Wicks SR.
\newblock Mutations of the {Caenorhabditis} {elegansBrain}-{Specific}
  {Inorganic} {Phosphate} {Transporter} eat-4Affect {Habituation} of the
  {Tap}–{Withdrawal} {Response} without {Affecting} the {Response} {Itself}.
\newblock Journal of Neuroscience.  2000 Jun; 20(11):4337--4344.
\newblock \urlprefix\url{http://www.jneurosci.org/content/20/11/4337}.

\bibitem[{Rankin et~al.(1990)Rankin, Catherine H. and Beck, Christine D. O. and
  Chiba, Catherine M.}]{rankin_caenorhabditis_1990}
\textbf{\color{eLifeMediumGrey} Rankin CH}, Beck CDO, Chiba CM.
\newblock Caenorhabditis elegans: {A} new model system for the study of
  learning and memory.
\newblock Behavioural Brain Research.  1990 Feb; 37(1):89--92.
\newblock
  \urlprefix\url{http://www.sciencedirect.com/science/article/pii/016643289090074O},
  \href{10.1016/0166-4328(90)90074-O}{\doiprefix 10.1016/0166-4328(90)90074-O}.

\bibitem[{Ringach and Shapley(2004)Ringach, Dario and Shapley,
  Robert}]{ringach_reverse_2004}
\textbf{\color{eLifeMediumGrey} Ringach D}, Shapley R.
\newblock Reverse correlation in neurophysiology.
\newblock Cognitive Science.  2004 Mar; 28(2):147--166.
\newblock
  \urlprefix\url{http://www.sciencedirect.com/science/article/pii/S0364021303001174},
  \href{10.1016/j.cogsci.2003.11.003}{\doiprefix 10.1016/j.cogsci.2003.11.003}.

\bibitem[{Sanyal et~al.(2004)Sanyal, Suparna and Wintle, Richard F. and Kindt,
  Katie S. and Nuttley, William M. and Arvan, Rokhand and Fitzmaurice, Paul and
  Bigras, Eve and Merz, David C. and Hébert, Terence E. and Kooy, Derek van
  der and Schafer, William R. and Culotti, Joseph G. and Tol, Hubert H. M.
  Van}]{sanyal_dopamine_2004}
\textbf{\color{eLifeMediumGrey} Sanyal S}, Wintle RF, Kindt KS, Nuttley WM,
  Arvan R, Fitzmaurice P, Bigras E, Merz DC, Hébert TE, Kooy Dvd, Schafer WR,
  Culotti JG, Tol HHMV.
\newblock Dopamine modulates the plasticity of mechanosensory responses in
  {Caenorhabditis} elegans.
\newblock The EMBO Journal.  2004 Jan; 23(2):473--482.
\newblock \urlprefix\url{http://emboj.embopress.org/content/23/2/473},
  \href{10.1038/sj.emboj.7600057}{\doiprefix 10.1038/sj.emboj.7600057}.

\bibitem[{Schwartz et~al.(2006)Schwartz, Odelia and Pillow, Jonathan W. and
  Rust, Nicole C. and Simoncelli, Eero P.}]{schwartz_spike-triggered_2006}
\textbf{\color{eLifeMediumGrey} Schwartz O}, Pillow JW, Rust NC, Simoncelli EP.
\newblock Spike-triggered neural characterization.
\newblock Journal of Vision.  2006 Jul; 6(4):484--507.
\newblock \href{10.1167/6.4.13}{\doiprefix 10.1167/6.4.13}.

\bibitem[{Schwarz et~al.(2011)Schwarz, Juliane and Lewandrowski, Ines and
  Bringmann, Henrik}]{schwarz_reduced_2011}
\textbf{\color{eLifeMediumGrey} Schwarz J}, Lewandrowski I, Bringmann H.
\newblock Reduced activity of a sensory neuron during a sleep-like state in
  {Caenorhabditis} elegans.
\newblock Current Biology.  2011 Dec; 21(24):R983--R984.
\newblock
  \urlprefix\url{http://www.sciencedirect.com/science/article/pii/S0960982211012073},
  \href{10.1016/j.cub.2011.10.046}{\doiprefix 10.1016/j.cub.2011.10.046}.

\bibitem[{Stephens et~al.(2008)Stephens, Greg J. and Johnson-Kerner, Bethany
  and Bialek, William and Ryu, William S.}]{stephens_dimensionality_2008}
\textbf{\color{eLifeMediumGrey} Stephens GJ}, Johnson-Kerner B, Bialek W, Ryu
  WS.
\newblock Dimensionality and {Dynamics} in the {Behavior} of {C}. elegans.
\newblock PLoS Computational Biology.  2008 Apr; 4(4):e1000028.
\newblock \urlprefix\url{http://dx.doi.org/10.1371%2Fjournal.pcbi.1000028},
  \href{10.1371/journal.pcbi.1000028}{\doiprefix 10.1371/journal.pcbi.1000028}.

\bibitem[{Stirman et~al.(2011)Stirman, Jeffrey N and Crane, Matthew M and
  Husson, Steven J and Wabnig, Sebastian and Schultheis, Christian and
  Gottschalk, Alexander and Lu, Hang}]{stirman_real-time_2011}
\textbf{\color{eLifeMediumGrey} Stirman JN}, Crane MM, Husson SJ, Wabnig S,
  Schultheis C, Gottschalk A, Lu H.
\newblock Real-time multimodal optical control of neurons and muscles in freely
  behaving {Caenorhabditis} elegans.
\newblock Nature methods.  2011 Feb; 8(2):153--158.
\newblock \urlprefix\url{http://www.ncbi.nlm.nih.gov/pubmed/21240278},
  \href{10.1038/nmeth.1555}{\doiprefix 10.1038/nmeth.1555}.


\bibitem[{Swierczek et~al.(2011)Swierczek, Nicholas A. and Giles, Andrew C. and
  Rankin, Catharine H. and Kerr, Rex A.}]{swierczek_high-throughput_2011}
\textbf{\color{eLifeMediumGrey} Swierczek NA}, Giles AC, Rankin CH, Kerr RA.
\newblock High-throughput behavioral analysis in \textit{{C}. elegans}.
\newblock Nature Methods.  2011 Jun; 8(7):592.
\newblock \urlprefix\url{https://www.nature.com/articles/nmeth.1625},
  \href{10.1038/nmeth.1625}{\doiprefix 10.1038/nmeth.1625}.

\bibitem[{Timbers et~al.(2013)Timbers, Tiffany A. and Giles, Andrew C. and
  Ardiel, Evan L. and Kerr, Rex A. and Rankin, Catharine
  H.}]{timbers_intensity_2013}
\textbf{\color{eLifeMediumGrey} Timbers TA}, Giles AC, Ardiel EL, Kerr RA,
  Rankin CH.
\newblock Intensity discrimination deficits cause habituation changes in
  middle-aged {Caenorhabditis} elegans.
\newblock Neurobiology of Aging.  2013 Feb; 34(2):621--631.
\newblock
  \urlprefix\url{http://www.sciencedirect.com/science/article/pii/S0197458012002291},
  \href{10.1016/j.neurobiolaging.2012.03.016}{\doiprefix
  10.1016/j.neurobiolaging.2012.03.016}.

\bibitem[{Wicks and Rankin(1995)Wicks, S. R. and Rankin, C.
  H.}]{wicks_integration_1995}
\textbf{\color{eLifeMediumGrey} Wicks SR}, Rankin CH.
\newblock Integration of mechanosensory stimuli in {Caenorhabditis} elegans.
\newblock Journal of Neuroscience.  1995 Mar; 15(3):2434--2444.
\newblock \urlprefix\url{http://www.jneurosci.org/content/15/3/2434}.

\bibitem[{Wiltschko et~al.(2015)Wiltschko, Alexander B. and Johnson,
  Matthew J. and Iurilli, Giuliano and Peterson, Ralph E. and Katon,
  Jesse M. and Pashkovski, Stan L. and Abraira, Victoria E. and Adams,
  Ryan P. and Datta, Sandeep Robert}]{wiltschko_mapping_2015}
\textbf{\color{eLifeMediumGrey} Wiltschko A}, Johnson M, Iurilli G, Peterson R,
  Katon J, Pashkovski S, Abraira V, Adams R, Datta S.
\newblock Mapping {Sub}-{Second} {Structure} in {Mouse} {Behavior}.
\newblock Neuron.  2015 Dec; 88(6):1121--1135.
\newblock
  \urlprefix\url{http://www.sciencedirect.com/science/article/pii/S0896627315010375},
  \href{10.1016/j.neuron.2015.11.031}{\doiprefix 10.1016/j.neuron.2015.11.031}.

\bibitem[{Yemini et~al.(2013)Yemini, Eviatar and Jucikas, Tadas and Grundy,
  Laura J. and Brown, André E. X. and Schafer, William
  R.}]{yemini_database_2013}
\textbf{\color{eLifeMediumGrey} Yemini E}, Jucikas T, Grundy LJ, Brown AEX,
  Schafer WR.
\newblock A database of \textit{{Caenorhabditis} elegans} behavioral
  phenotypes.
\newblock Nature Methods.  2013 Jul; 10(9):877.
\newblock \urlprefix\url{https://www.nature.com/articles/nmeth.2560},
  \href{10.1038/nmeth.2560}{\doiprefix 10.1038/nmeth.2560}.

\bibitem[{Yilmaz and Meister(2013)Yilmaz, Melis and Meister,
  Markus}]{yilmaz_rapid_2013}
\textbf{\color{eLifeMediumGrey} Yilmaz M}, Meister M.
\newblock Rapid {Innate} {Defensive} {Responses} of {Mice} to {Looming}
  {Visual} {Stimuli}.
\newblock Current Biology.  2013 Oct; 23(20):2011--2015.
\newblock
  \urlprefix\url{http://www.sciencedirect.com/science/article/pii/S0960982213009913},
  \href{10.1016/j.cub.2013.08.015}{\doiprefix 10.1016/j.cub.2013.08.015}.

\end{thebibliography}

\end{document}